\documentclass[epsfig]{elsart}
\usepackage{amsfonts,amssymb,graphicx}
\begin{document}

\newcommand{\re}{\mathop{\mathrm{Re}}}
\newcommand{\im}{\mathop{\mathrm{Im}}}
\newcommand{\D}{\mathop{\mathrm{d}}}
\newcommand{\I}{\mathop{\mathrm{i}}}
\newcommand{\E}{\mathop{\mathrm{e}}}

\noindent {\Large DESY 05-239}

\noindent {\Large November 2005}

\bigskip

\begin{frontmatter}

\journal{NIM A}
\date{}

\title{Statistical properties of the radiation from VUV FEL at DESY
operating at 30~nm wavelength in the femtosecond regime}

\author{E.L.~Saldin},
\author{E.A.~Schneidmiller},
and
\author{M.V.~Yurkov}

\address{Deutsches Elektronen-Synchrotron (DESY),
Hamburg, Germany}

\begin{abstract}

Since spring, 2005 vacuum ultraviolet free electron laser (VUV FEL) at
DESY operates as user facility in the wavelength range around 30~nm.
Electron beam formation system at the VUV FEL is essentially nonlinear
and naturally results in a formation of a short high-current leading
peak in the density distribution that produces FEL radiation. The main
feature of the femtosecond mode of operation of the VUV FEL is the
production of short, down to 20~fs radiation pulses with GW-level peak
power. This paper is devoted to detailed studies of the short-pulse
effects in the VUV FEL at DESY operating in the femtosecond regime, and
to analysis of first experimental results.

\end{abstract}

\end{frontmatter}

\clearpage
\setcounter{page}{1}

\section{Introduction}

The project of the VUV FEL at DESY is realized in two phases. Phase I
(1999-2002) served for a proof-of-principle of SASE FEL operation and
for system tests of the hardware. Phase II of the VUV FEL has been
built as an extension of phase I to shorter wavelengths (down to 6~nm)
and is used as the first VUV FEL user facility starting in spring,
2005. The nominal design of the VUV FEL assumes a linearized
compression of the electron bunch in two bunch compressors with the
help of the 3rd harmonic superconducting RF cavity \cite{ttf2-update}.
This would allow to prepare electron bunch with relatively long lasing
part and to reach original specifications for the output radiation in
terms of pulse duration, of about 200~fs FWHM. Longer pulse duration is
required for user applications exploiting high average brilliance, or
photon flux. On the other hand, there is a strong interest of the user
community in shorter radiation pulses for time-resolved experiments.

VUV FEL Phase I demonstrated unique femtosecond mode of operation
\cite{ttf-sat-prl,ttf-sat-epj} which was not considered at an early
design stage of the project \cite{design-rep-ttf}. Thorough analysis
has shown that due to nonlinear compression and small local energy
spread the short high-current (3~kA) leading peak (spike) in the bunch
density distribution was produced by beam formation system. Despite
strong collective effects (of which the most critical was the
longitudinal space charge after compression) this spike was bright
enough to drive FEL process up to the saturation for the wavelengths
around 100 nm \cite{s2e-nim}. In addition to the possibility for
production of high-power femtosecond pulses this mode of FEL operation
demonstrated high stability with respect to drifts of machine
parameters.

Successful operation of the VUV FEL Phase I in the femtosecond regime
encouraged us to extend such a mode of operation for shorter
wavelengths. Relevant theoretical study has been performed in
\cite{vuvfel-th}. It has been found that the beam formation system of
the linac can be tuned for production of bunches with a
high-peak-current spike capable for effective driving of the FEL
process such that the VUV FEL can safely saturate even at the shortest
design wavelength of 6 nm with a GW level of the peak power in short
pulses of 15-100 fs duration.

Optimum range of parameters suggested in \cite{vuvfel-th} has been
chosen for the commissioning of the VUV FEL. First experimental results
obtained at the VUV FEL operating at the radiation wavelength around
30~nm have shown good agreement with predictions \cite{vuvfel-exp}.
Commissioning of the VUV FEL proceeded in parallel with first user
experiments. Our contacts with user community shows that planning of
future user experiments at the VUV FEL requires more detailed knowledge
of the expected statistical properties of the radiation source, and
present paper fills this gap.

\section{Formation of the electron bunch}

\begin{figure}[b]

\begin{center}
\includegraphics[width=0.95\textwidth]{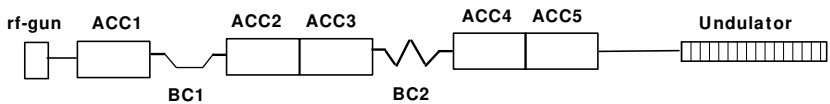}
\end{center}

\caption{
Schematic layout of the VUV FEL (status for year 2005).
Abbreviations ACC and BC stand for accelerating module and
bunch compressor, respectively
}
\label{fig:ttf2-2005}
\end{figure}

Schematic layout of the VUV FEL (status for the year 2005) is shown in
Fig.~\ref{fig:ttf2-2005}  \cite{ttf2-update}. The electron beam is
produced in a rf gun and accelerated up to 700 MeV by five cryomodules
ACC1-ACC5. At energy levels of 130 and 380 MeV the electron bunches are
compressed in the bunch compressors BC1 and BC2. The undulator is a
fixed 12 mm gap permanent magnet device, period length is 2.73~cm and
peak magnetic field is 0.486~T. The undulator system is subdivided into
six segments, each 4.5 m long. An important difference from the final
setup \cite{ttf2-update} is that the 3rd harmonic rf cavity is not
installed yet. Thus, beam formation system is essentially based on the
use of nonlinear compression. This naturally leads to formation of
asymmetrically shaped electron bunches with a sharp high current spike
on a 100~fs scale duration. At a proper optimization of the bunch
compression system it is possible to preserve small value of the
emittance in the current spike. Bunch formation system of the VUV FEL
operating in the femtosecond regime has been studied in details in
\cite{vuvfel-th}. We considered two possible options of operation: with
a nominal charge of 0.5~nC, and with higher charge, of 1~nC. Both
options can be realized experimentally and provide different modes of
the VUV FEL operation in terms of output characteristics of the
radiation.

\begin{figure}[tb]

\includegraphics[width=0.5\textwidth]{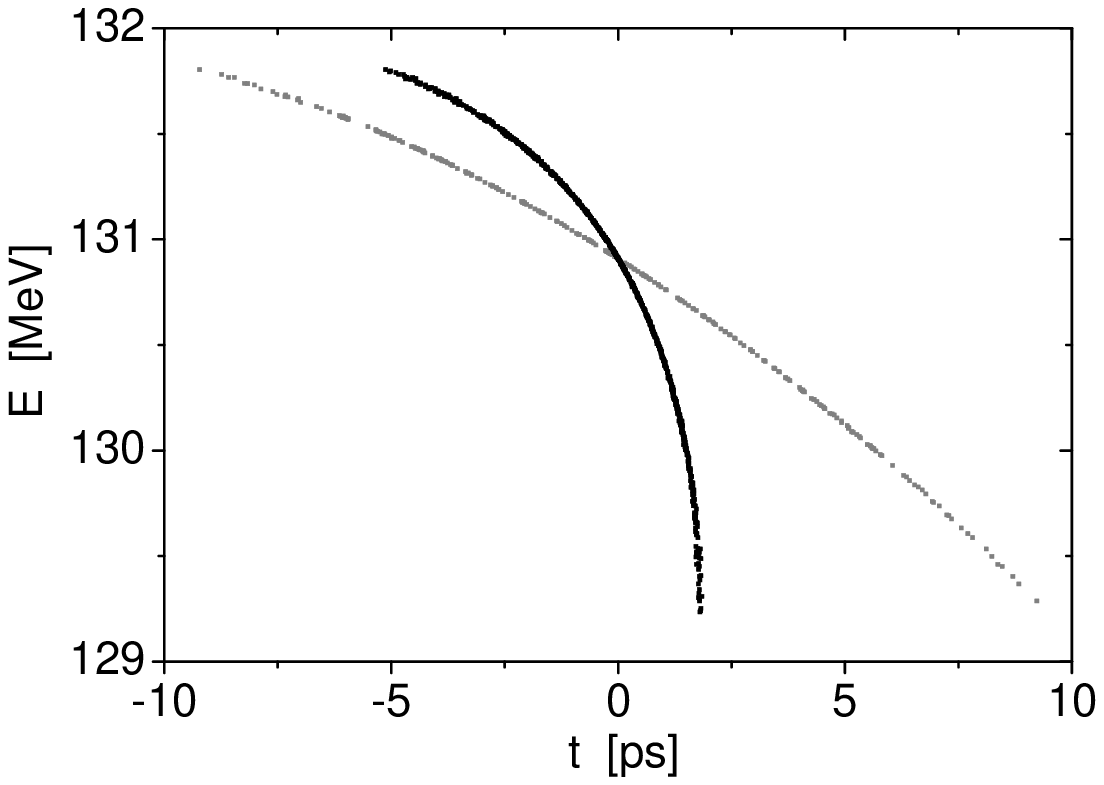}
\includegraphics[width=0.5\textwidth]{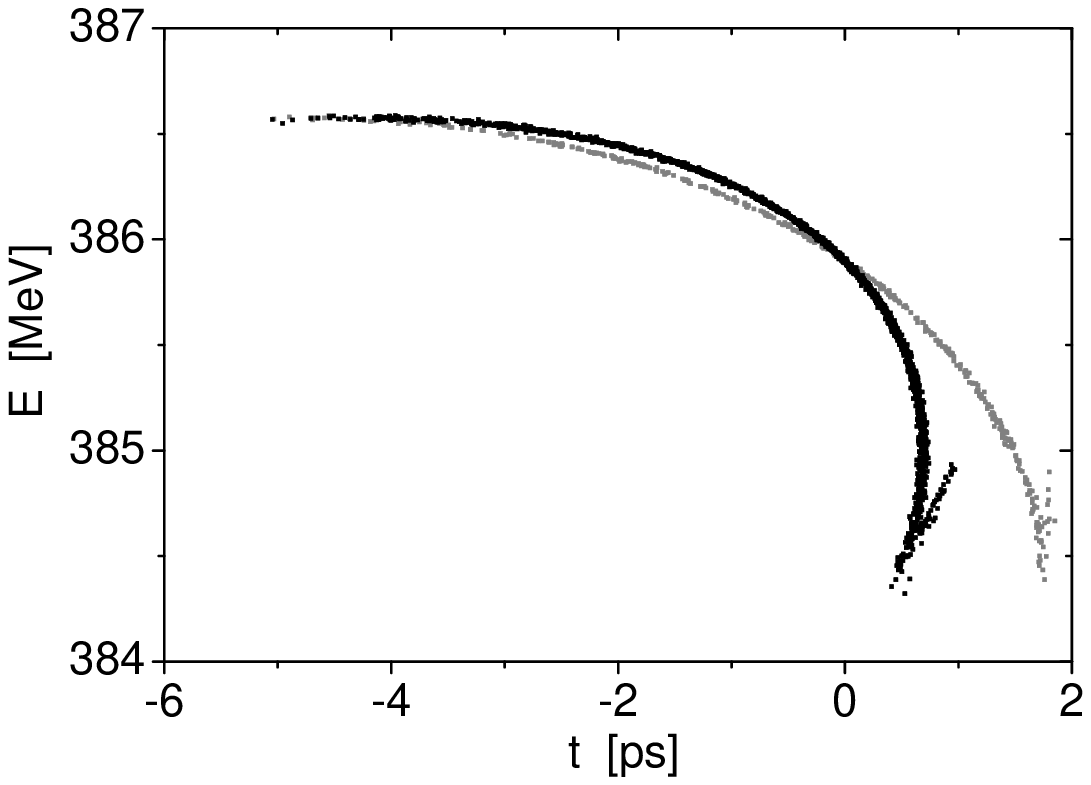}

\caption{
Schematic illustration of two-stage nonlinear bunch compression.
Phase space distribution of the electrons before
(grey dots) and after bunch compressors BC1 (left plot) and BC2 (right
plot).
Bunch head is at the right side
}
\label{fig:bunch-comp-ps}
\end{figure}

\begin{figure}[tb]

\includegraphics[width=0.32\textwidth]{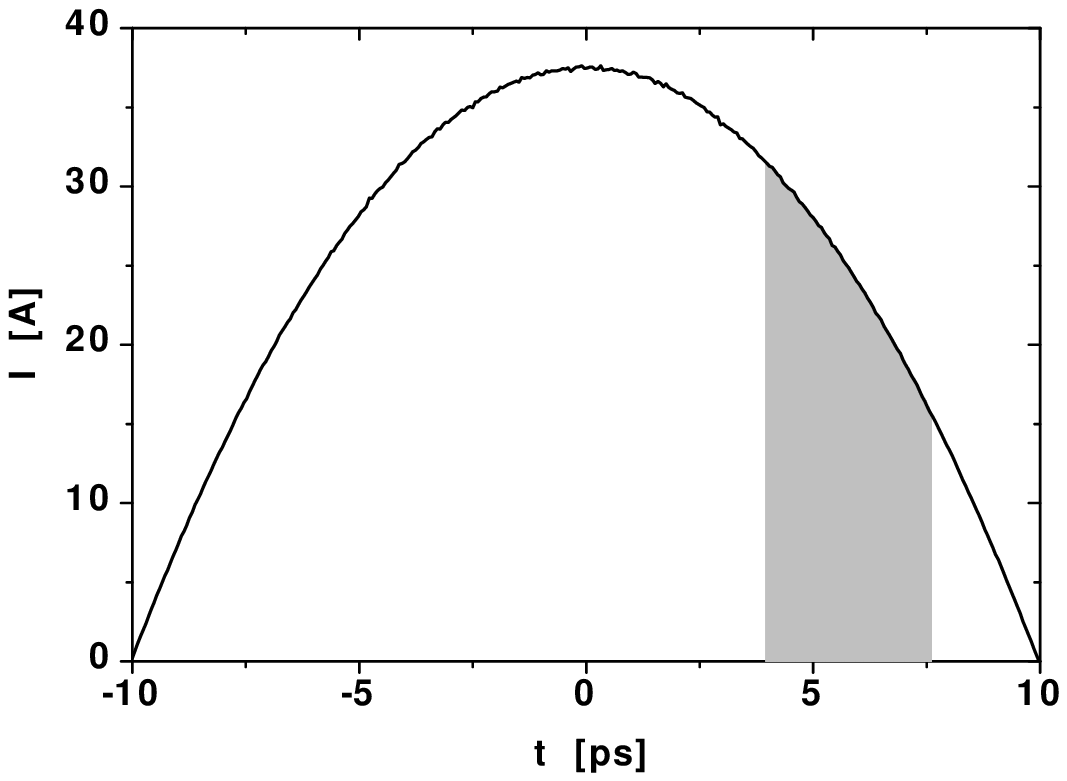}
\includegraphics[width=0.32\textwidth]{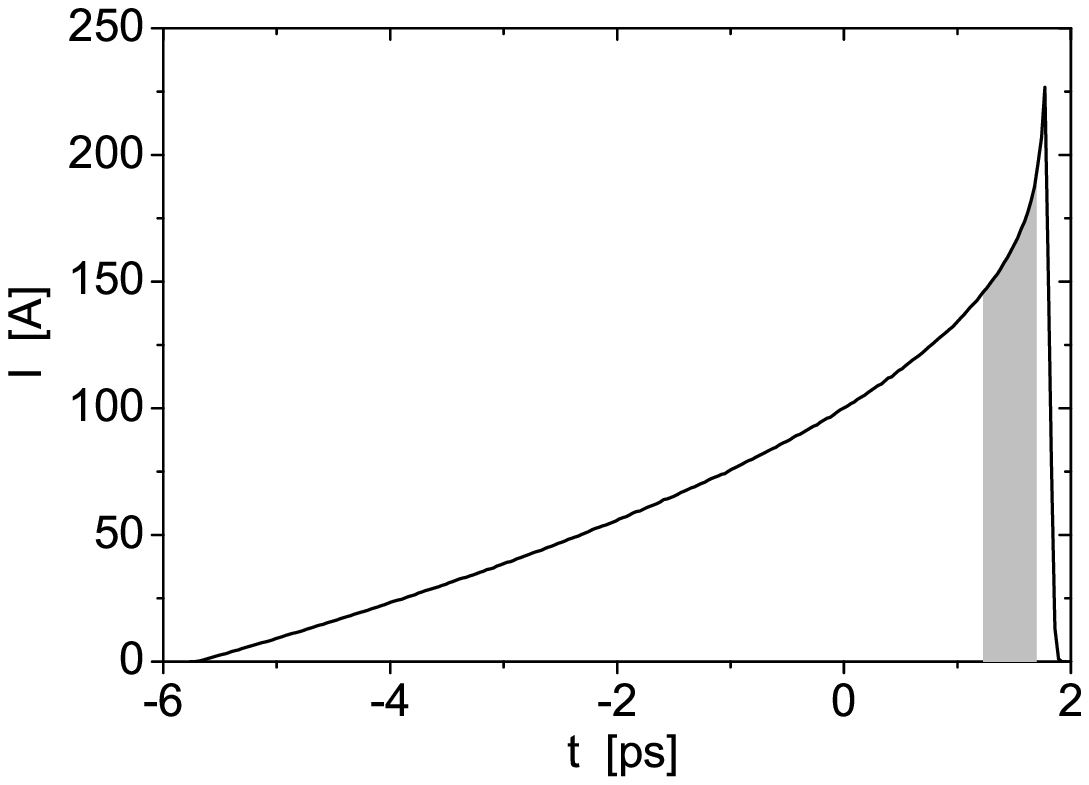}
\includegraphics[width=0.32\textwidth]{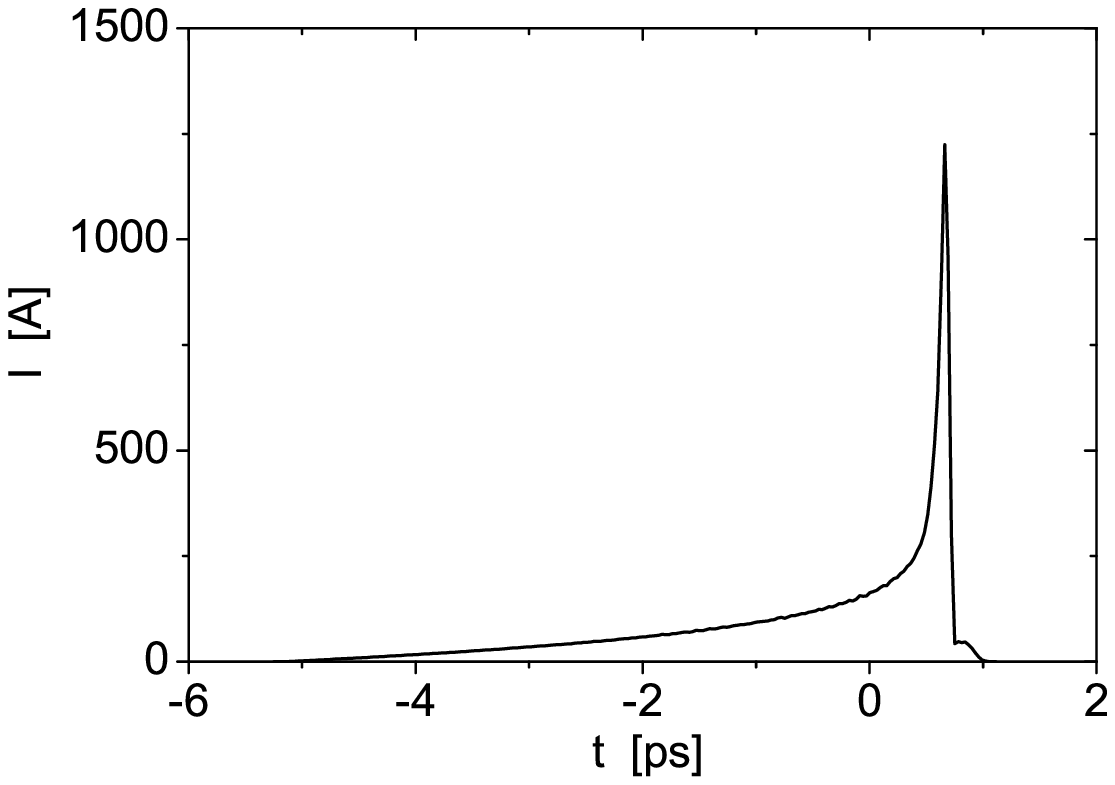}

\caption{
Schematic illustration of two-stage nonlinear bunch compression.
Current distribution along the electron bunch after
accelerating module ACC1, bunch compressors BC1, and BC2 (left, middle,
and right plot, respectively). Grey area on the left and middle
plots shows part of the bunch compressed to high current spike after
BC2 (right plot).
Bunch head is at the right side
}
\label{fig:bunch-comp-i}
\end{figure}

Operation of the nonlinear bunch compression system realized at the VUV
FEL is illustrated schematically with plots presented in
Figs.~\ref{fig:bunch-comp-ps} and \ref{fig:bunch-comp-i}. The electron
beam originating from the electron gun is relatively long ($1.5-2$~mm
rms) with respect to the rf-wave length (23~cm). When  the bunch is
accelerated off-crest in the accelerating module ACC1, longitudinal
phase space accumulates an rf-induced curvature. This distortion
results, downstream of the bunch compressor, in a non-Gaussian
distribution with a local charge concentration.

In fact, collective effects play significant role in the bunch
compression process for short pulses. For high-current part of the
bunch with rms length $\sigma_z$ and peak current $I$ coherent
synchrotron radiation (CSR) and longitudinal space charge (LSC) effects
scale as $\sigma_z^{-1/3}$ and $\sigma_z^{-1}$, respectively. For
instance, LSC-induced energy chirp along the high-current spike of
the bunch grows when the bunch travels from the bunch compressor to the
undulator entrance \cite{s2e-nim}:

\begin{equation}
\frac{d (\Delta \gamma)}{d z} \simeq \ 2.4 \ \frac{I}{I_{\mathrm{A}}} \
\frac{\ln (\gamma \sigma_z/\sigma_{\bot})}
{\sigma_z \gamma^2} \ ,
\label{eq:chirp}
\end{equation}

\noindent where $I_{\mathrm{A}} = 17 $kA is the Alfven current,
$\gamma$ the relativistic factor and $\sigma_{\bot}$ the rms transverse
size of the electron bunch.

\begin{figure}[tb]

\includegraphics[width=0.5\textwidth]{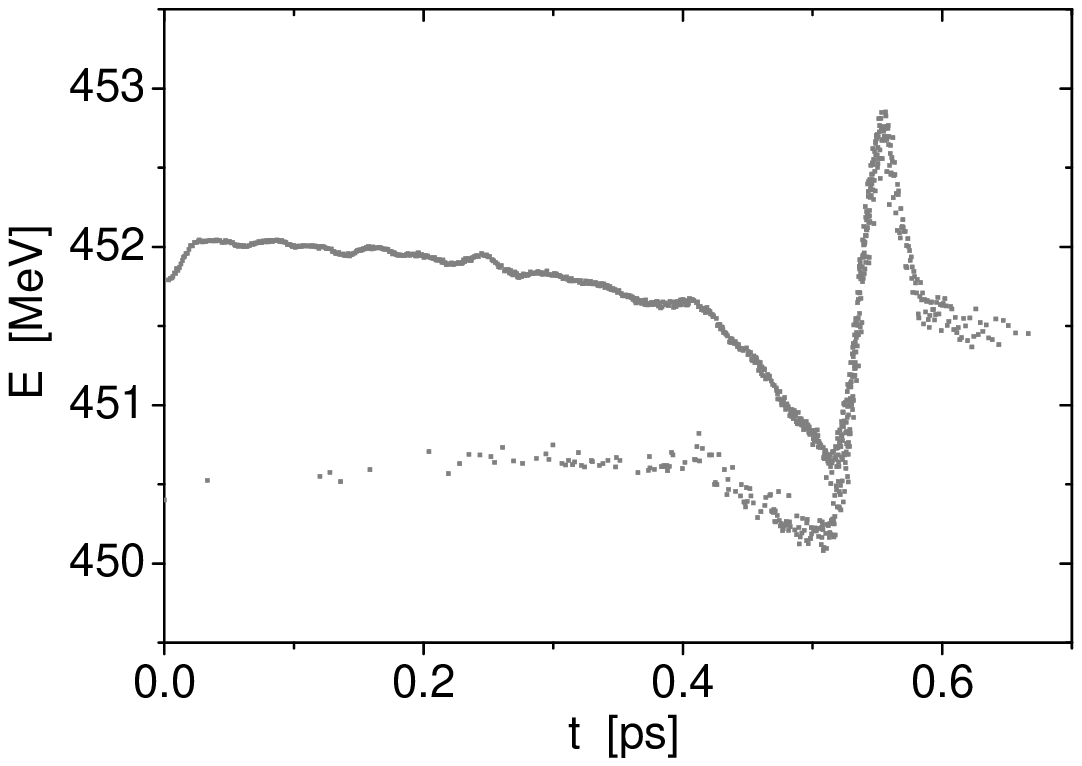}
\includegraphics[width=0.5\textwidth]{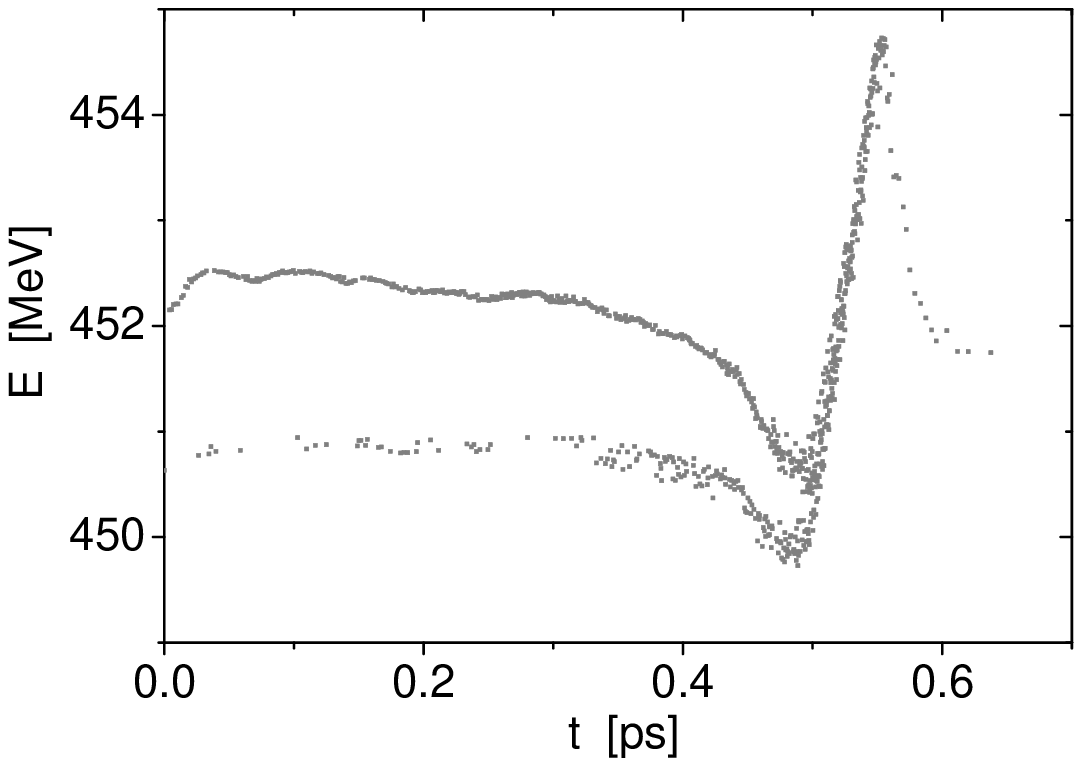}

\vspace*{-5mm}

\includegraphics[width=0.5\textwidth]{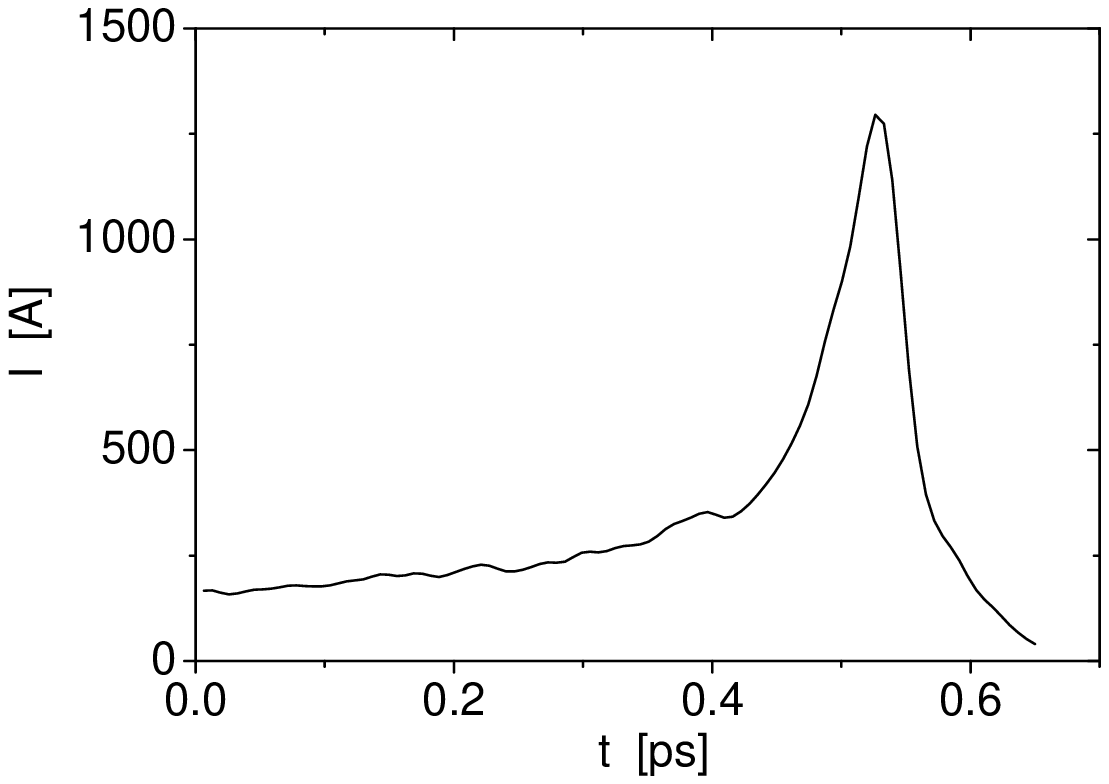}
\includegraphics[width=0.5\textwidth]{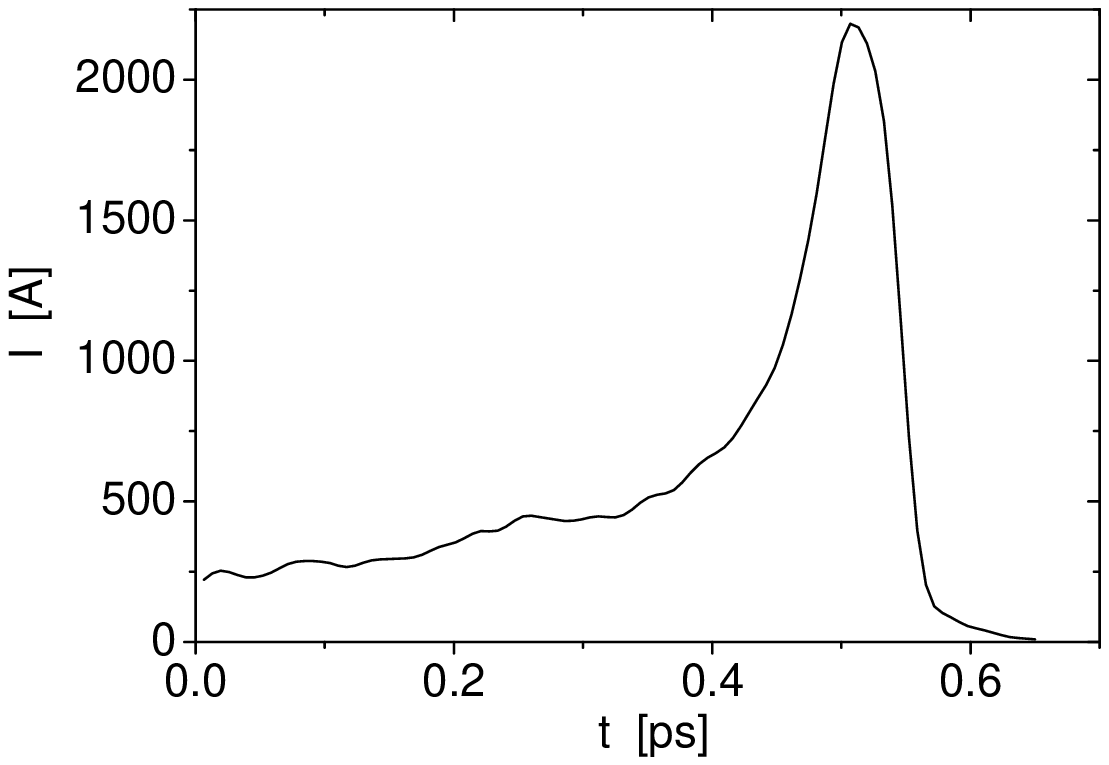}

\vspace*{-5mm}

\includegraphics[width=0.5\textwidth]{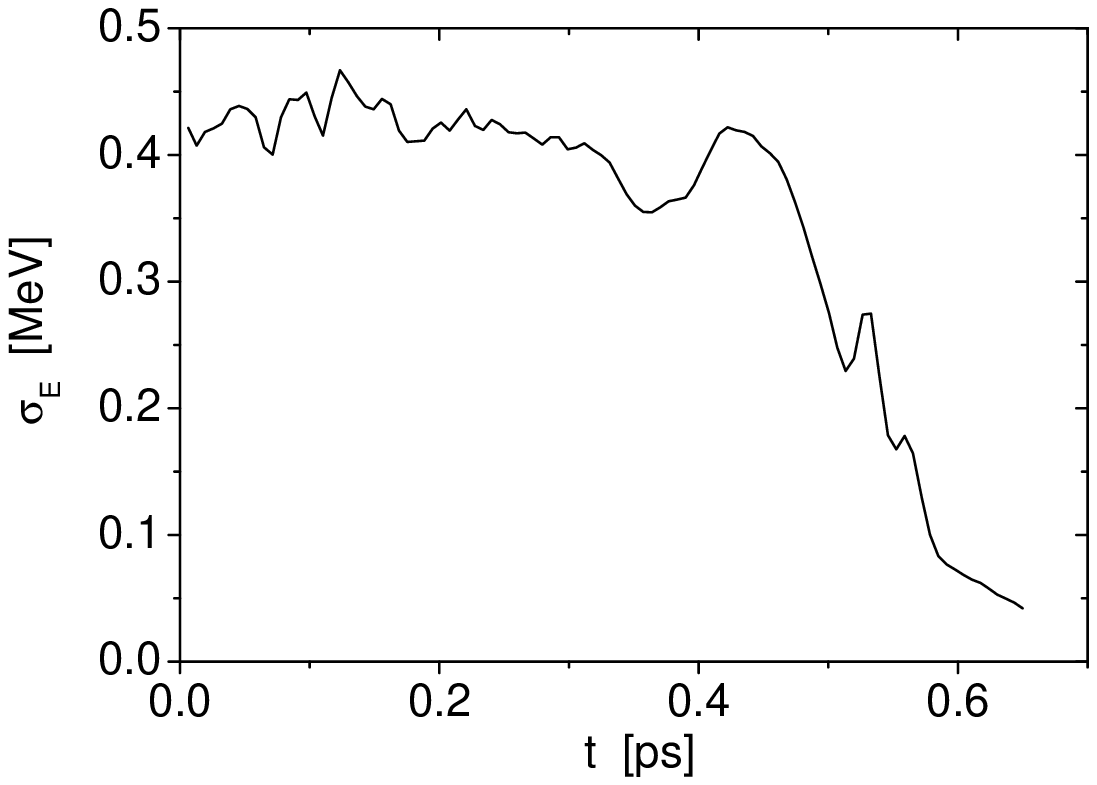}
\includegraphics[width=0.5\textwidth]{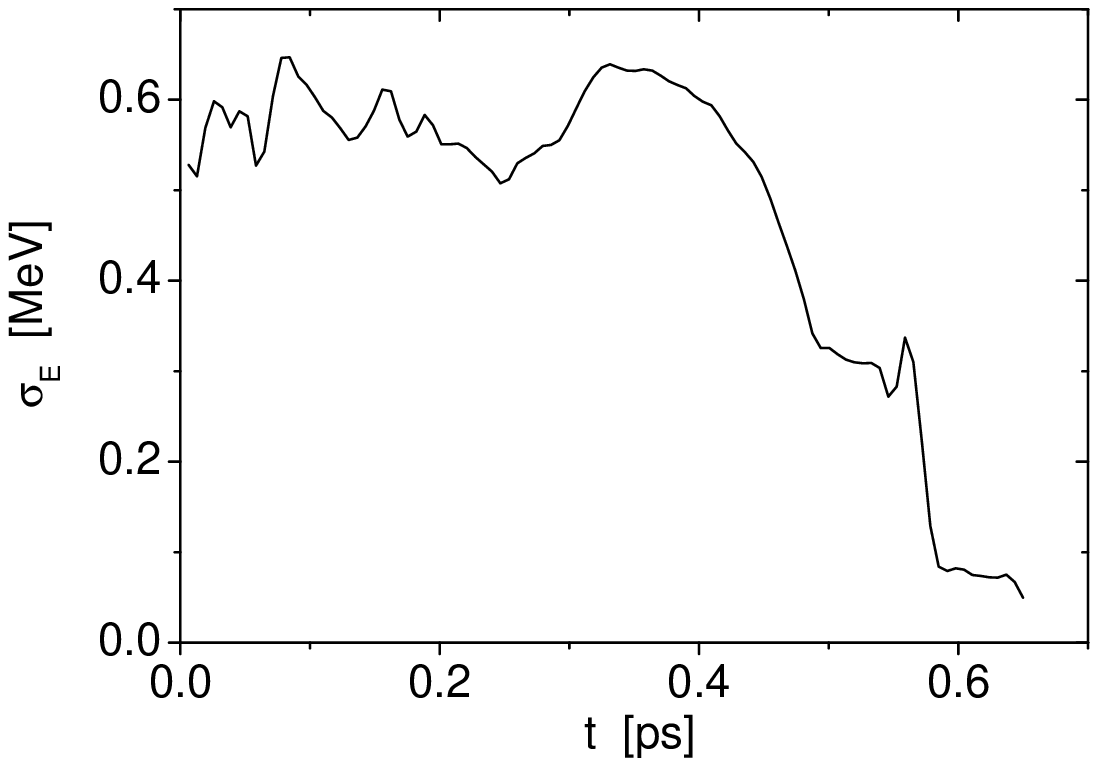}

\vspace*{-5mm}

\includegraphics[width=0.5\textwidth]{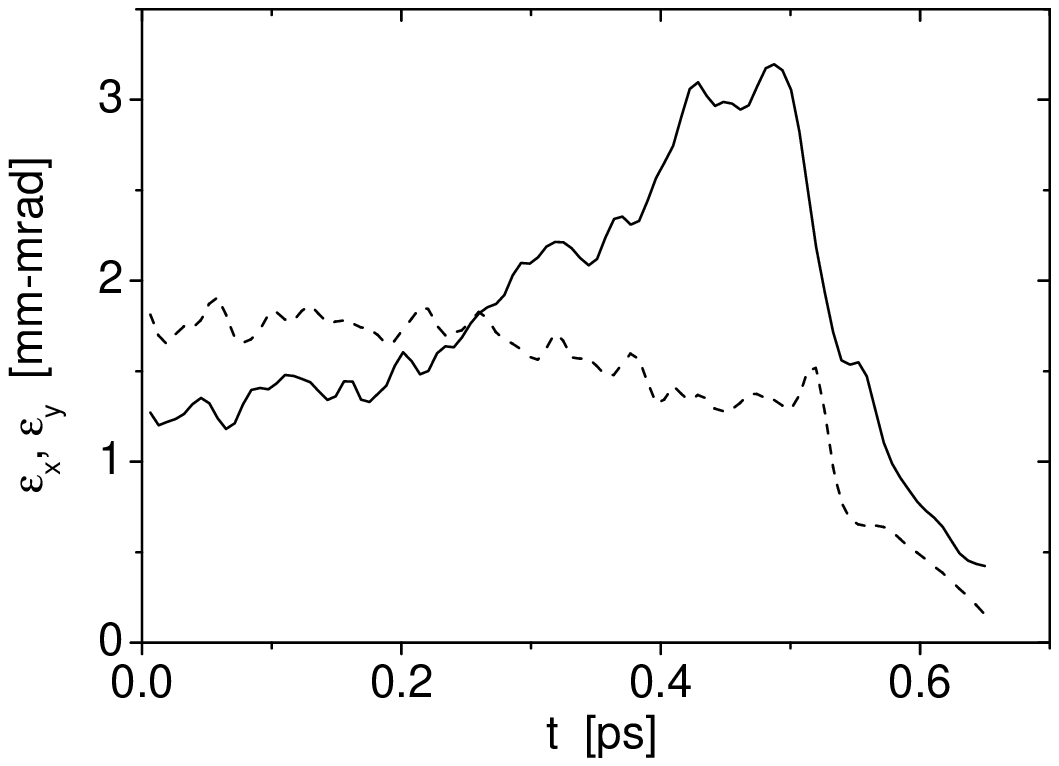}
\includegraphics[width=0.5\textwidth]{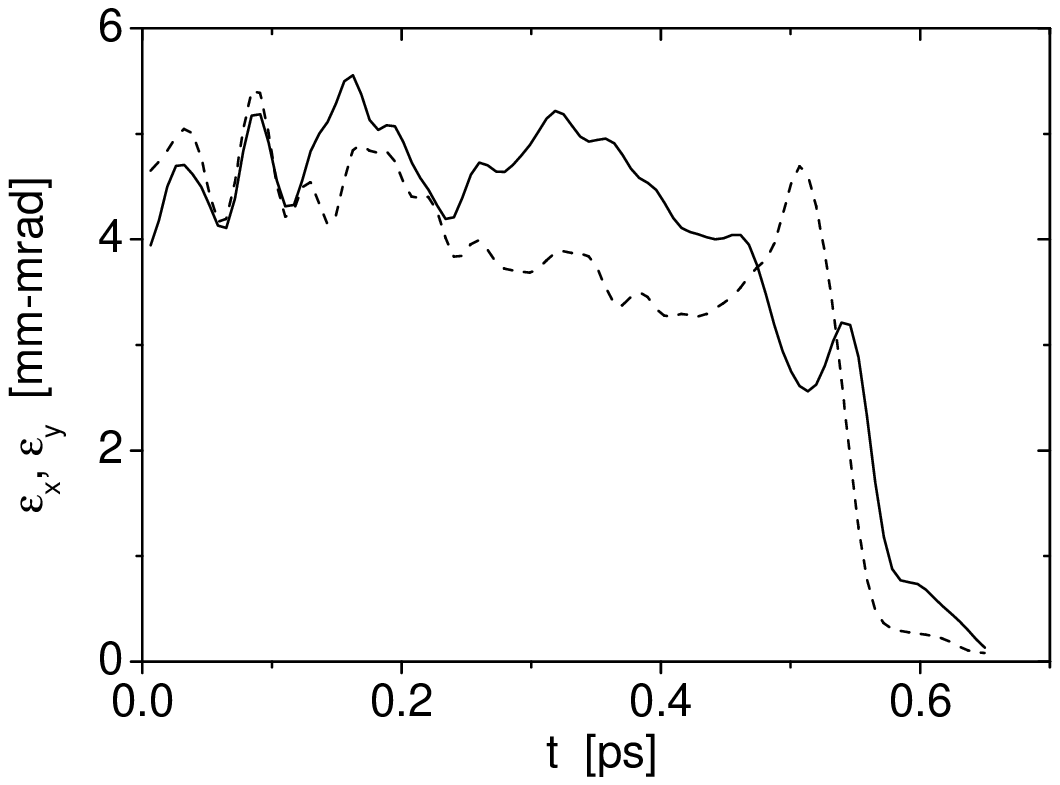}

\vspace*{-5mm}

\caption{Phase space distribution,
current, slice emittance and slice
energy spread along the bunch at the undulator entrance.
Left and right columns correspond to the of 0.5 and 1~nC,
respectively.
Bunch head is at the right side
}
\label{fig:bunch-ue}
\end{figure}

Target goal for the FEL is high value of peak current keeping slice
emittance and energy spread at a low level. Also, energy chirp along
the bunch should not be too large in order to prevent gain degradation
and increase of the spectrum width. In order to tolerate collective
effects, a procedure of two-stage compression has been developed which
allows to produce electron bunches with good lasing properties (see
Figs.~\ref{fig:bunch-comp-ps} and \ref{fig:bunch-comp-i})
\cite{vuvfel-th}. Calculations have been performed with the codes Astra
\cite{astra} and {\tt elegant} \cite{elegant}. Starting from peak
current of 30-50~A at the exit of accelerating module ACC1, a mild
compression is performed in the bunch compressor BC1, and then a high-current
spike (containing about $10-15$\% of the total charge) is formed in the bunch
compressor BC2. Combination of rf phases in accelerating modules ACC1, ACC2,
and ACC3 defines a slice in the initial distribution of which the spike is
formed (see Fig.~\ref{fig:bunch-comp-i}). Optimum parameters of the electron
bunch at the undulator entrance are shown in Fig.~\ref{fig:bunch-ue}. It is the
head of the bunch with a high peak current which is capable to drive the lasing
process. Complicated phase space distribution of the electrons in the head of
the bunch is mainly due to the action of the longitudinal space charge field.

\section{Parameter space of the VUV FEL}

General overview of expected output characteristics of the VUV FEL has
been presented in our previous paper \cite{vuvfel-th}. Here we perform
more detailed analysis of the features of the radiation related to
ultra-short pulse duration. Slice properties of the bunch at the
undulator entrance demonstrate rather complicated behavior (see
Fig.~\ref{fig:bunch-ue}). Evidently, a high current spike in the head
of the bunch has preference for light amplification because of higher
value of peak current and smaller emittance. Parameters of the spike
are as follows: peak current is $1-2$~kA, FWHM length of high
current spike is $20-30 \ \mu $m, and normalized emittance
$1.5-3.5$~mm-mrad for bunch charge $0.5-1$~nC.
Prior presentation of simulation results it is useful to perform
analysis of parameter space of the VUV FEL. As a zero order
approximation we use simple one-dimensional estimations in terms
of the FEL parameter $\rho $ \cite{bnp,book}:

\begin{displaymath}
\rho = \left[
\frac{I}{I_{\mathrm{A}}}
\frac{A_{JJ}^2 K^2 \lambda_{\mathrm{w}}^2}
{32 \pi ^2 \gamma ^2 \epsilon_n \beta _{\mathrm{f}} }
\right]^{1/3} \ .
\end{displaymath}

\noindent Here $\lambda_{\mathrm{w}}$ is undulator period, $K$ is rms
undulator parameter, $\gamma $ is relativistic factor, $I_{A} = mc^3/e
\simeq 17$~kA, $(-e)$ and $m$ are charge and mass of electron,
$\epsilon_n $ is normalized rms emittance, and $\beta _{\mathrm{f}}$ is
focusing beta function in the undulator. Coupling factor is $A_{JJ} =
[J_{0}(Q) - J_{1}(Q)]$, where $Q = K^2/[2(1+K^2)]$, and $K$ is rms
undulator parameter.
For focusing beta
function in the undulator of 4.5~m slice parameters in the current
spike result in the value of the FEL parameter $\rho = 2.5 - 3 \times
10^{-3}$. Thus, spectrum width generated by a slice of the electron
beam is expected to be about $\Delta \omega /\omega \simeq \rho \simeq
0.5$\%. Lasing part of the bunch has relatively strong energy chirp.
First consequence of the energy chirp is extra widening of the
radiation spectrum by $0.5-1$\% according to simple relation $\Delta
\omega /\omega = 2 \Delta E/E$. Another consequence of the energy chirp
is suppression of the FEL gain. The merit of the influence of the
energy chirp on the gain is $(\rho E/ \tau _{\mathrm{c}} )^{-1} \D E/\D
t$. It becomes to play significant role when relative energy change on
a scale of coherence length $c\tau _{\mathrm{c}}$ becomes to be
comparable with FEL parameter $\rho $ \cite{fel05-chirp}. This effect
is not small in the case under study, and results in significant
correction to the FEL gain within the leading current spike.
Suppression of the gain is stronger when slice current becomes to be
less. In the high gain linear regime the effect of energy chirp leads
to further shortening of the lasing part of the electron bunch.
Estimations show that with all effects taken into account the lasing
part of the electron bunch is only 3-5 times larger than coherence
length. As a result, we can expect specific features of the radiation
from SASE FEL driven by ultra-short electron bunch \cite{short-bunch}.
These features are: ultra-short radiation pulse duration (comparable
with coherence length), and strong suppression of the energy
fluctuations in the nonlinear regime.

We see that physical effects influencing operation of VUV FEL in the
femtosecond regime are rather transparent. However, quantitative
description of the output radiation can be obtained only with numerical
simulations. To extract more detailed information about radiation
properties, we performed 500 statistically independent runs with
three-dimensional, time-dependent FEL simulation code FAST \cite{fast}.
The result of each run contains parameters of the output radiation
(field and phase) stored on three-dimensional mesh. At the next stage
of the numerical experiment the data arrays are handled with
postprocessor codes to extract different properties of the SASE FEL
radiation such as averaged values, spectra, and statistical
distributions.

\section{Temporal structure of the radiation}

\begin{figure}[tb]

\includegraphics[width=0.5\textwidth]{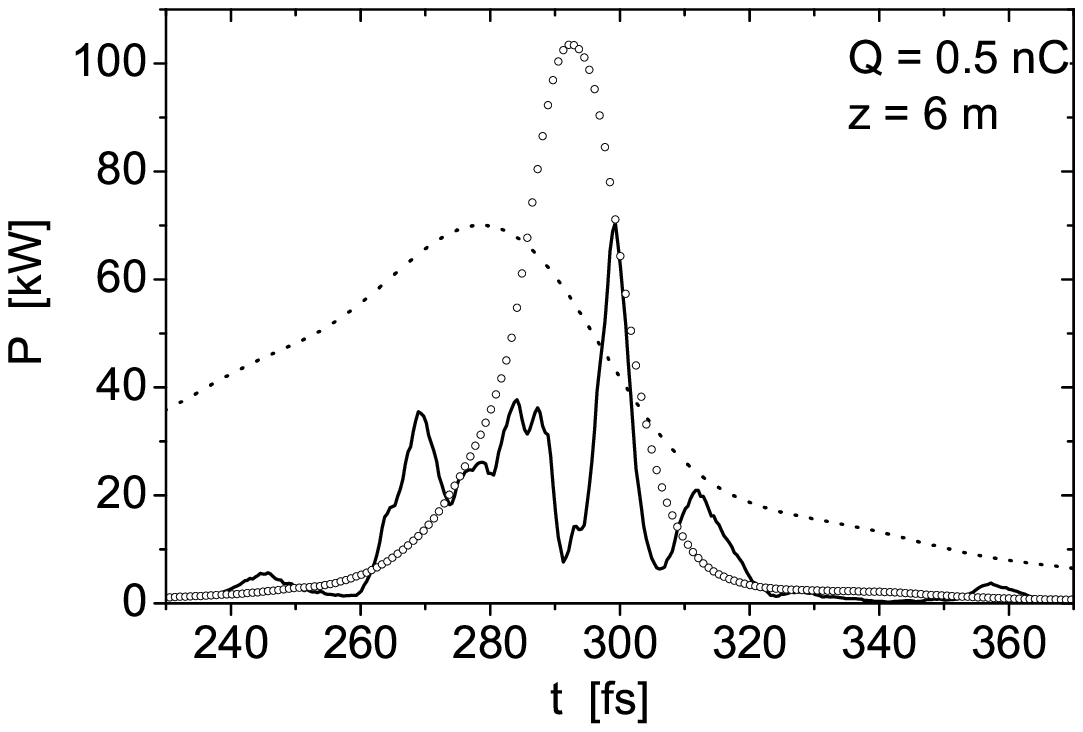}
\includegraphics[width=0.5\textwidth]{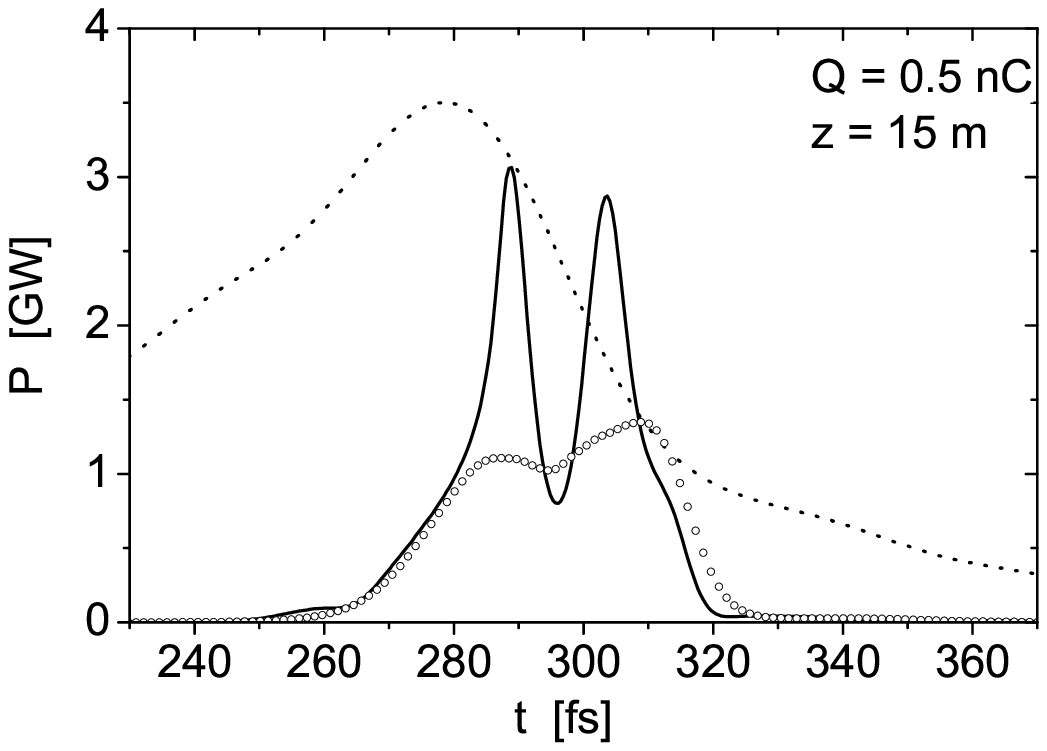}

\vspace*{-5mm}

\includegraphics[width=0.5\textwidth]{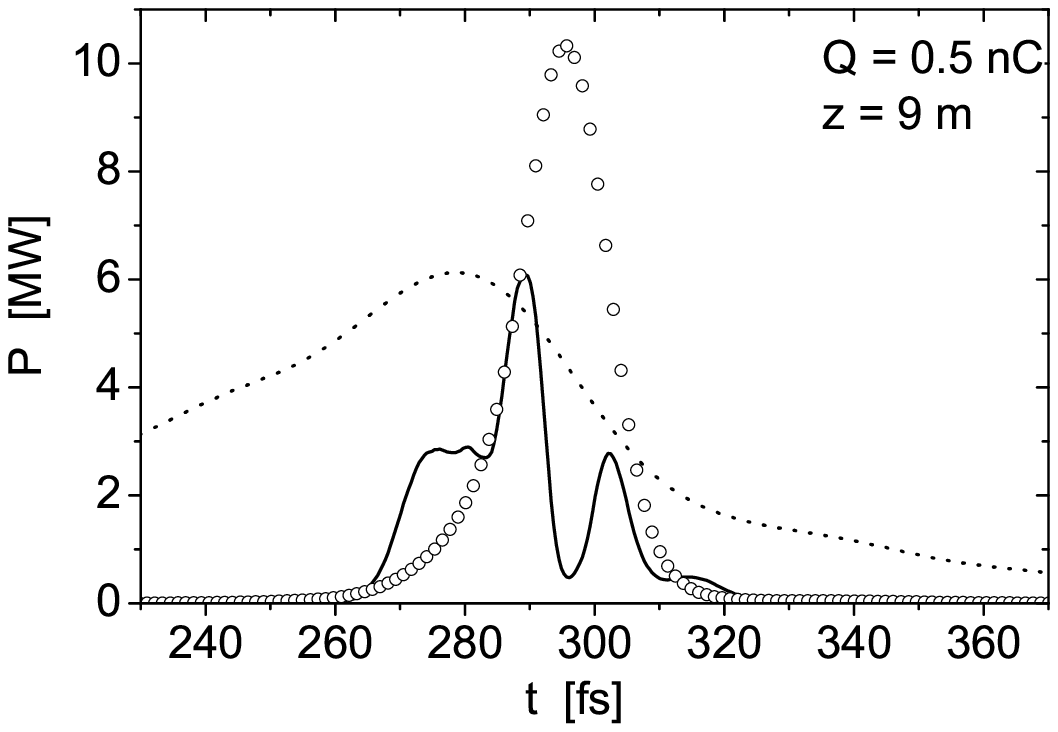}
\includegraphics[width=0.5\textwidth]{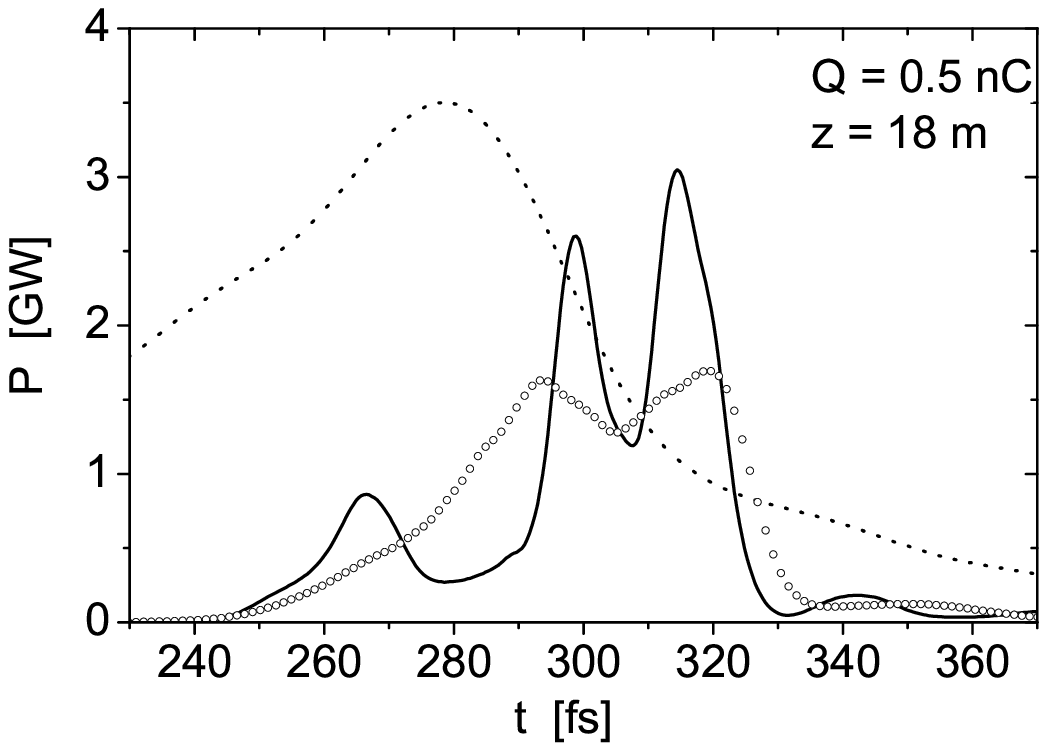}

\vspace*{-5mm}

\includegraphics[width=0.5\textwidth]{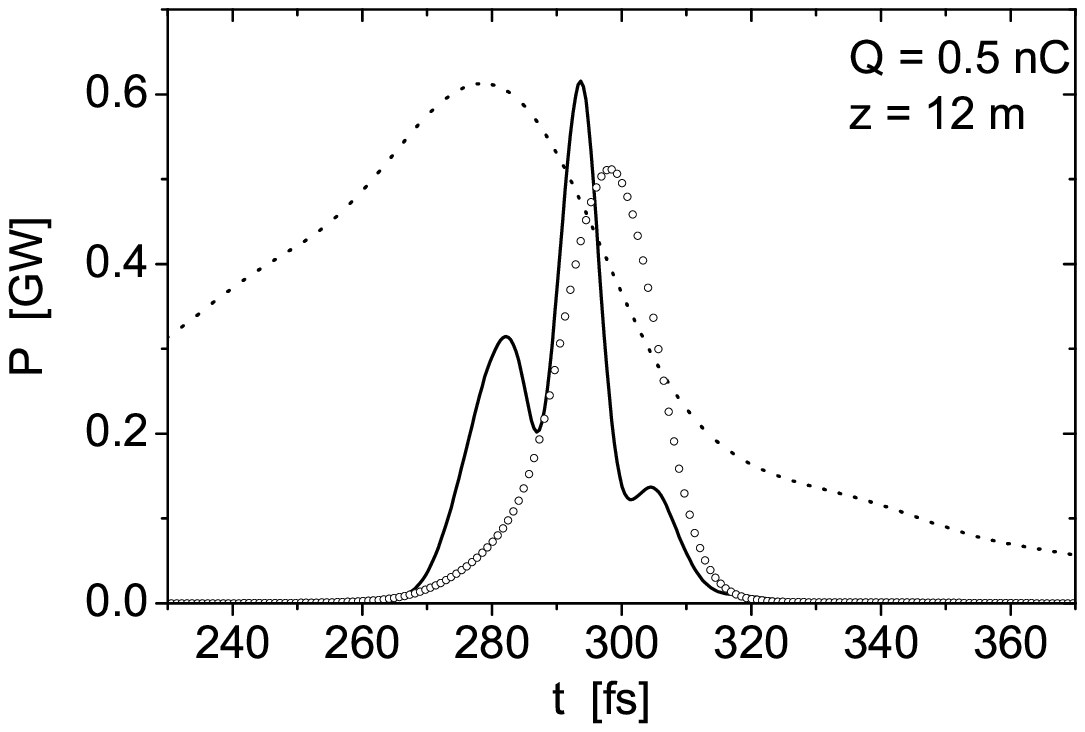}
\includegraphics[width=0.5\textwidth]{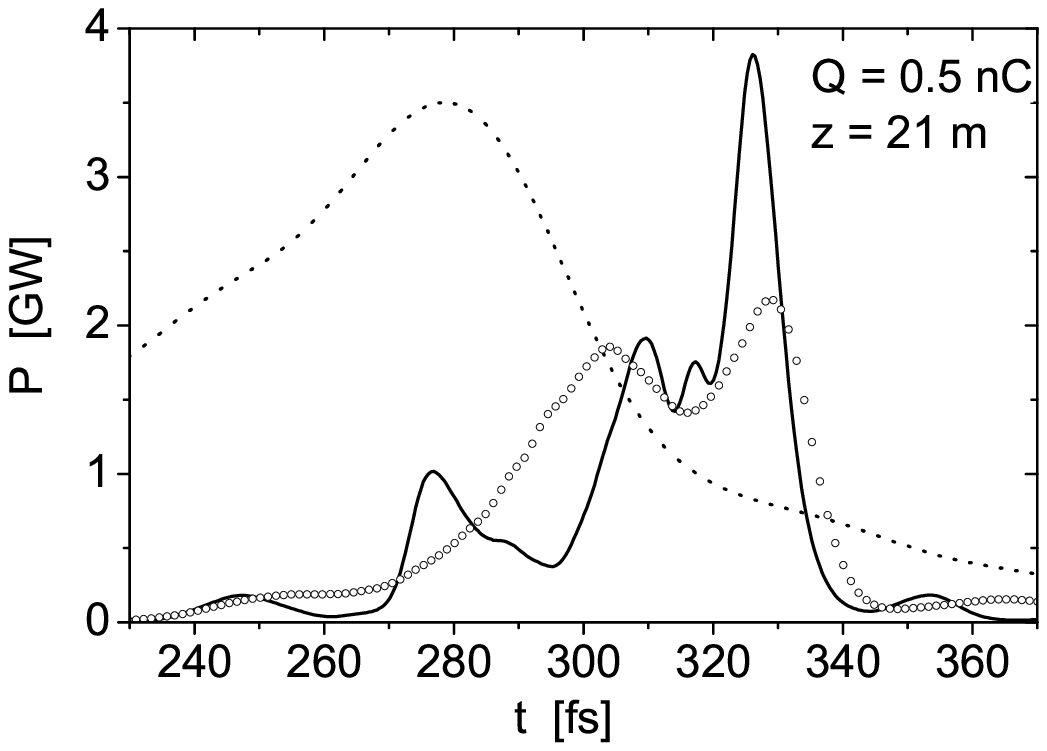}

\vspace*{-5mm}

\caption{
Evolution of temporal structure of the radiation pulse.
along the undulator.
Solid and circles correspond to a single pulse and
circles averaged profile, respectively.
Left and right columns correspond to linear and nonlinear mode of
operation, respectively. Bunch charge is 0.5~nC.
Dashed line shows profile of the electron bunch.
Bunch head is at the right side
}
\label{fig:ptemp-05}
\end{figure}

\begin{figure}[tb]

\includegraphics[width=0.5\textwidth]{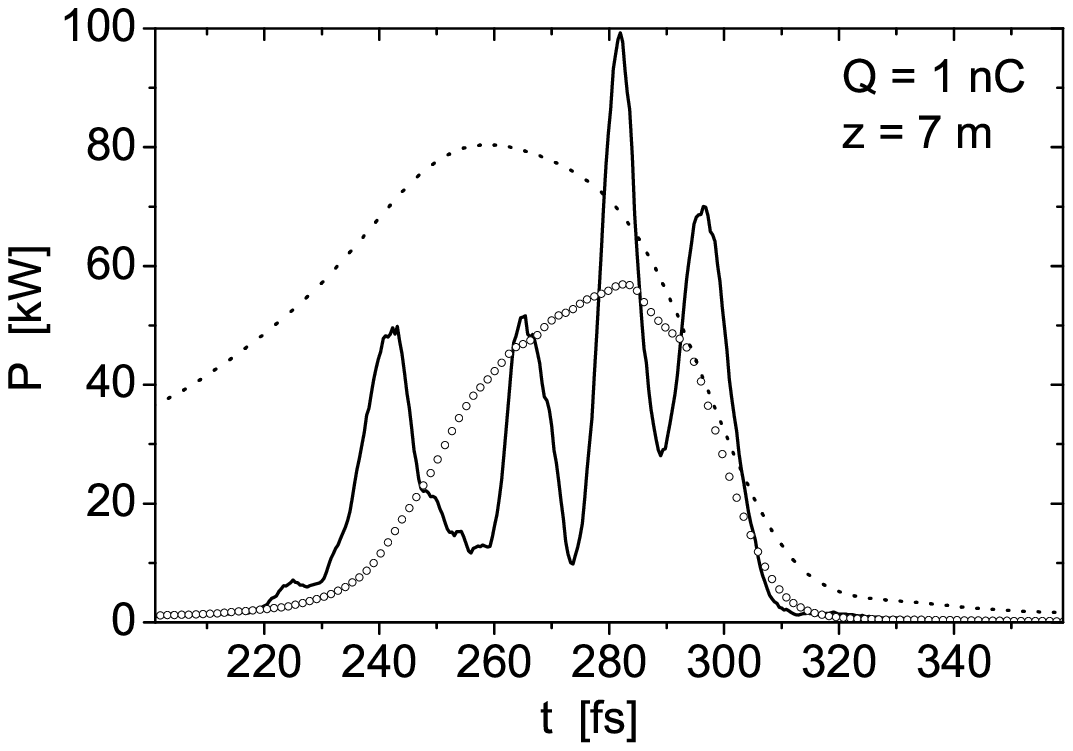}
\includegraphics[width=0.5\textwidth]{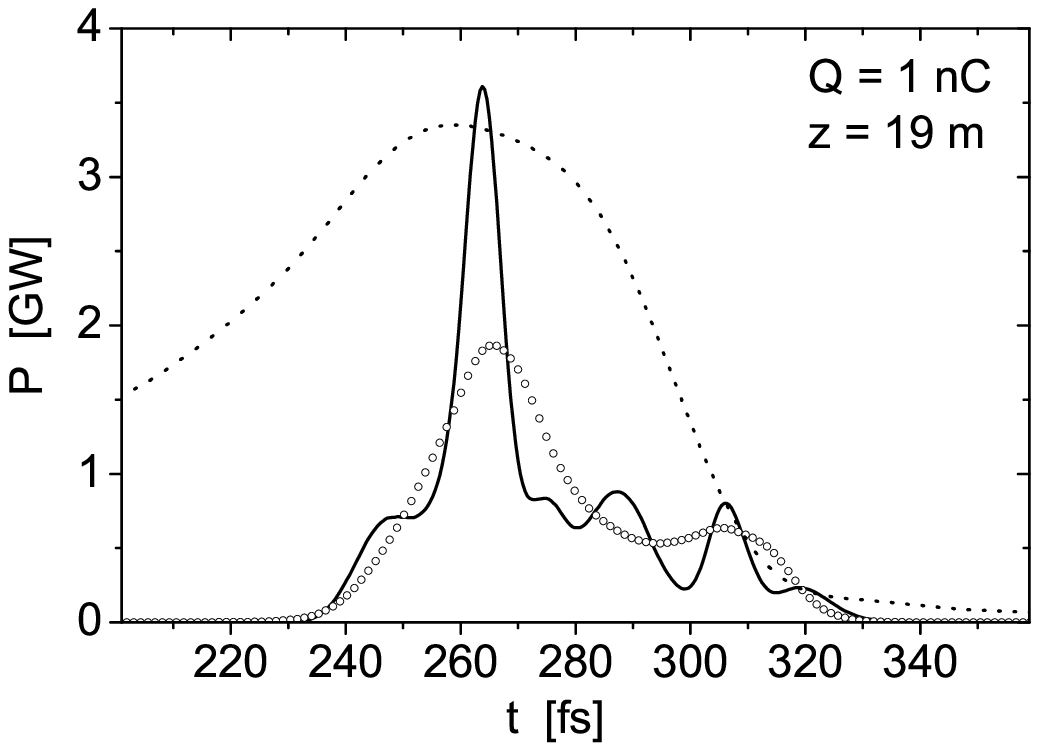}

\vspace*{-5mm}

\includegraphics[width=0.5\textwidth]{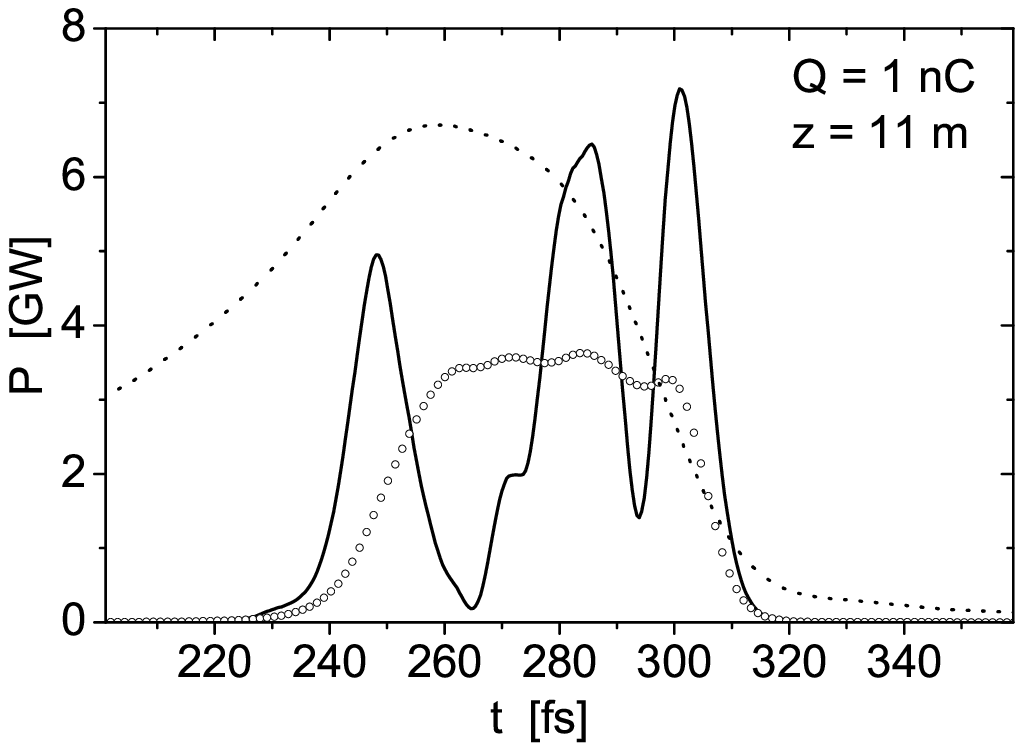}
\includegraphics[width=0.5\textwidth]{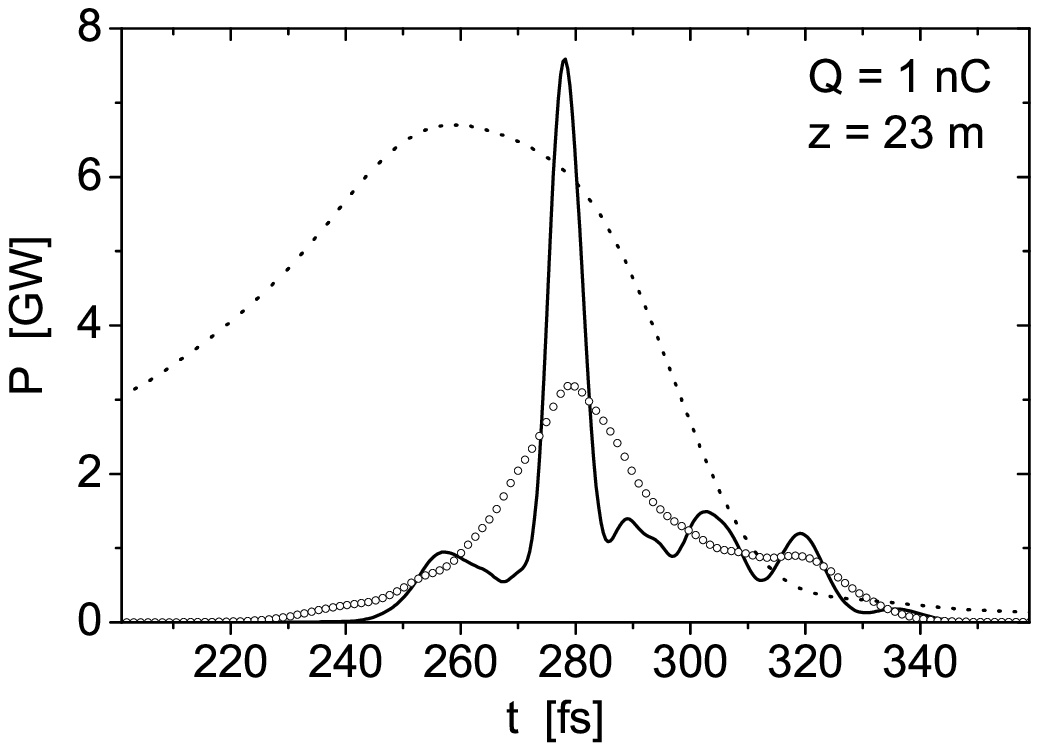}

\vspace*{-5mm}

\includegraphics[width=0.5\textwidth]{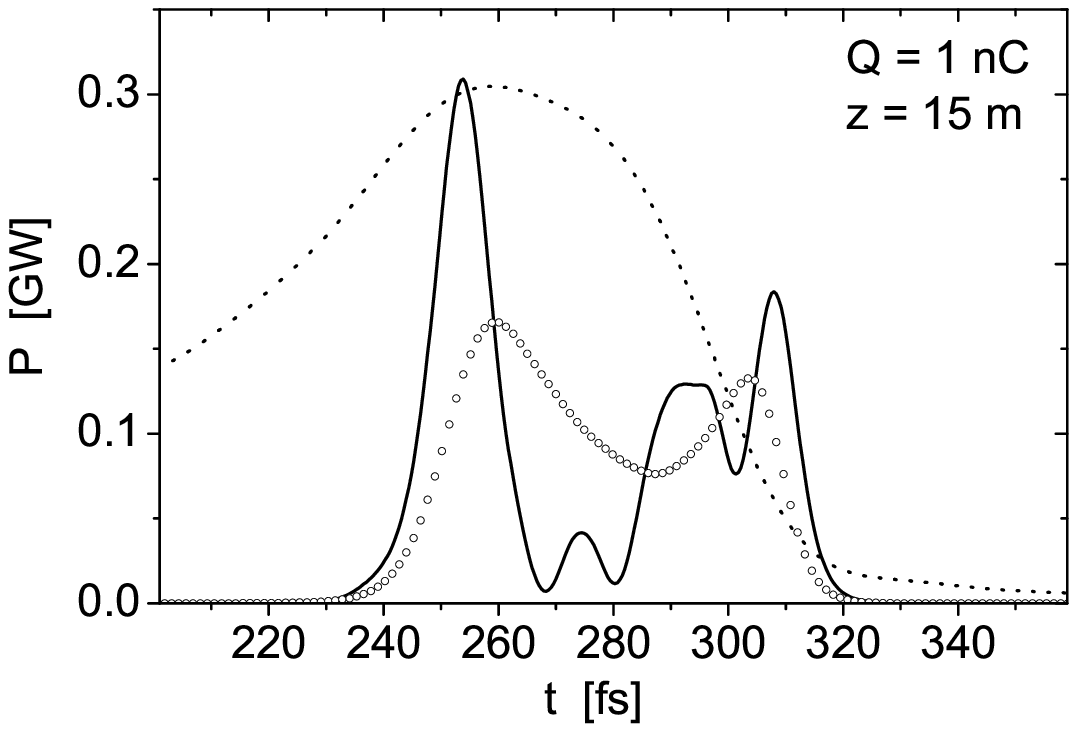}
\includegraphics[width=0.5\textwidth]{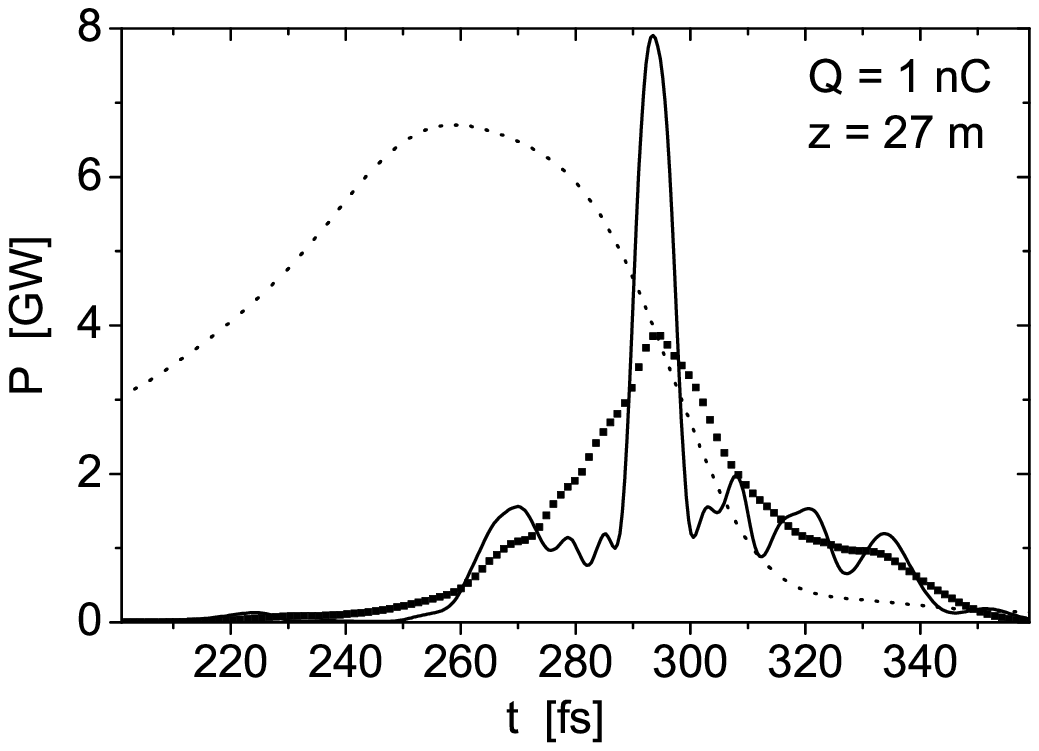}

\vspace*{-5mm}

\caption{
Evolution of temporal structure of the radiation pulse.
along the undulator.
Solid line and circles correspond to a single pulse and
averaged profile, respectively.
Left and right columns correspond to linear and nonlinear mode of
operation, respectively. Bunch charge is 1~nC.
Dashed line shows profile of the electron bunch.
Bunch head is at the right side
}
\label{fig:ptemp-1}
\end{figure}

\begin{figure}[tb]

\includegraphics[width=0.5\textwidth]{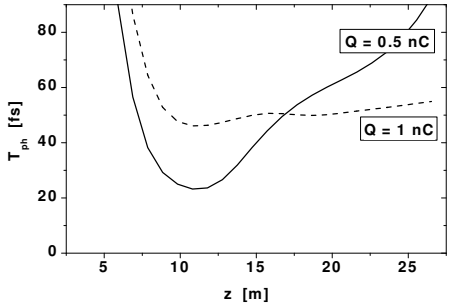}

\caption{
Evolution of the FWHM radiation pulse length along undulator.
Solid and dashed line correspond to the charge of 0.5 and 1~nC,
respectively
}
\label{fig:lpulse}
\end{figure}

\begin{figure}[tb]

\includegraphics[width=0.5\textwidth]{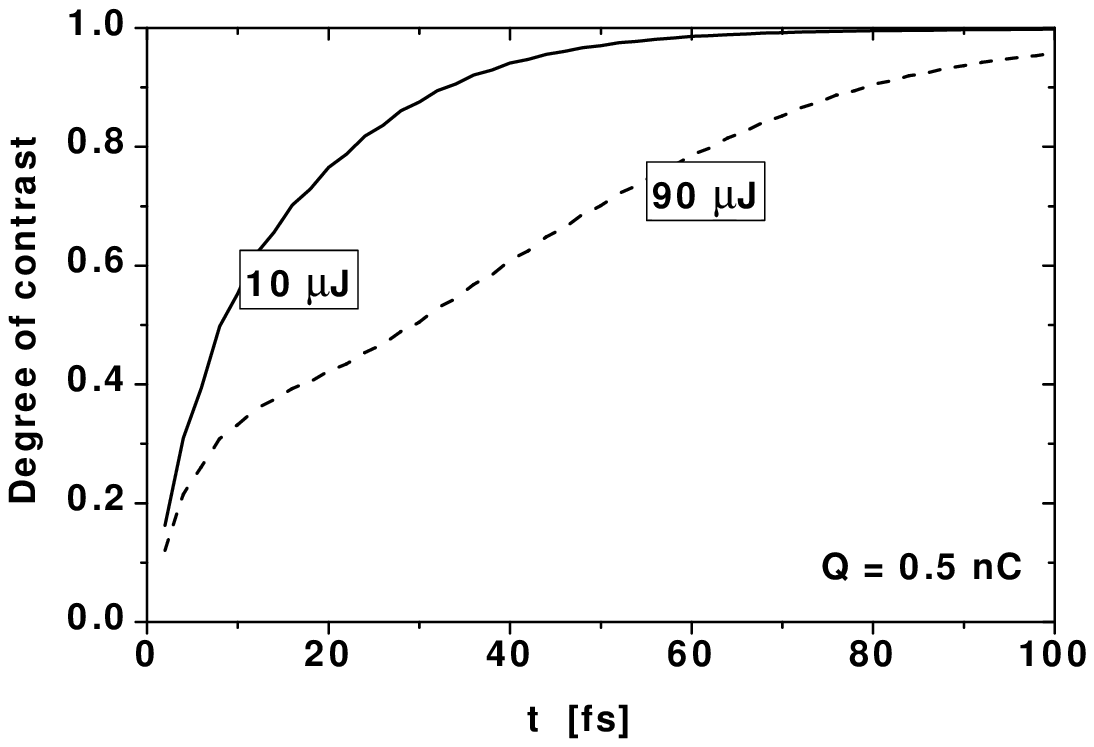}
\includegraphics[width=0.5\textwidth]{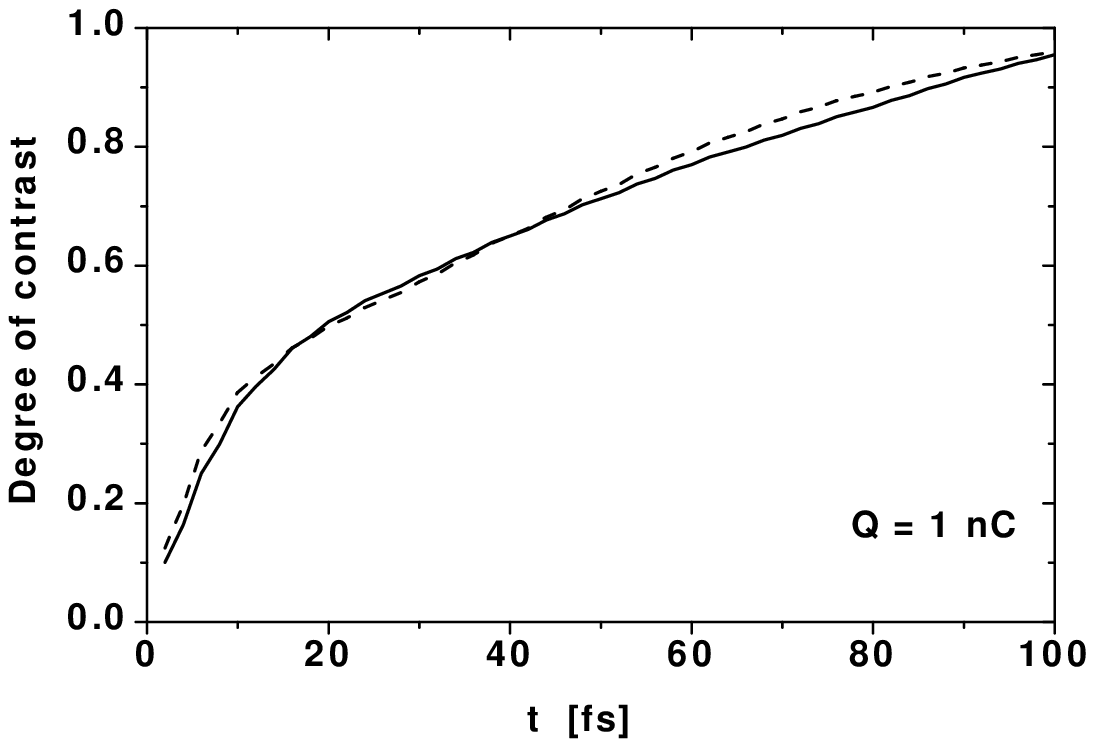}

\caption{
Degree of contrast versus time gate around a spike with maximum
intensity.
Left: bunch charge is 0.5~nC.
right: bunch charge is 1~nC.
Solid and dashed line corresponds to the end of the linear regime
and deep nonlinear regime, respectively
}
\label{fig:c05}
\end{figure}

Figures~\ref{fig:ptemp-05} and \ref{fig:ptemp-1} shows evolution of
temporal structure of the radiation pulse (single pulse and averaged)
along the undulator. Analysis of these results tells us that we deal
with nonstationary process of light amplification which occurs due to
short duration of lasing part of the electron bunch. Note that in the
linear mode of operation we obtain significant suppression of the
radiation in the tail of the pulse. Actually, there are two effects
leading to the sharp decrease of the gain. The first one is decrease of
the current. And the second effect, as we mentioned above, is the
energy chirp along the electron bunch. Its influence grows strongly
when the current drops down. Thus, only half of the electron spike
contribute to the light amplification in the linear regime. Numerical
simulations shows that in the end of the linear regime the VUV FEL
driven by 0.5~nC electron bunches produces short, down to 20~fs (FWHM)
radiation pulse. VUV FEL driven by 1~nC bunch produces twice longer
pulses. The reason for this is that 0.5~nC bunch has more narrow lasing
part than 1~nC bunch (see Fig.~\ref{fig:bunch-ue}). It is relevant to
note that ultra-short pulse duration of 20~fs occurs not only due to
ultra-short duration of the lasing part of the electron bunch. Another
effect is suppression of the slippage effect in the linear regime
\cite{book}. In the free electron laser we deal with the wave
propagating not in the free space, but with the wave which dynamically
interacts with the electron bunch. Slippage of the electromagnetic wave
with respect to the electron beam is given by the value of the group
velocity, $v_{\mathrm{g}} = \D \omega /\D k$. Group velocity of the
electromagnetic wave in the high gain FEL regime is less than velocity
of light $c$, and this results in significant reduction of the slippage
(about factor of 3 in our case) \cite{book}.

In the linear mode of operation pulse duration is permanently reduced
with the undulator length. When amplification process enters nonlinear
stage, we obtain opposite situation. Radiation pulse lengthening occurs
mainly due to two effects. First, suppression of the group velocity in
the nonlinear regime becomes to be much less than in the high gain
linear regime. Thus, radiation starts to slip forward faster. Second,
the tail of the electron bunch starts to produce visible amount of
radiation. As a result we find that in the nonlinear regime the 0.5~nC
case has no benefit in terms of pulse duration. Figure~\ref{fig:lpulse}
shows evolution of the FWHM pulse length along the undulator. It is
seen that shortest pulse duration occurs in the end of linear regime
when average radiation energy is about a few $\mu $J.

For the experiments utilizing the property of ultra short radiation
pulse duration it is important to know the degree of contrast of the
radiation pulse. The figure of merit for the degree of contrast is the
ratio of the radiation energy within a time window $\tau $ around a
spike with maximum peak power to the total energy in the radiation
pulse:

\begin{equation}
C(\tau) =  \frac{\int _{-\tau/2}^{\tau/2} P(t) \D t }
{\int _{-\infty }^{\infty } P(t) \D t } \ .
\label{eq:contrast}
\end{equation}

\noindent The results for the average degree of contrast shown in
Fig.~\ref{fig:c05} have been derived from 500 statistically independent
runs. For fixed time window $\tau $ around the spike with maximum
intensity we calculate the value of $C(\tau)$ for each pulse and then
perform averaging over ensemble. Solid and dashed curves in
Fig.~\ref{fig:c05} correspond to the end of the linear regime and deep
nonlinear regime, respectively. We see that considered modes of the VUV
FEL operation (driven by bunches with 0.5~nC and 1~nC charge) provide
quite different properties for the contrast of the radiation pulse. The
VUV FEL driven by 0.5~nC bunch is capable to produce short, down to
20~fs radiation pulses with GW-level peak power and contrast of 80\%.
This happens due to shorter lasing part of the bunch. Maximum contrast
of the radiation pulses occurs in the end of the linear mode of
operation. Average energy in the radiation pulse is about 10~$\mu $J in
this case. Increase of the radiation energy above this value leads to
gradual decrease of the contrast due to radiation pulse lengthening in
the nonlinear regime. The case of 1~nC demonstrate the contrast of the
radiation pulse nearly independent on the pulse energy: 80\% at 50~fs.

\clearpage

\section{Temporal correlation functions}

    The shot noise in the electron beam causes fluctuations of the beam
density which are random in time and space. As a result, the radiation,
produced by such a beam, has random amplitudes and phases in time and
space. Figures~\ref{fig:ptemp-05} and \ref{fig:ptemp-1} show examples
of single radiation pulse in terms of radiation power. Qualitative
analysis of these plots allows to estimate coherence time of the
radiation, but complete description of coherence properties can be
obtained only in terms of correlation functions. Radiation from SASE
FEL has a narrow band around resonant frequency $\omega $, and electric
field of the wave can be presented as

\begin{figure}[b]

\includegraphics[width=0.5\textwidth]{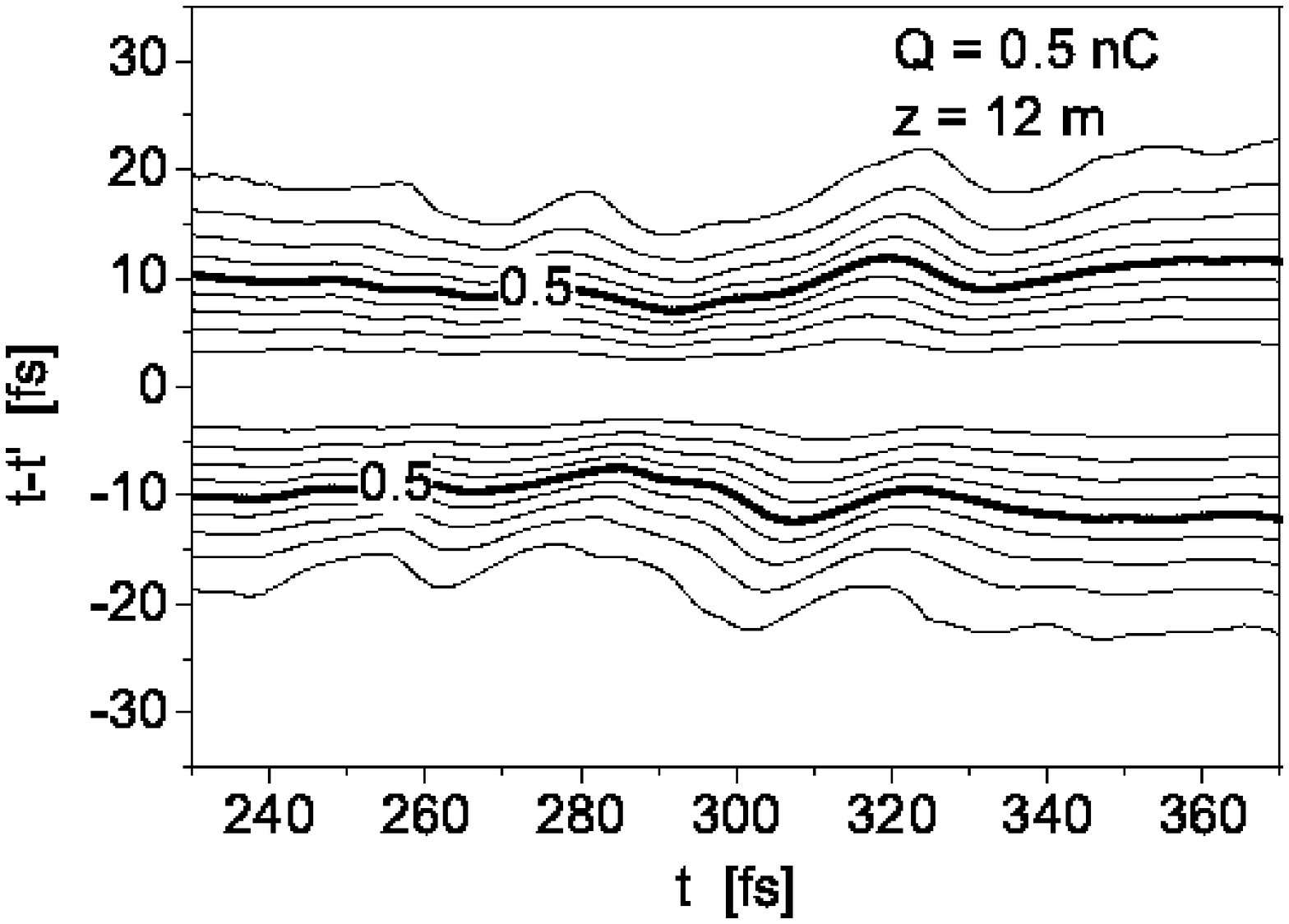}
\includegraphics[width=0.5\textwidth]{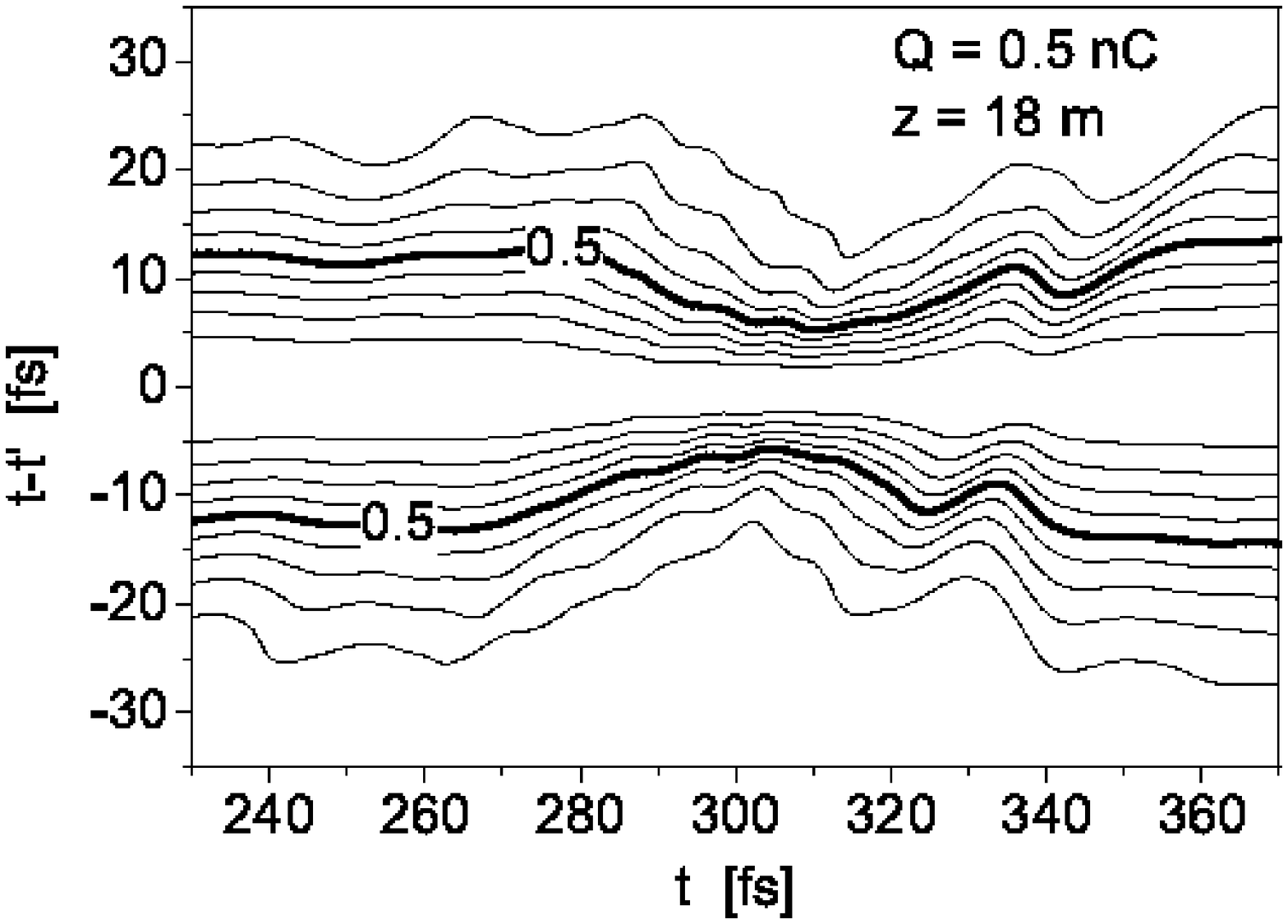}

\vspace*{-5mm}

\includegraphics[width=0.5\textwidth]{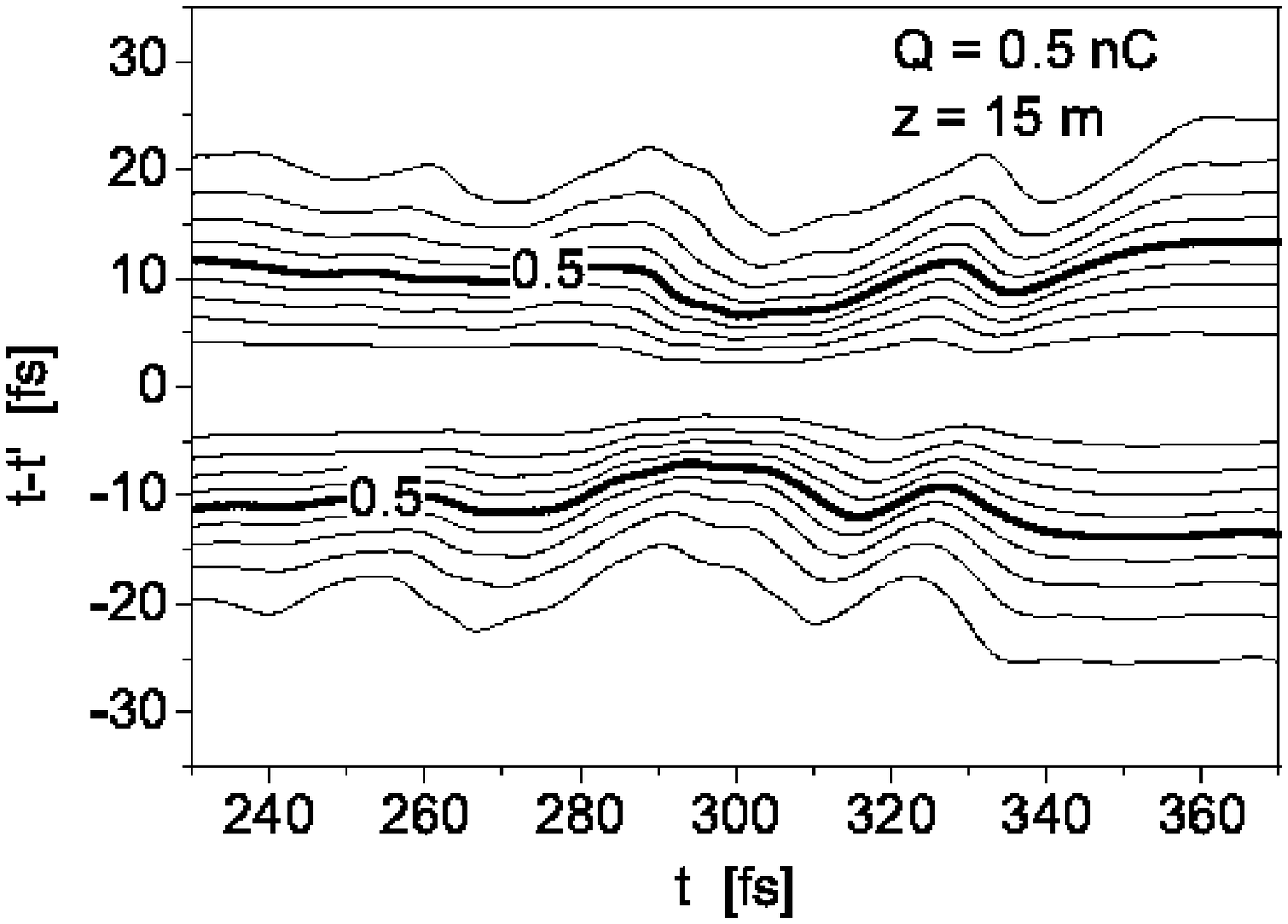}
\includegraphics[width=0.5\textwidth]{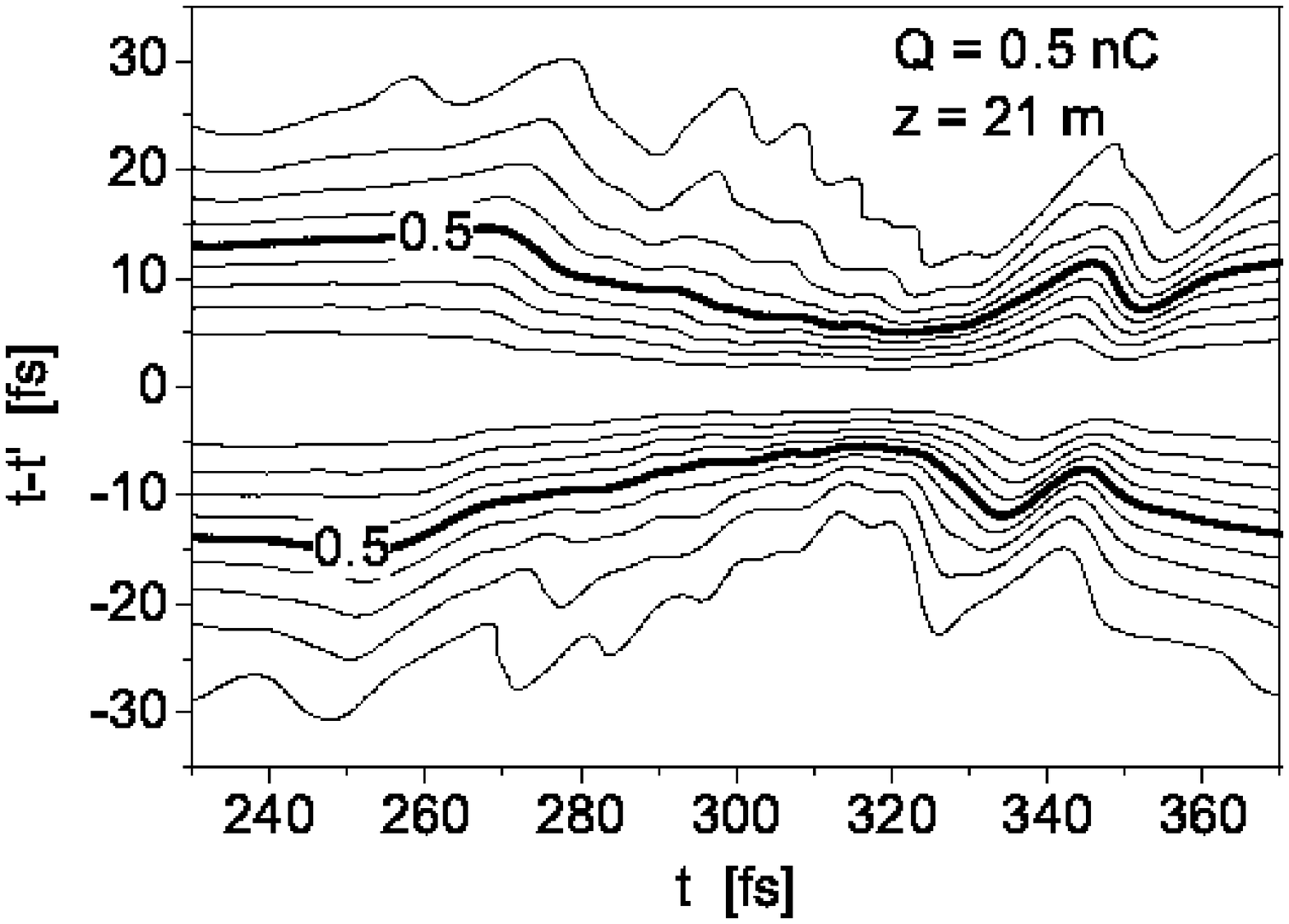}

\vspace*{-5mm}

\caption{
Contour plot of the module of the first order correlation function
for different undulator length.
Bold lines trace the FWHM level.
Bunch charge is 0.5~nC.
Time interval corresponds to that of temporal structure
shown in Fig.~\ref{fig:ptemp-05}.
Bunch head is at the right side
}
\label{fig:corf-05}
\end{figure}

\begin{figure}[tb]

\includegraphics[width=0.5\textwidth]{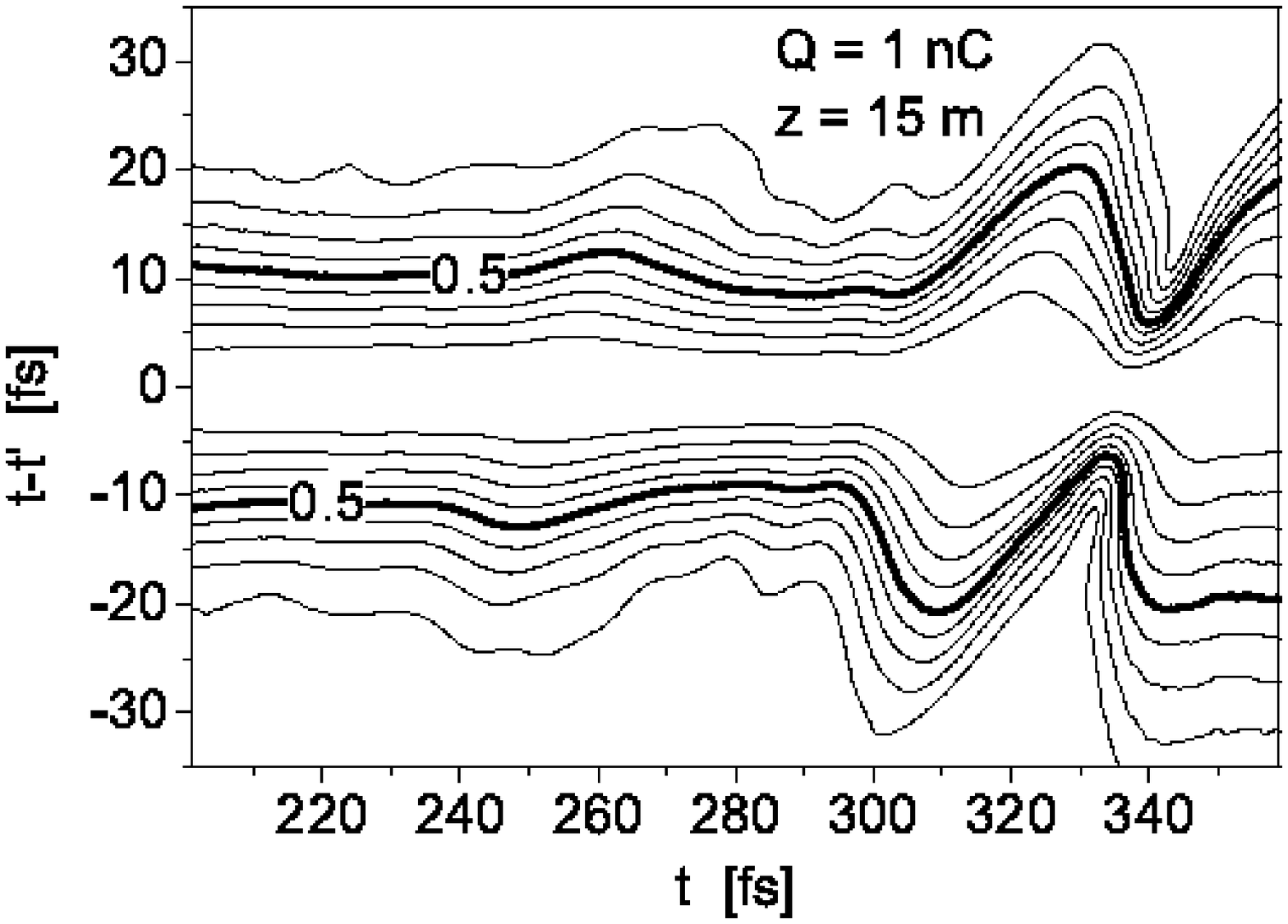}
\includegraphics[width=0.5\textwidth]{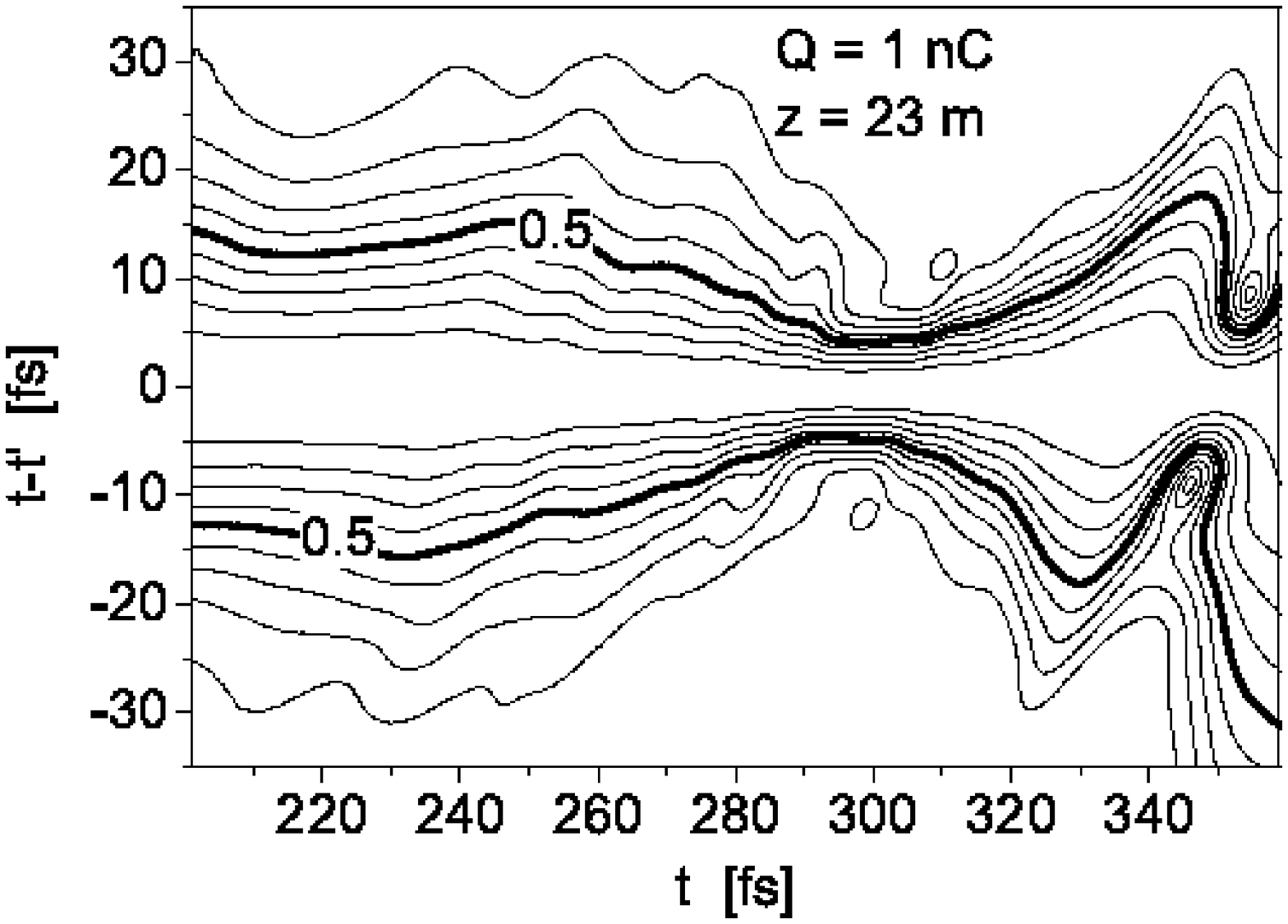}

\vspace*{-5mm}

\includegraphics[width=0.5\textwidth]{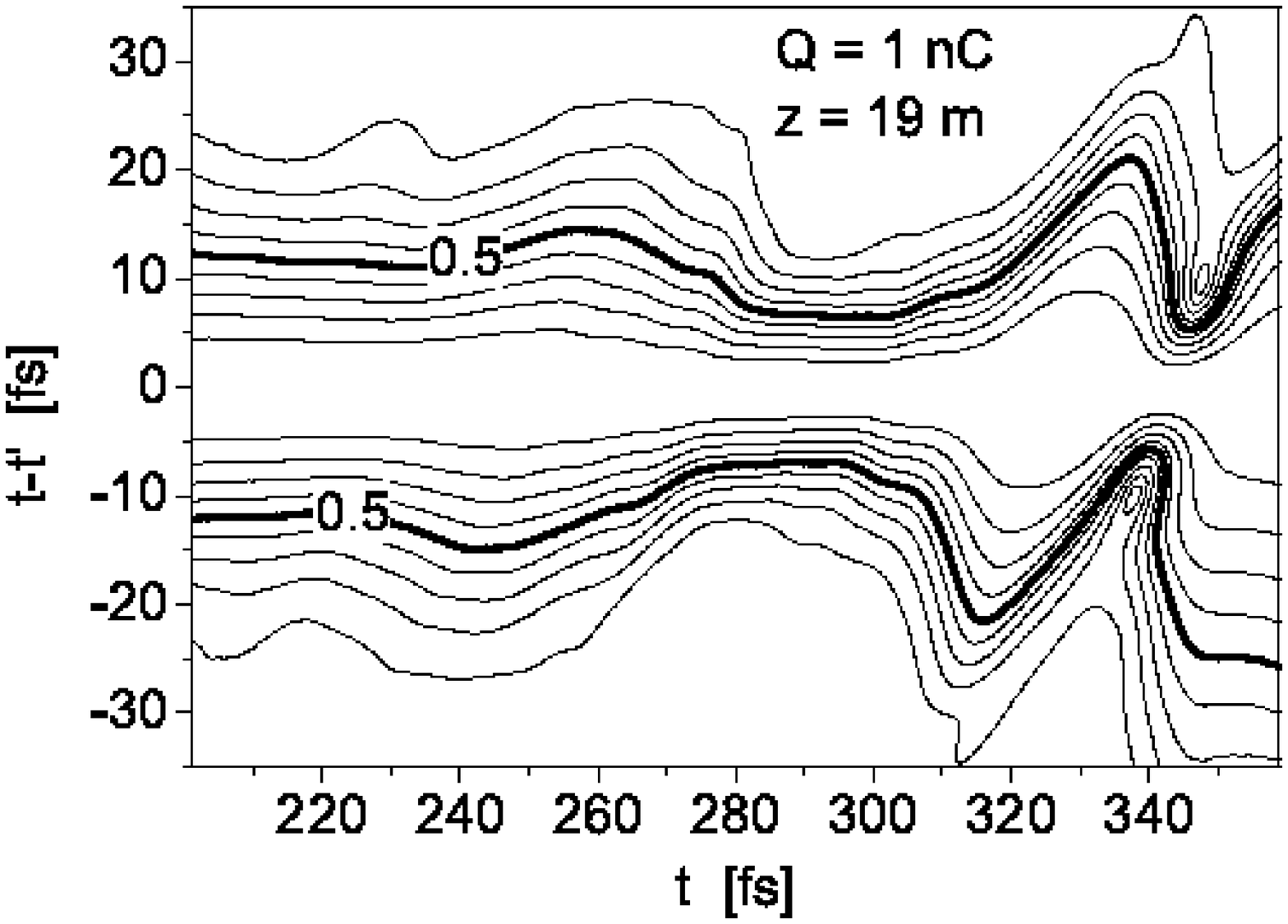}
\includegraphics[width=0.5\textwidth]{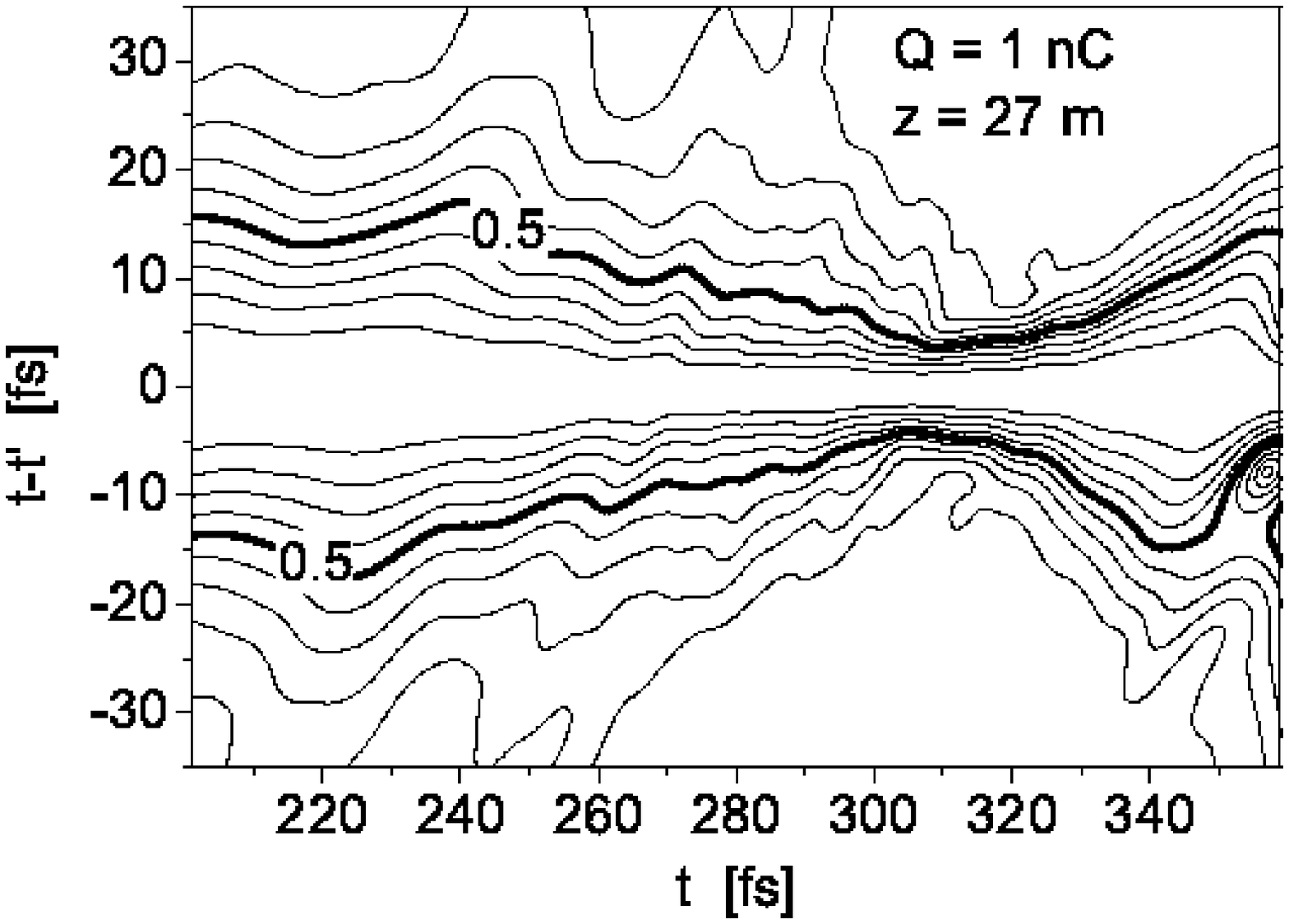}

\vspace*{-5mm}

\caption{
Contour plot of the module of the first order correlation function
for different undulator length.
Bold lines trace the FWHM level.
Bunch charge is 1~nC.
Time interval corresponds to that of temporal structure
shown in Fig.~\ref{fig:ptemp-1}.
Bunch head is at the right side
}
\label{fig:corf-1}
\end{figure}

\begin{figure}[tb]

\includegraphics[width=0.5\textwidth]{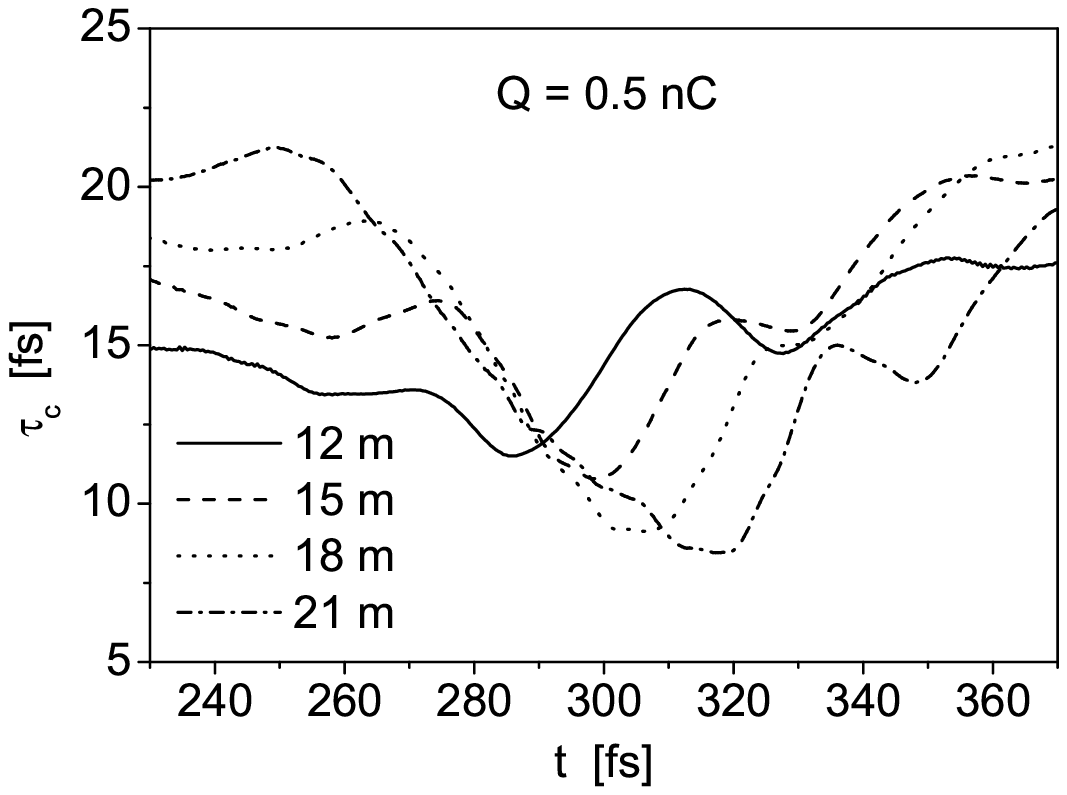}
\includegraphics[width=0.5\textwidth]{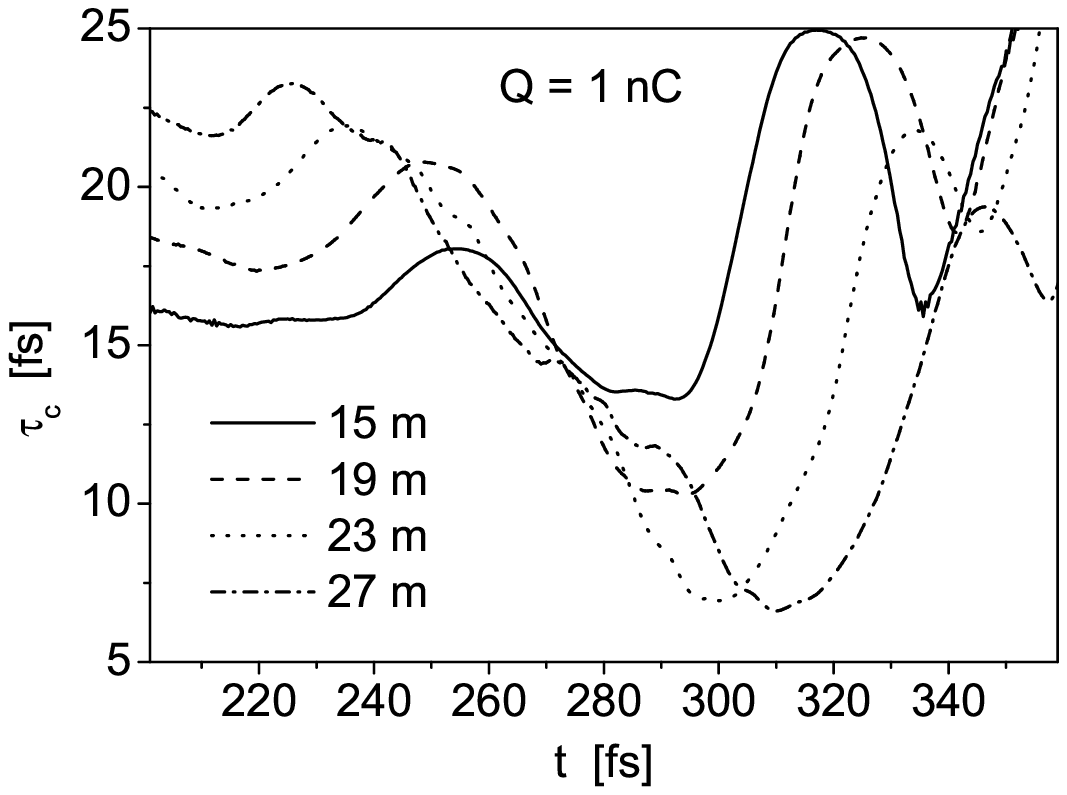}

\caption{Coherence time $\tau _{\mathrm{c}}$ along the radiation pulse
for different undulator length.
Left and right columns correspond to the of 0.5 and 1~nC,
respectively.
Time interval corresponds to that of temporal structure
shown in Figs.~\ref{fig:ptemp-05} and \ref{fig:ptemp-1}.
Bunch head is at the right side
}
\label{fig:tcor}
\end{figure}

\begin{equation}
E_{y}(z,t) =
\tilde{E}(z,t)e^{\I\omega _0(z/c-t)}
+ {\mathrm{C.C.}} \ ,
\label{eq:field-real}
\end{equation}

\noindent where $\tilde{E}(z,t)$ is slowly varying amplitude. In the
case under study we deal with non-stationary random process, and use
general definition for the first-order time correlation function:

\begin{equation}
g_1(t,t') =
\frac{\langle \tilde{E}(t)\tilde{E}^*(t')\rangle }
{\left[\langle |\tilde{E}(t)|^2\rangle
\langle |\tilde{E}(t')|^2\rangle \right]^{1/2}} \ .
\label{def-corfunction}
\end{equation}

\noindent Here $ \langle \ldots \rangle $ means averaging over ensemble
(shots). Numerical simulation code produces arrays for complex
amplitudes for the slowly varying amplitude of the radiation field in
the near zone. Then we propagate this field into the far zone, and
calculate coherence properties for the radiation field at zero
observation angle. Statistical accuracy of 500 events is sufficient for
reliable description of correlation functions.
Figures~\ref{fig:corf-05} and \ref{fig:corf-1} show the evolution of
the first order correlation function from the end of linear regime down
to deep nonlinear regime. Bold lines trace the FWHM level. Time window
of the plots is the same as for temporal structure of the radiation
shown in Figs.~\ref{fig:ptemp-05} and \ref{fig:ptemp-1}.
Following general definition we calculate coherence time as

\begin{equation}
\tau_{\mathrm{c}} (t) = \int \limits^{\infty}_{-\infty}
| g_1(t,t') |^2 \D t' \ .
\label{coherence-time-def}
\end{equation}

\noindent Figure~\ref{fig:tcor} shows coherence time $\tau
_{\mathrm{c}} (t) $ along the radiation pulse for different undulator
length. Calculation are performed from the data shown in
Figs.~\ref{fig:corf-05} and \ref{fig:corf-1}. We see that behavior of
the correlation function and correlation time along the radiation pulse
is rather complicated. From practical point of view the value of
interest is the coherence time in the core of the radiation pulse
(maximum of the averaged radiation power). We see that coherence time
for the core of the radiation pulse has maximum value in the end of
the linear regime (solid line), and drops down gradually when
amplification process enters deep nonlinear regime (dash-dot line).
Such a behavior is quite natural, and is well known from studies of SASE
FEL driven by long pulses \cite{book,stat-oc}. Origin of the reduction
of coherence time is in the growth of the sidebands in the nonlinear
media.

\section{Statistics of the radiation energy}

\begin{figure}[b]

\includegraphics[width=0.5\textwidth]{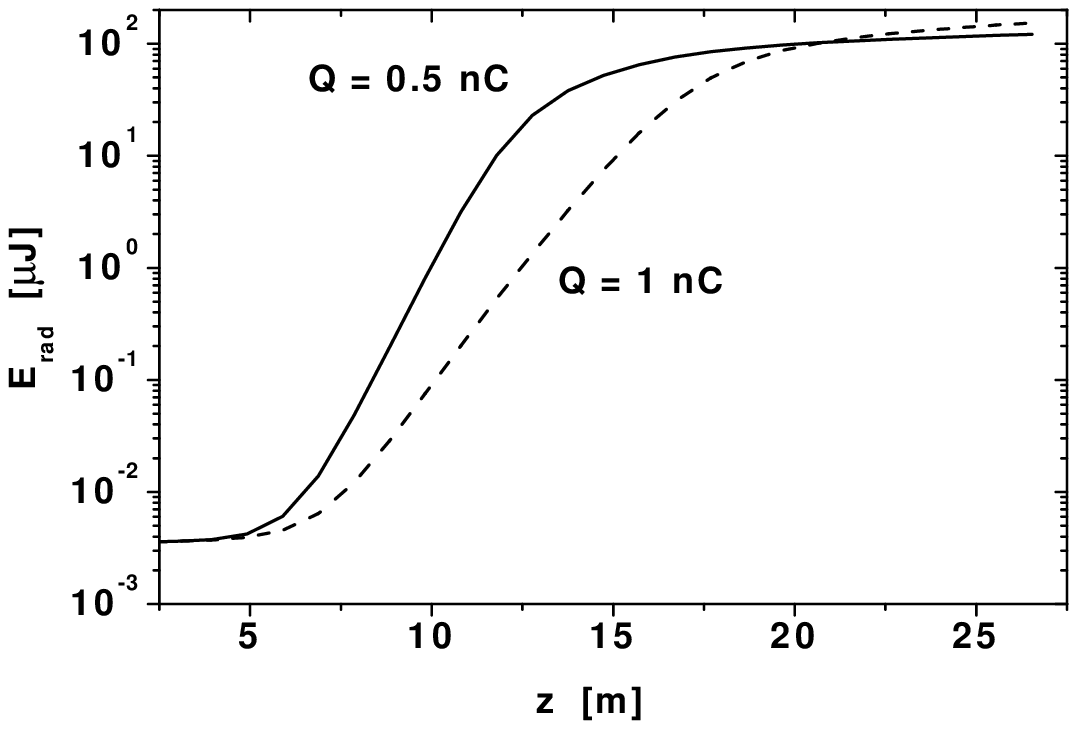}
\includegraphics[width=0.5\textwidth]{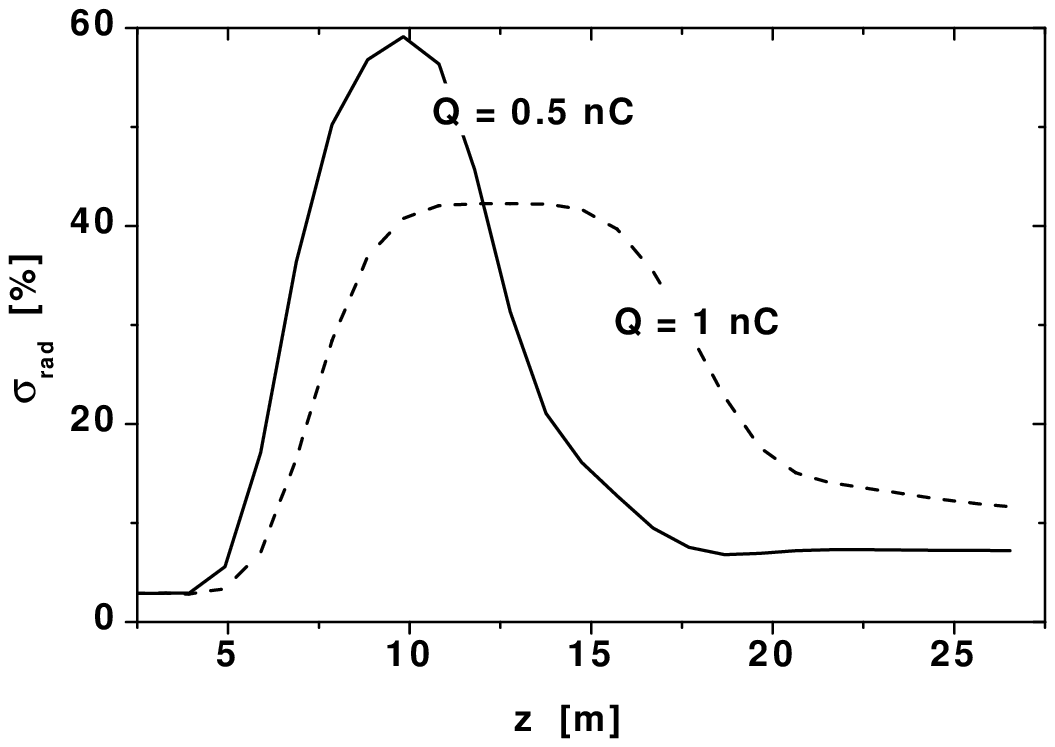}

\caption{
Average energy in the radiation pulse (left plot) and
fluctuations of the energy in the radiation pulse
(right plot) versus
undulator length.
Solid and dashed lines refer to bunch charge 0.5 nC and 1~nC,
respectively
}
\label{fig:pz}
\end{figure}

Figure~\ref{fig:pz} shows average energy in the radiation pulse and rms
fluctuations as functions of position along the undulator. Expected
level of the output energy at saturation is approximately the same for
0.5 and 1~nC bunch charge, of about 100~$\mu $J, while the saturation
length is a little bit shorter for the case of 0.5~nC.
More pronounced difference is in the behavior of the fluctuations of
the radiation energy. Plots presented in Fig.~\ref{fig:pz} show the
value $\sigma_E = \sqrt{\langle (E -\langle E \rangle )^2
\rangle}/\langle E \rangle $. These fluctuations reach maximum value in
the end of the linear regime. It is known that the radiation from SASE
FEL operating in the linear regime possesses properties of completely
chaotic polarized light \cite{book,stat-oc}. One important property is
that probability distribution of the energy in the radiation pulse must
follow gamma-distribution:

\begin{equation}
p(E) = \frac{M^M}{\Gamma (M)}
\left( \frac{E}{\langle E\rangle }\right)^{M-1} \frac{1}{\langle E\rangle }
\exp \left( -M \frac{E}{\langle E\rangle } \right) \ ,
\label{gamma}
\end{equation}

\noindent where $\Gamma (M)$ is the gamma function, and $M = 1/
\sigma_E^2$. The parameter $M$ can be interpreted as
the average number of ``degrees of freedom'' or ``modes'' in the
radiation pulse. According to Fig.~\ref{fig:pz} number of modes for the
VUV FEL is expected to be in the range $3-6$. Smaller number of modes
for the case of 0.5~nC is clear indication for shorter pulse duration.

\begin{figure}[tb]

\includegraphics[width=0.5\textwidth]{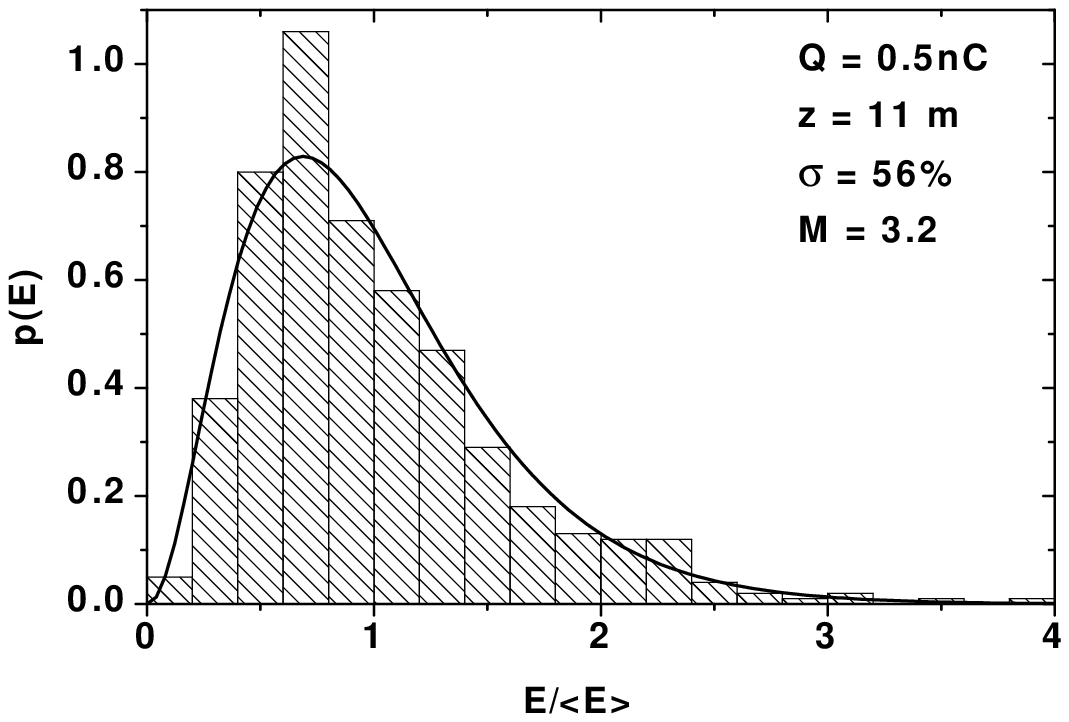}
\includegraphics[width=0.5\textwidth]{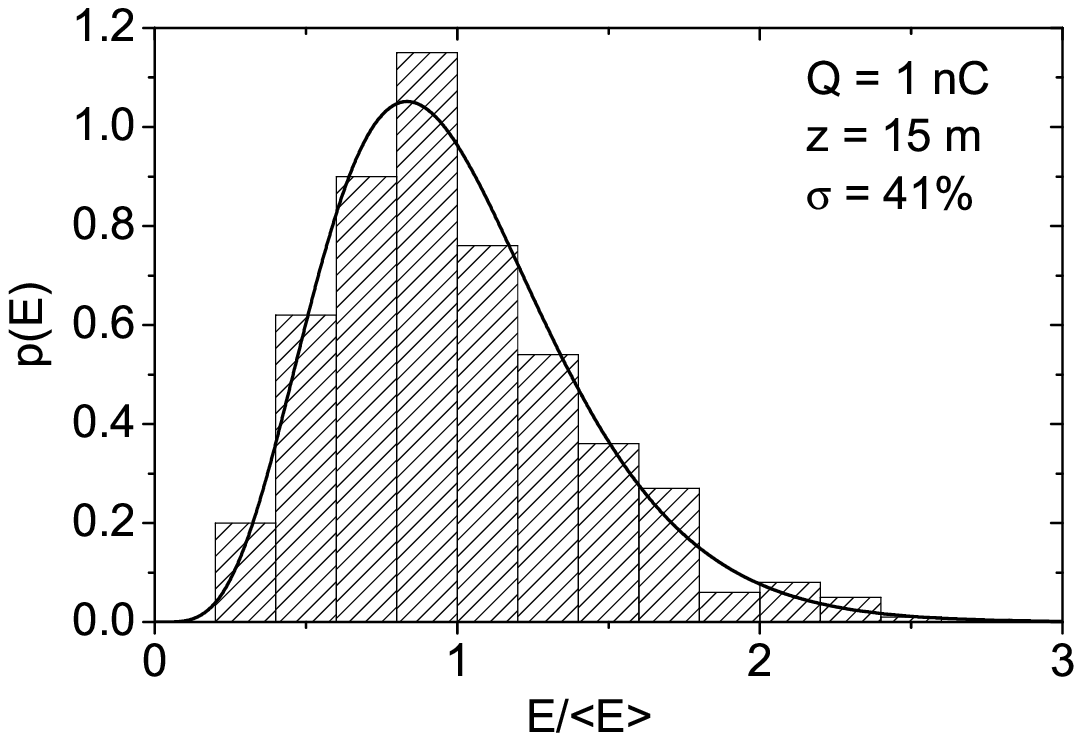}

\vspace*{-5mm}

\includegraphics[width=0.5\textwidth]{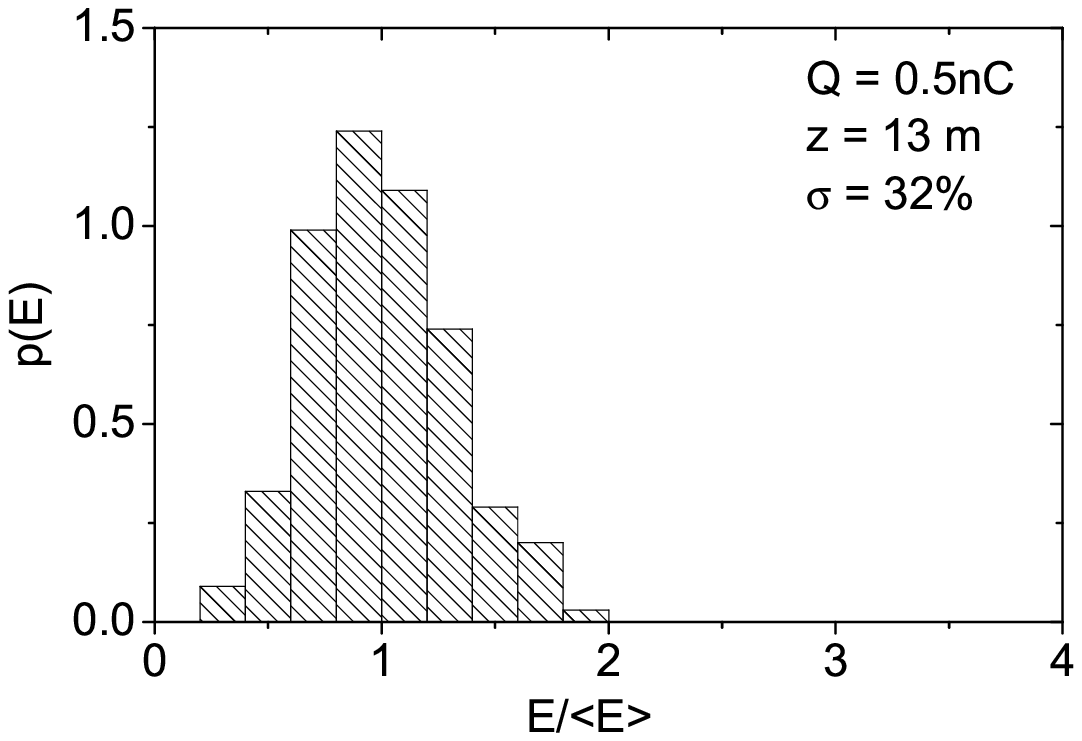}
\includegraphics[width=0.5\textwidth]{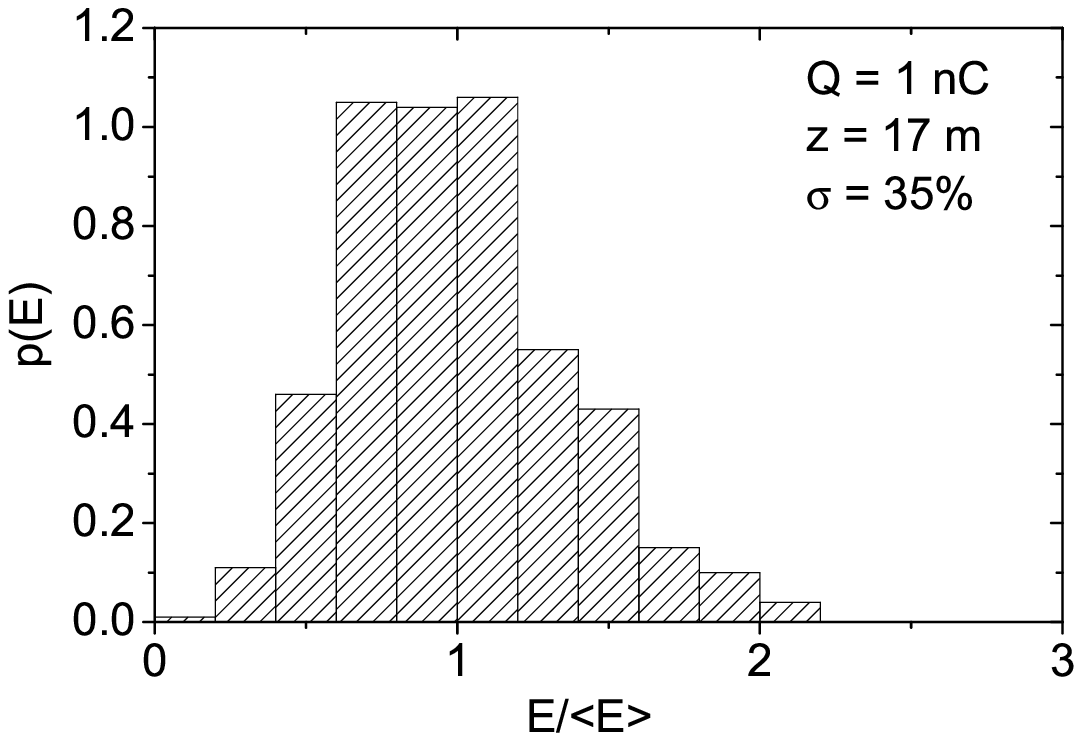}

\vspace*{-5mm}

\includegraphics[width=0.5\textwidth]{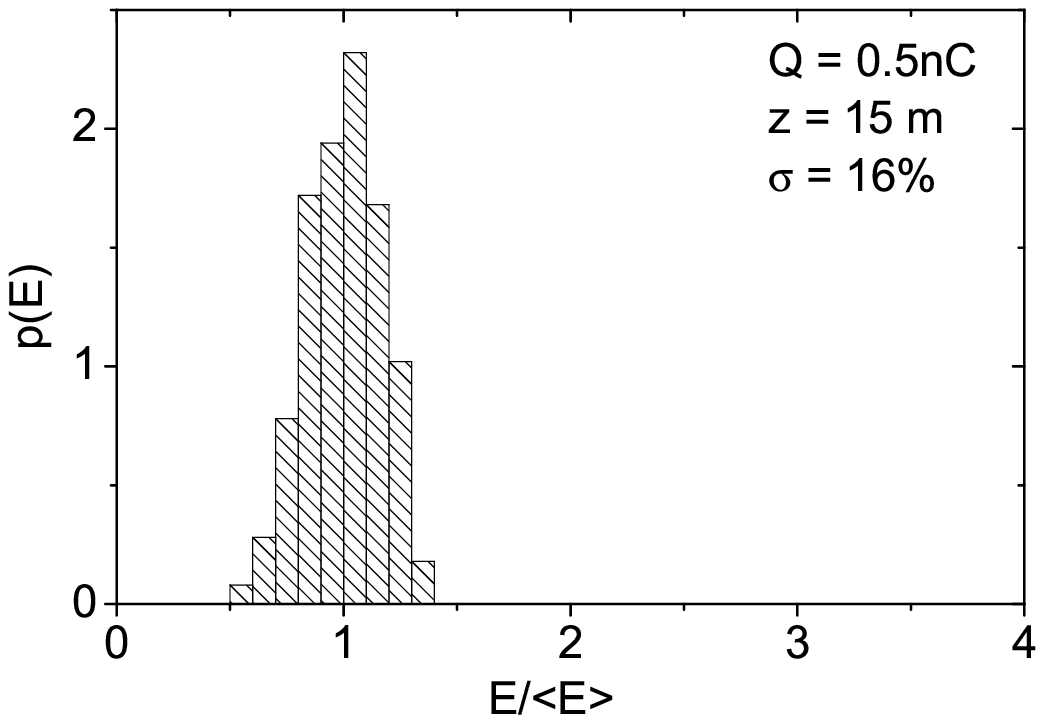}
\includegraphics[width=0.5\textwidth]{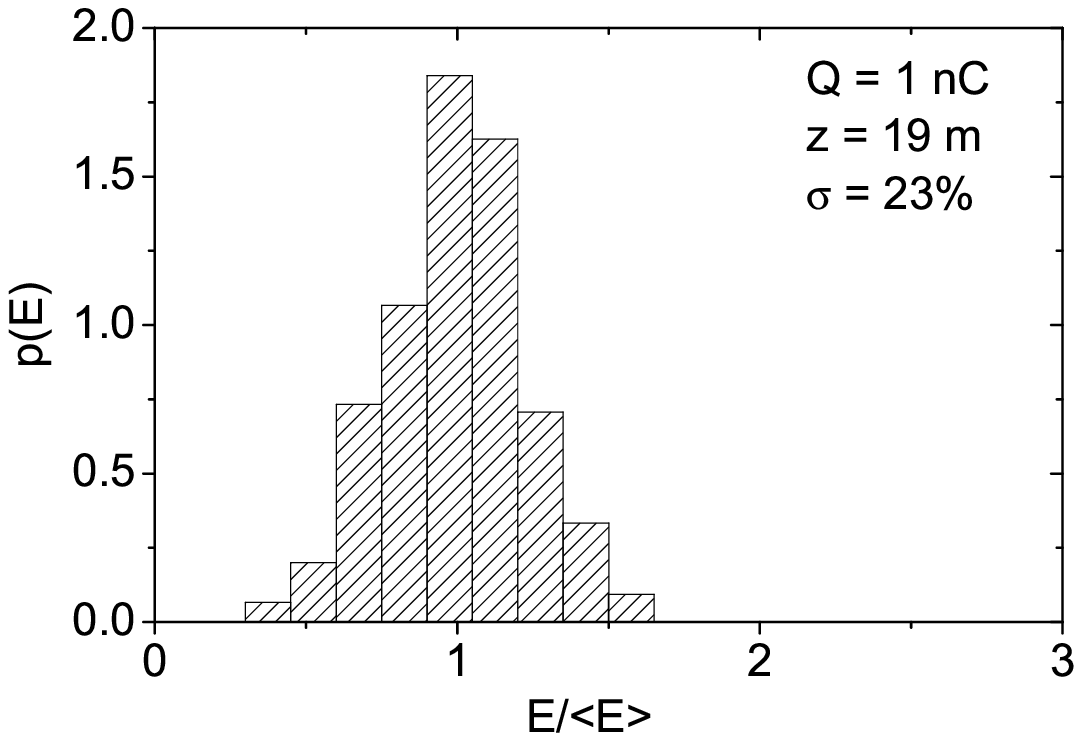}

\vspace*{-5mm}

\includegraphics[width=0.5\textwidth]{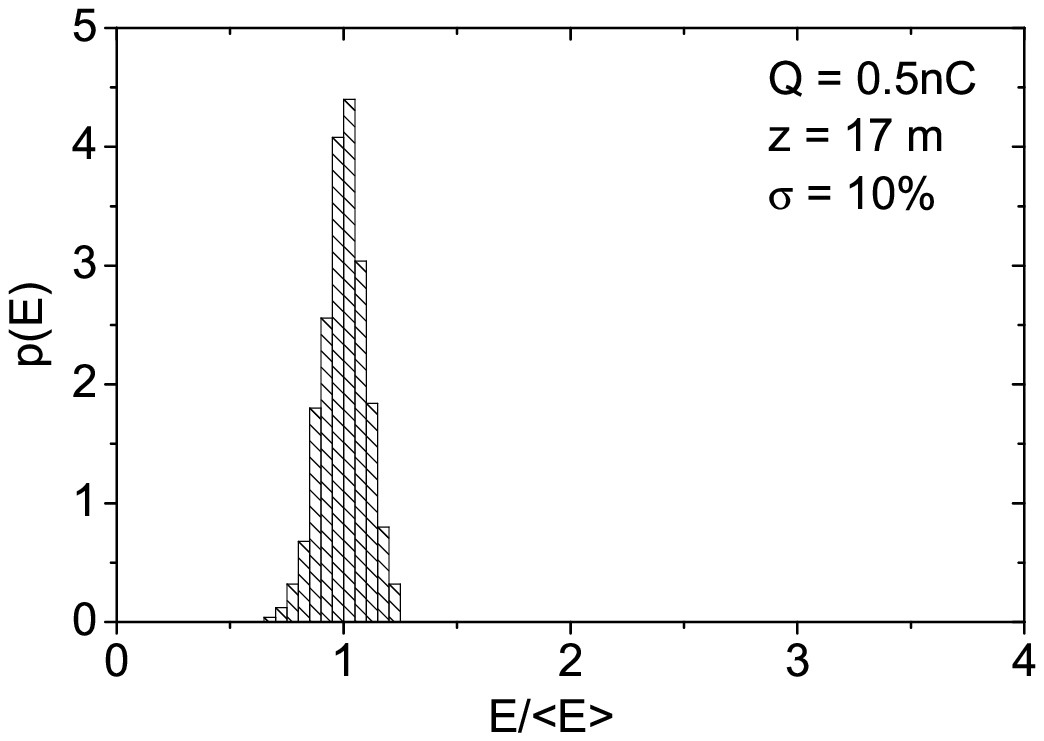}
\includegraphics[width=0.5\textwidth]{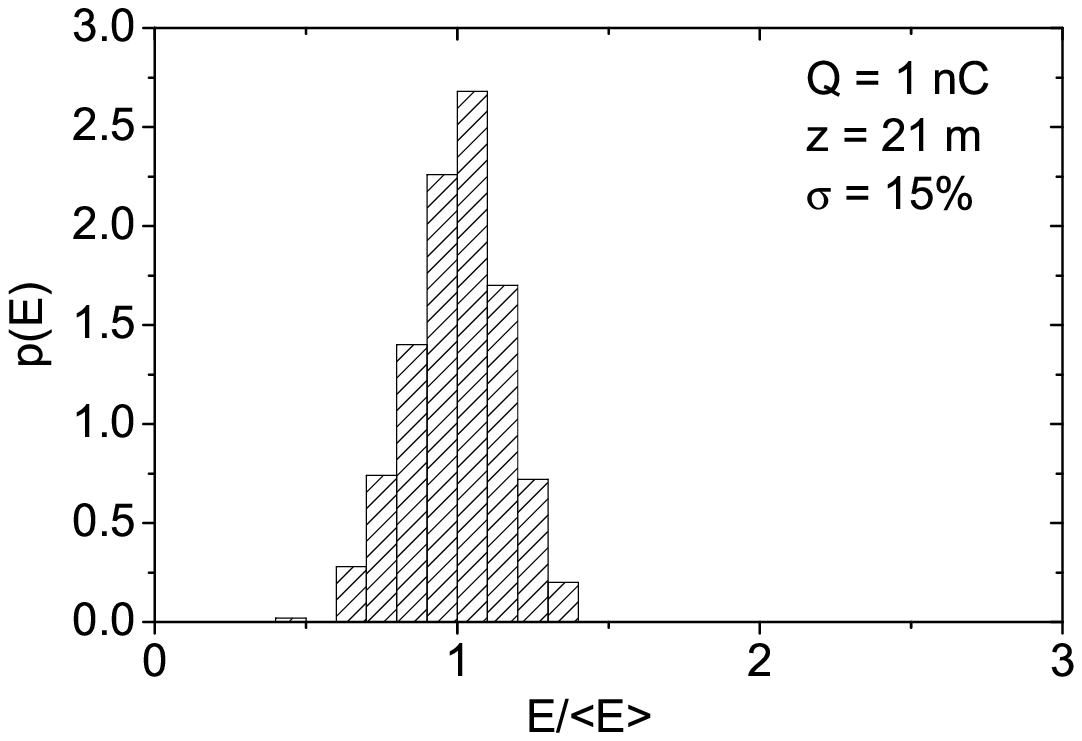}

\vspace*{-5mm}

\caption{
Evolution of the probability distribution from end of the
linear regime down to deep nonlinear regime.
Left and right columns correspond to the case of 0.5 and 1~nC,
respectively.
Solid lines show gamma distribution
}
\label{fig:hist-05}

\end{figure}

\begin{figure}[tb]

\includegraphics[width=0.5\textwidth]{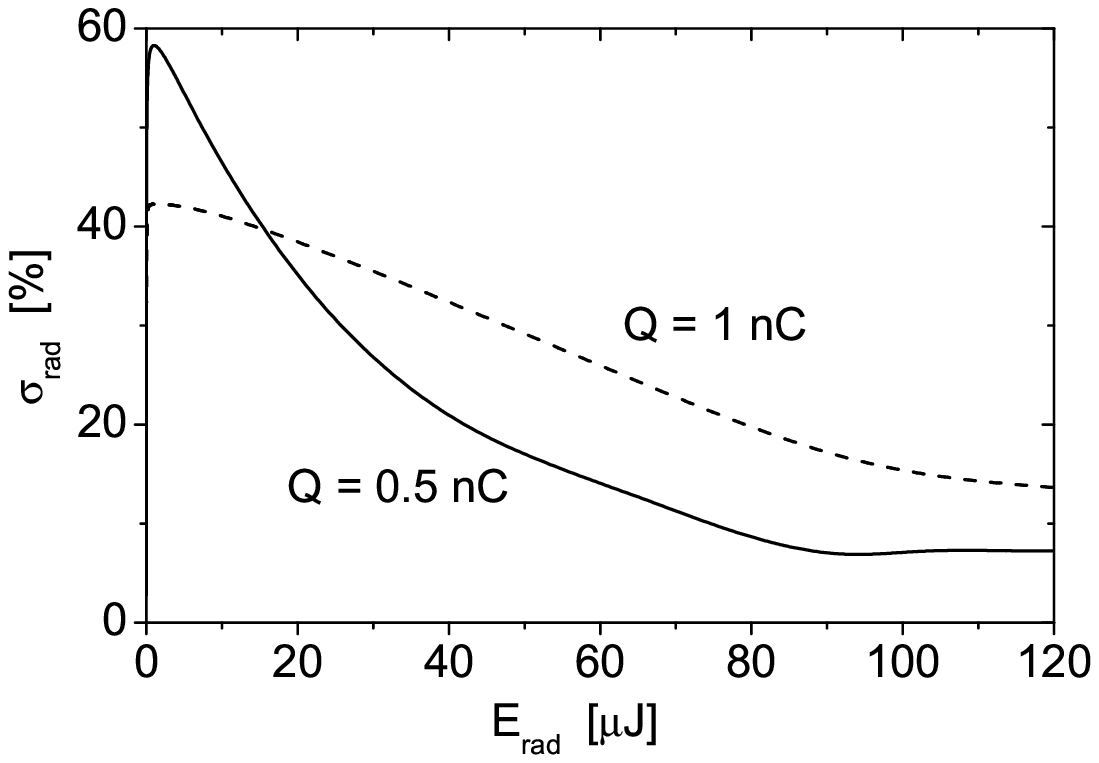}
\includegraphics[width=0.5\textwidth]{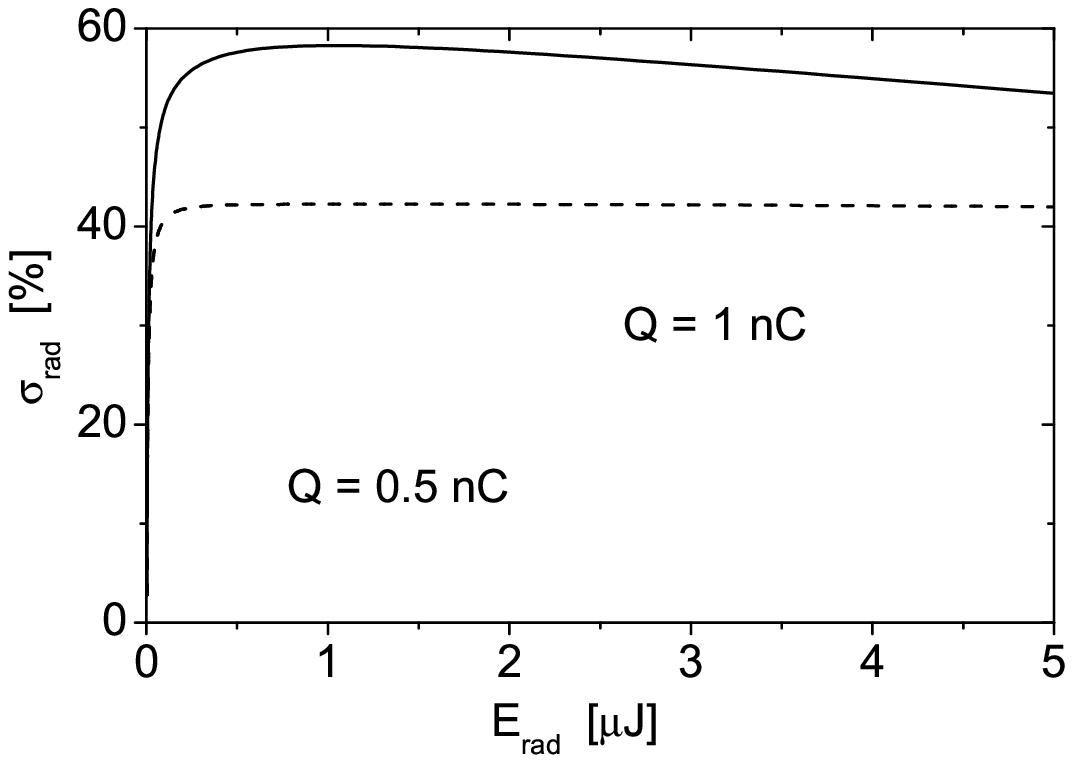}

\vspace*{-5mm}

\caption{
Fluctuations of the energy in the radiation pulse versus
average energy in the radiation pulse.
Right plot shows zoomed area of the left plot.
Solid and dashed lines refer to bunch charge 0.5 nC and 1~nC,
respectively
}
\label{fig:sigversuse}

\end{figure}

Amplification process in the SASE FEL starts from shot noise in the
electron beam, and the radiation pulse initially consists of a large
number of longitudinal and transverse modes. In the linear stage of
amplification the number of transverse modes in the radiation pulse
drops down due to the mode selection process. First this process passes
the stage exponential decay, and then asymptotically approaches to
single-mode case. Final degree of transverse coherence is limited by
finite spectrum width of the SASE FEL \cite{oc-sase-coherence}. The
number of longitudinal modes in the end of the linear is settled
proportionally to the ratio of the length of the lasing part of the
bunch to the coherence length. Thus, in the linear regime we occur
permanent reduction of the number of radiation modes which leads to
growth of the fluctuations of the radiation energy (see
Fig.~\ref{fig:pz}).

Fluctuations of the radiation energy change
drastically when amplification process enters the nonlinear regime. One
can see from Fig.~\ref{fig:pz} that fluctuations of the radiation
energy drop down on a scale of about one gain length. Note that such a
fast drop of fluctuations is the feature of ultra-short pulse duration
\cite{short-bunch}. Nature of this phenomenon can be understood by
analyzing structure of the radiation pulse (see
Figs.~\ref{fig:ptemp-05} and \ref{fig:ptemp-1}). When amplification
process enters nonlinear stage, group velocity of the radiation becomes
to be close to the velocity of light, and saturated wavepackets start
to slip forward with respect to the electron bunch. Further growth of
the total radiation energy occurs due to the radiation of the bunched
electron beam. Since maximum bunching of the electron beam is limited
to the unity, this additional radiation is well stabilized, leading to
the overall stability of the total energy in the radiation pulse.

Figure~\ref{fig:hist-05} presents evolution of the probability
distributions of the radiation energy along the undulator As we
mentioned above the radiation from SASE FEL operating in the linear
regime possesses properties of completely chaotic polarized light
\cite{book,stat-oc}. In particular, an important property is that
probability distribution of the energy in the radiation pulse must
follow gamma-distribution (\ref{gamma}). In the nonlinear regime
probability distributions change significantly. Specific shape of the
distribution becomes to be dependent on undulator length and details of
the lasing part of the electron bunch (temporal profile of current,
energy, emittance). However, we note one specific feature of the
distributions: in the nonlinear regime they exhibit an asymmetry with a
tail spanning to lower radiation energies. Somehow the shape looks like
an mirrored gamma distribution. Similar qualitative behavior has been
obtained also at the VUV FEL, phase I, in the simulations and in the
experiment as well \cite{s2e-nim,ttf1-stat}. Our experimental
experience is that this feature can be used as an effective tool for
reliable detection of the nonlinear regime even in the case of strong
fluctuations of machine parameters.

One more practically important result can be derived from the data
shown in Fig.~\ref{fig:pz}. This is the dependency of fluctuations
of the energy in the radiation pulse versus average energy in the
radiation pulse (see Fig.~\ref{fig:sigversuse}). Note that this
dependency is not affected strongly by imperfections of the machine
parameters. The first order effect of machine imperfections
is the gain length which is excluded from the relation between
fluctuations and average energy. Thus, measuring energy fluctuations in
the linear regime and in the nonlinear regime we can derive rather
precise on-line estimation for the average energy in the
radiation pulse.

\section{Spectral properties of the radiation}

\begin{figure}[b]

\includegraphics[width=0.5\textwidth]{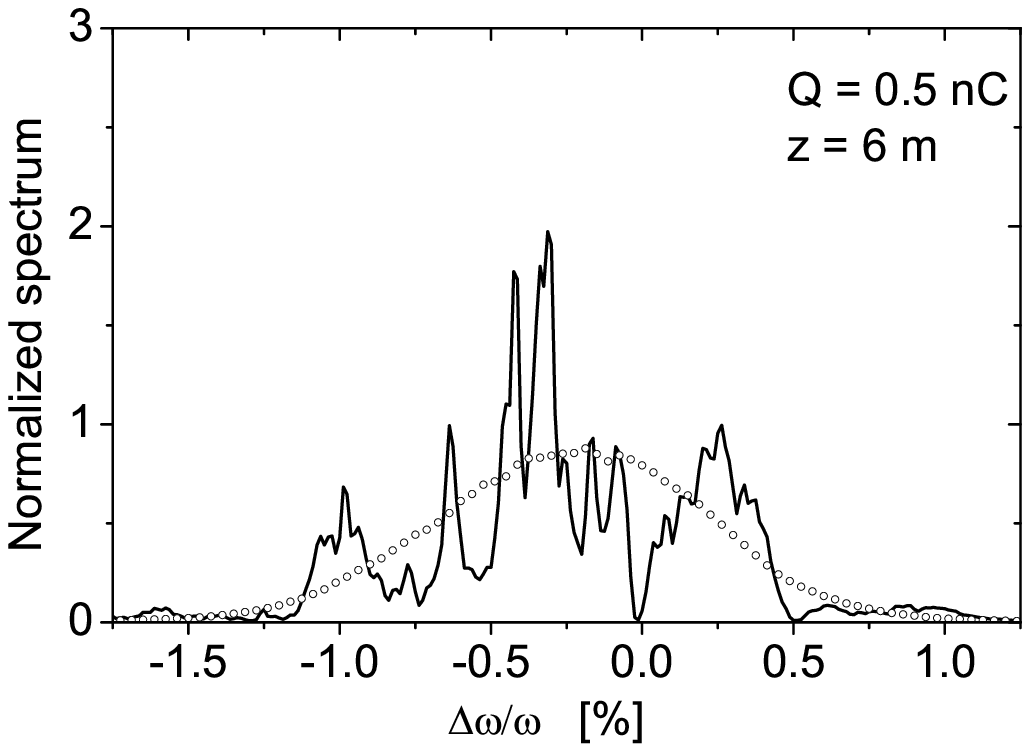}
\includegraphics[width=0.5\textwidth]{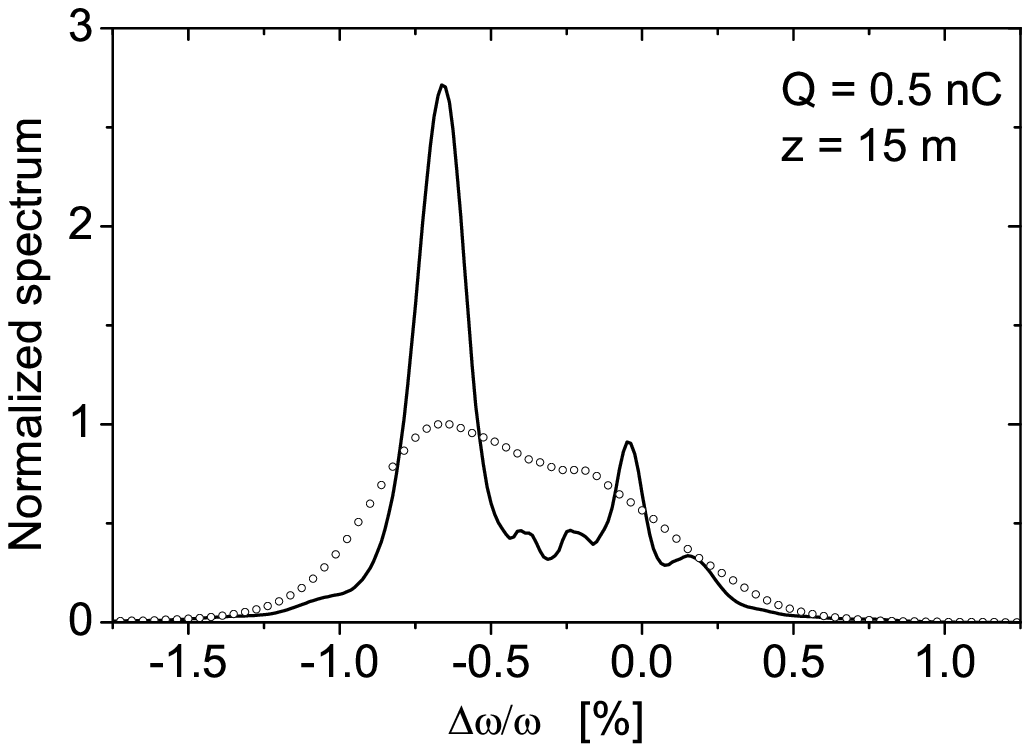}

\vspace*{-5mm}

\includegraphics[width=0.5\textwidth]{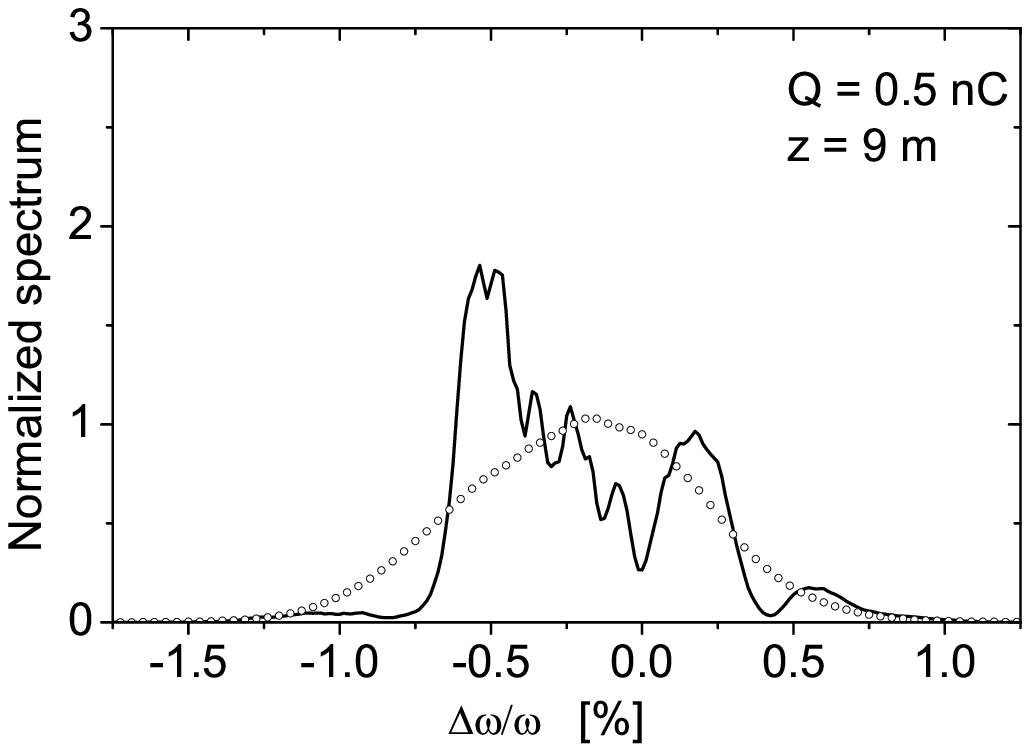}
\includegraphics[width=0.5\textwidth]{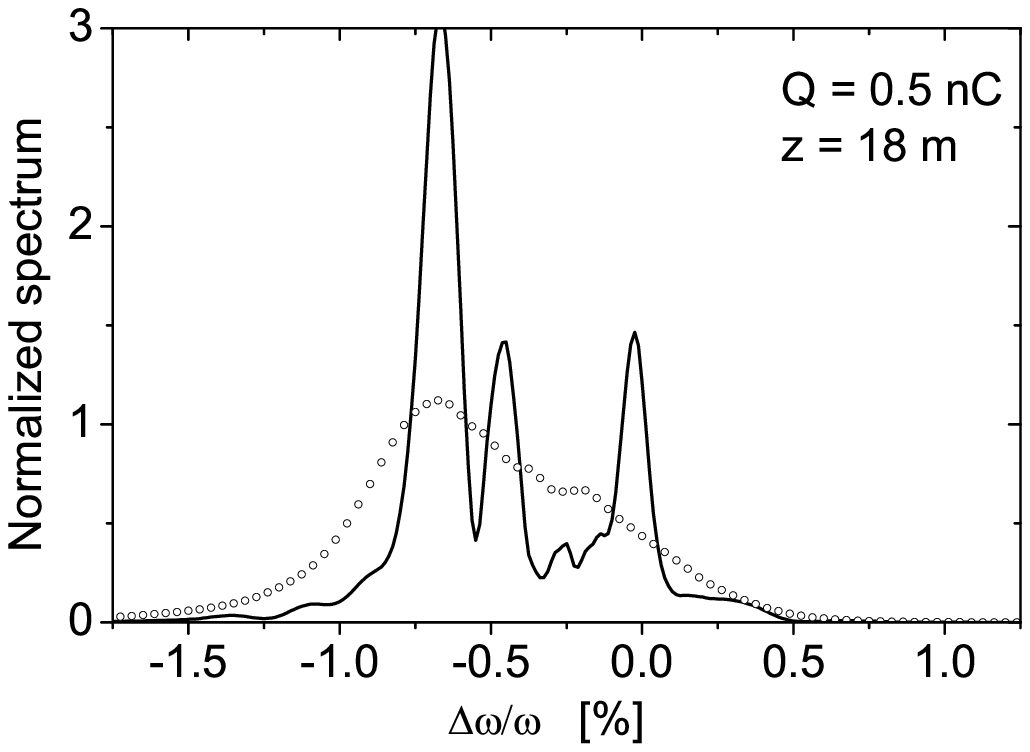}

\vspace*{-5mm}

\includegraphics[width=0.5\textwidth]{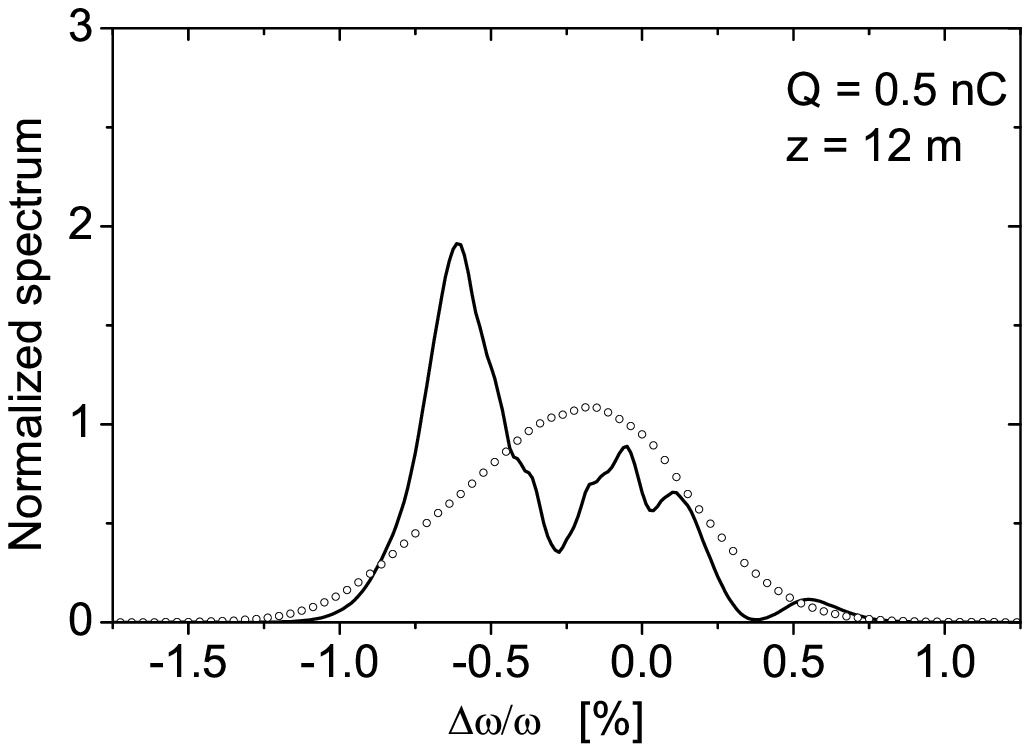}
\includegraphics[width=0.5\textwidth]{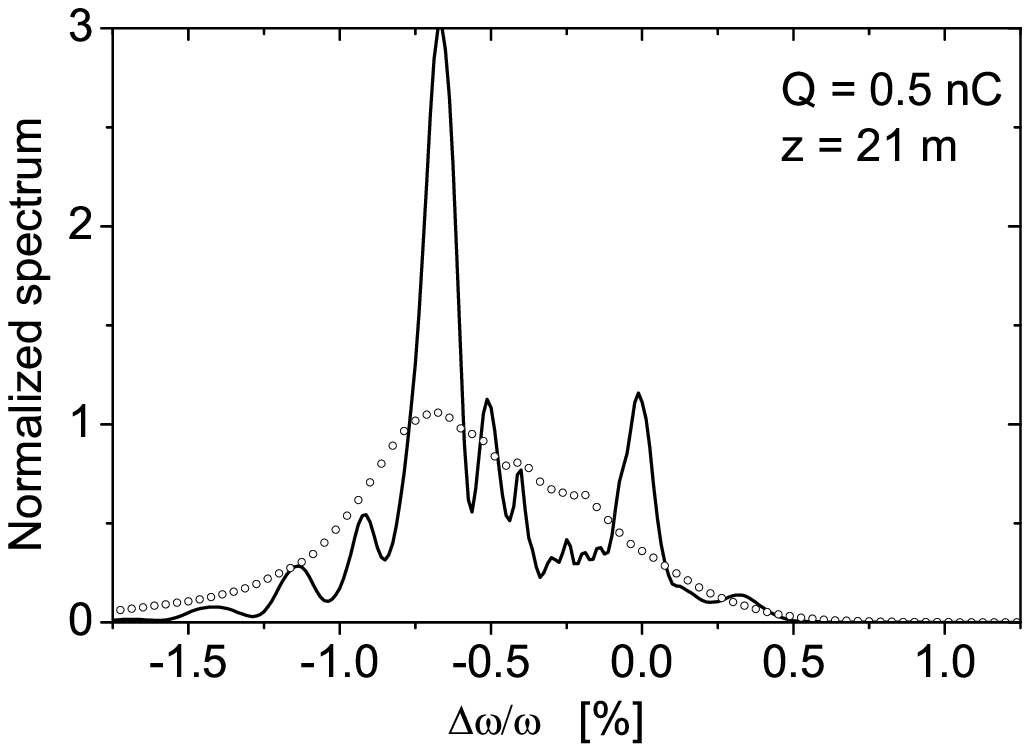}

\vspace*{-5mm}

\caption{
Evolution of spectral structure of the radiation pulse
along the undulator.
Solid line and circles correspond to a single pulse and
averaged profile, respectively.
Left and right columns correspond to linear and nonlinear mode of
operation, respectively. Bunch charge is 0.5~nC
}
\label{fig:spectr-05}
\end{figure}

\begin{figure}[tb]

\includegraphics[width=0.5\textwidth]{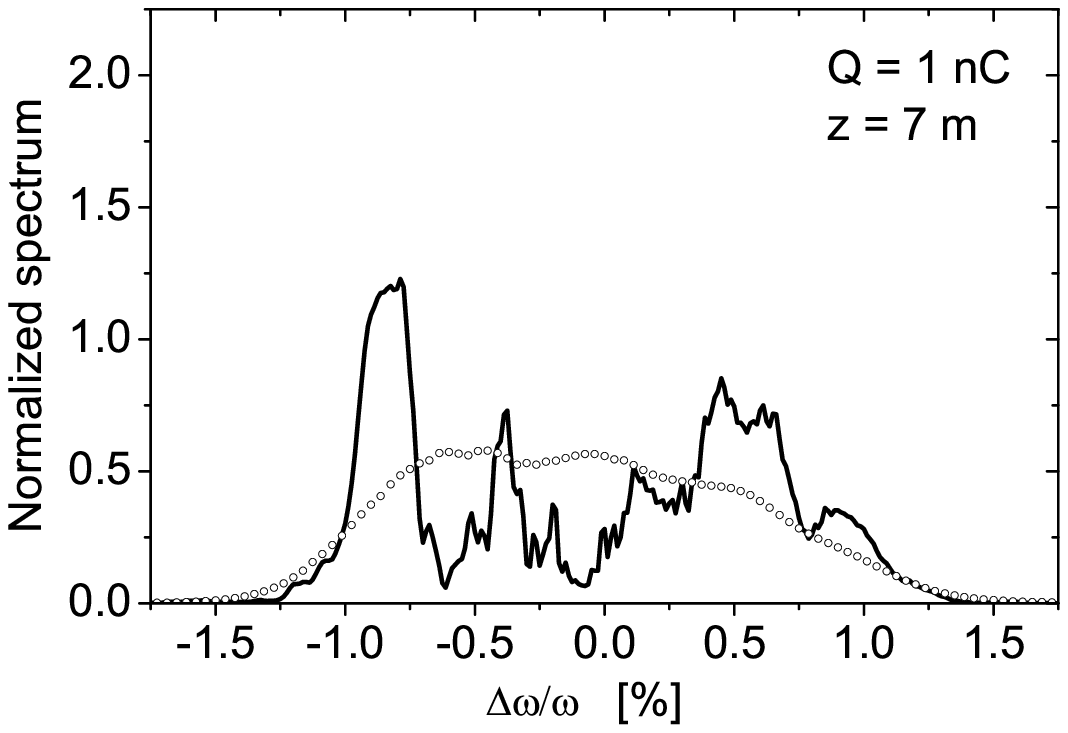}
\includegraphics[width=0.5\textwidth]{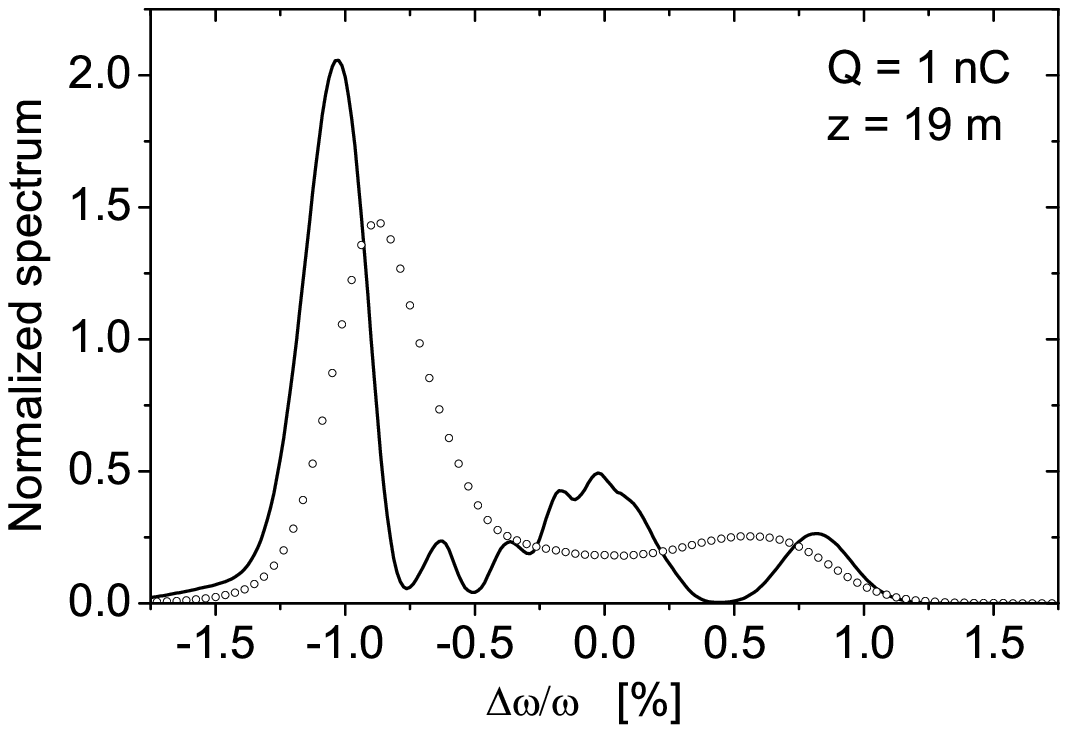}

\vspace*{-5mm}

\includegraphics[width=0.5\textwidth]{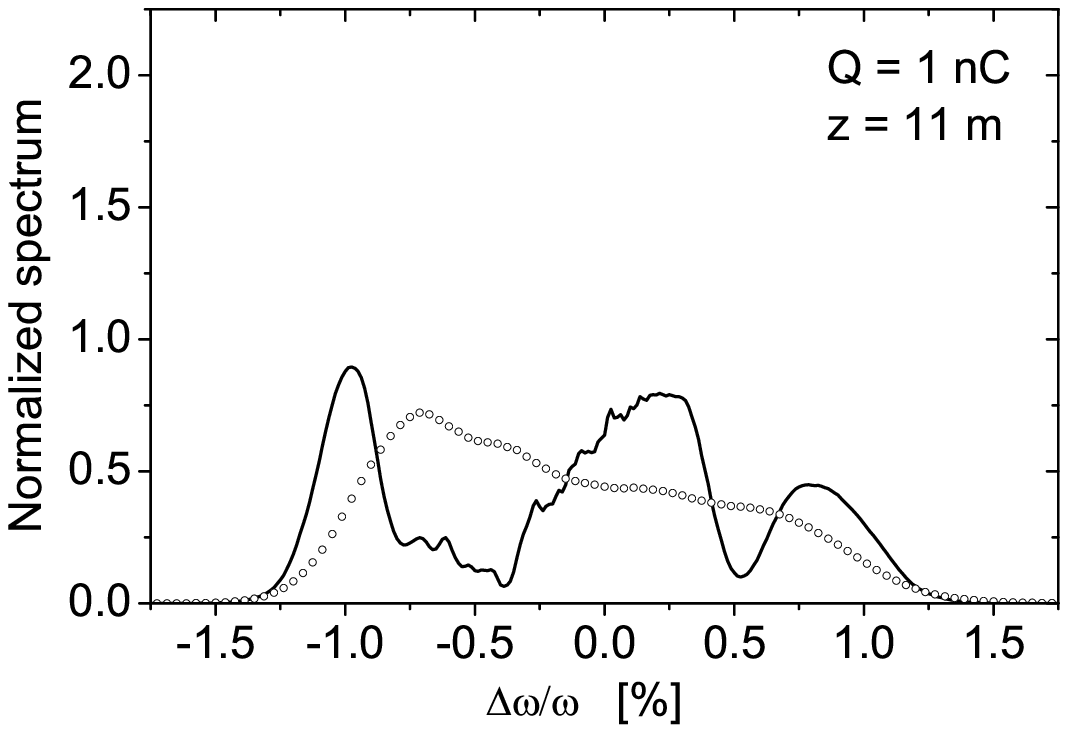}
\includegraphics[width=0.5\textwidth]{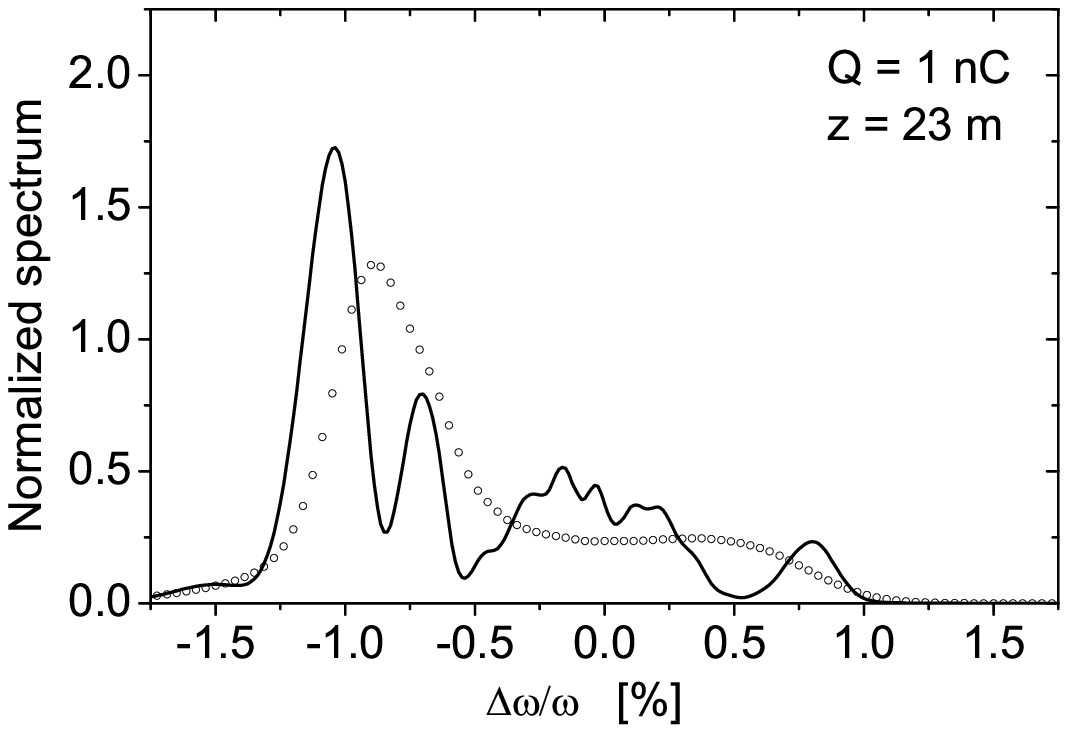}

\vspace*{-5mm}

\includegraphics[width=0.5\textwidth]{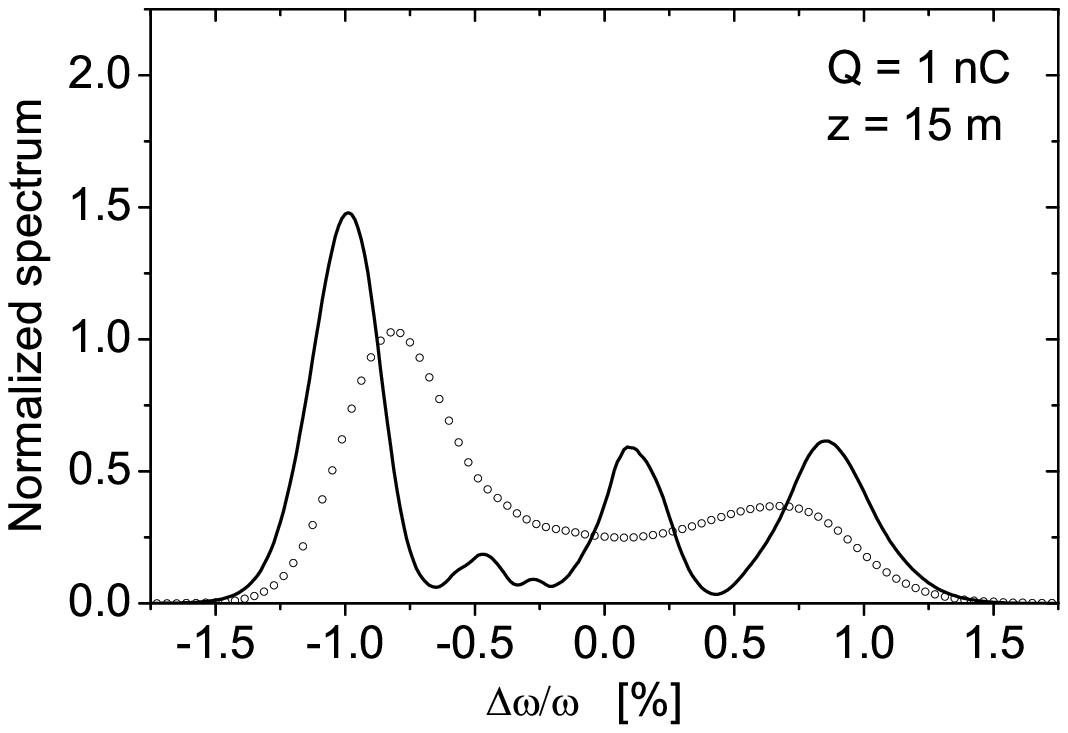}
\includegraphics[width=0.5\textwidth]{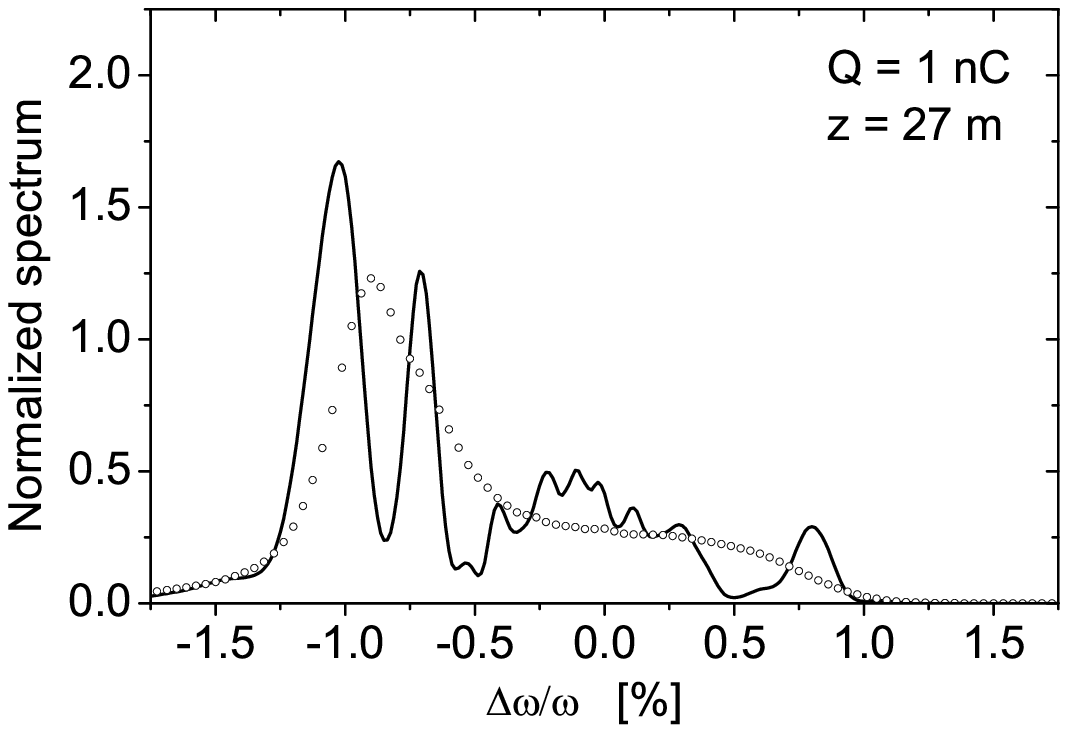}

\vspace*{-5mm}

\caption{
Evolution of spectral structure of the radiation pulse
along the undulator.
Solid line and circles correspond to a single pulse and
averaged profile, respectively.
Left and right columns correspond to linear and nonlinear mode of
operation, respectively. Bunch charge is 1~nC
}
\label{fig:spectr-1}
\end{figure}

\begin{figure}[tb]

\includegraphics[width=0.5\textwidth]{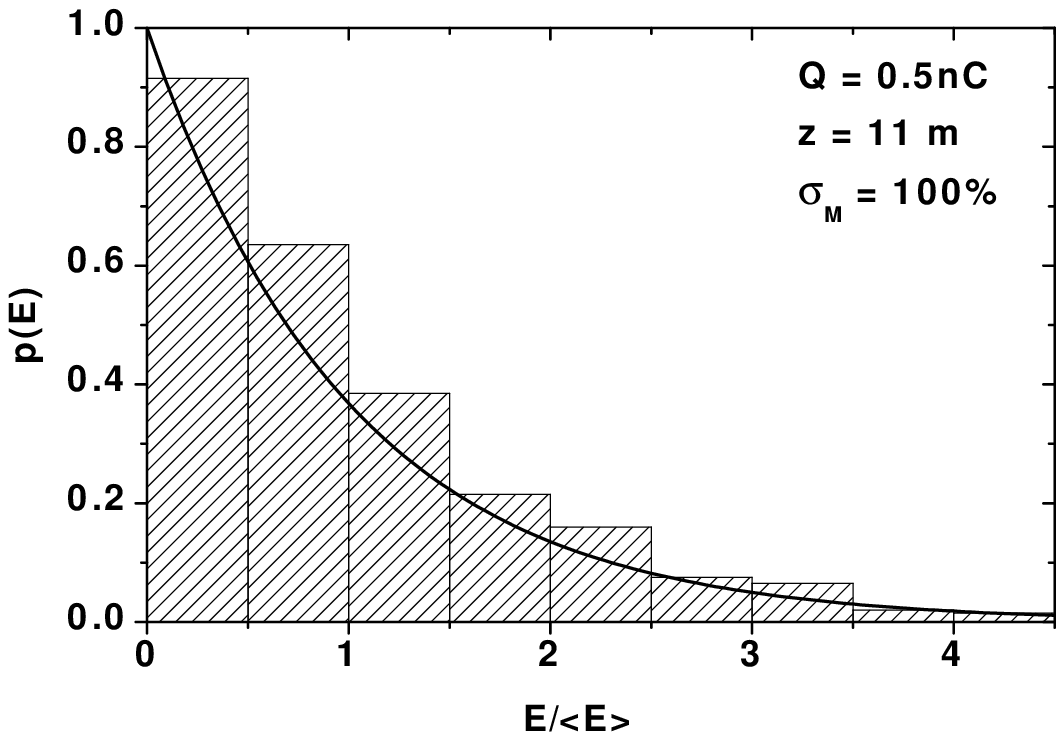}
\includegraphics[width=0.5\textwidth]{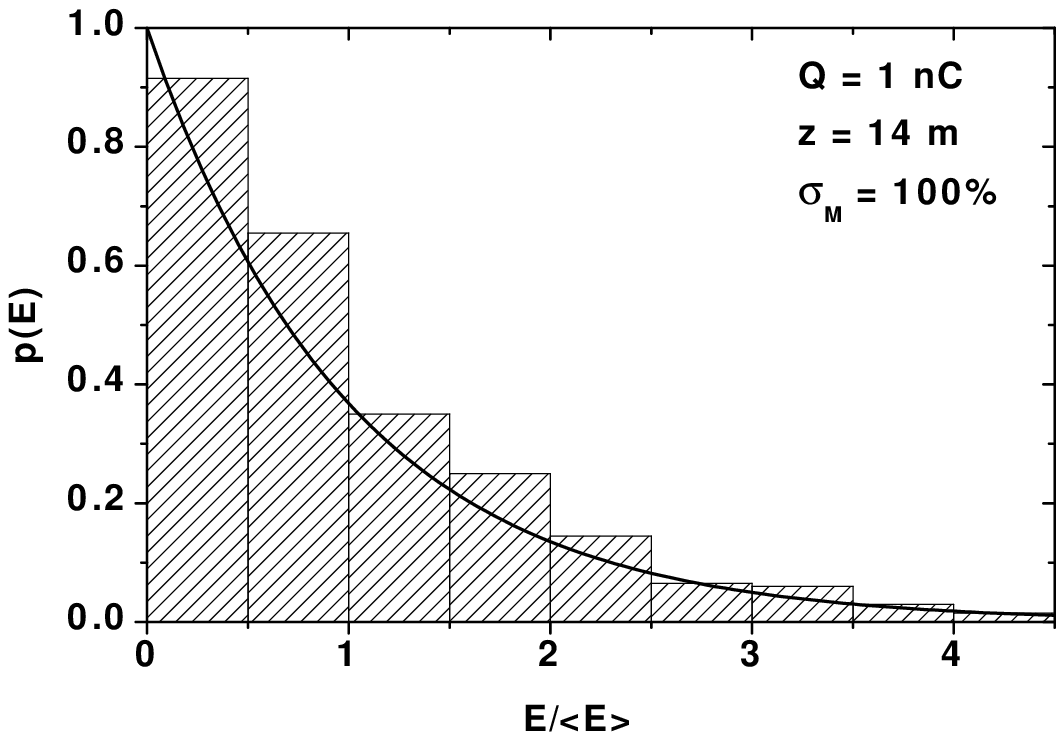}

\vspace*{-5mm}

\includegraphics[width=0.5\textwidth]{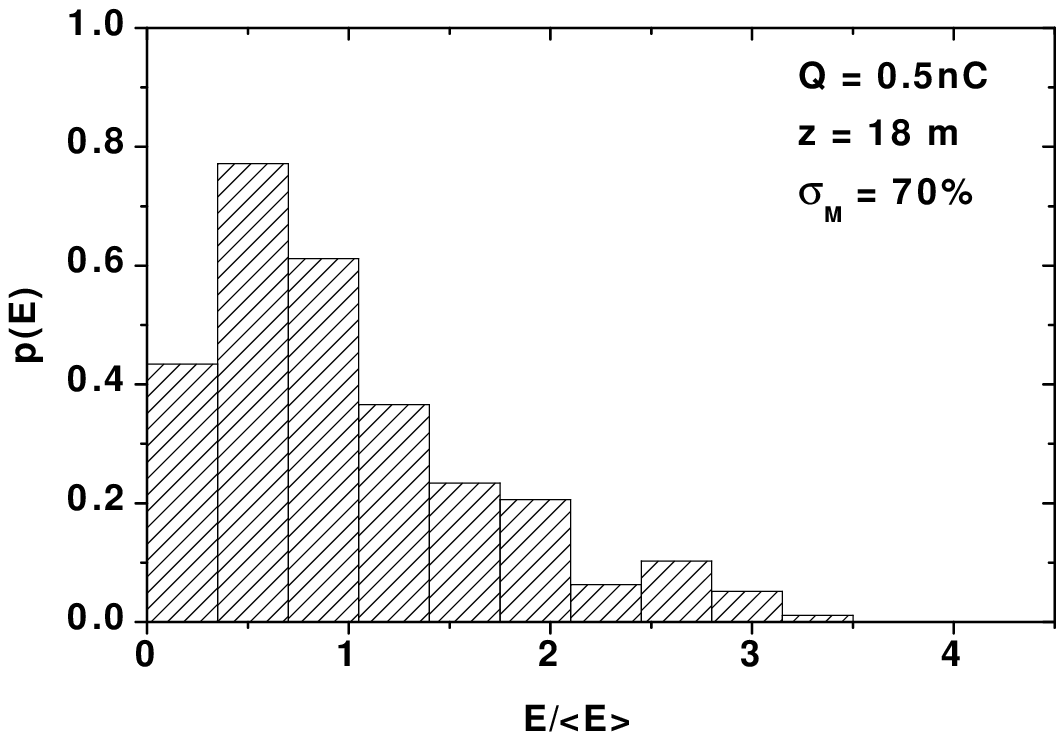}
\includegraphics[width=0.5\textwidth]{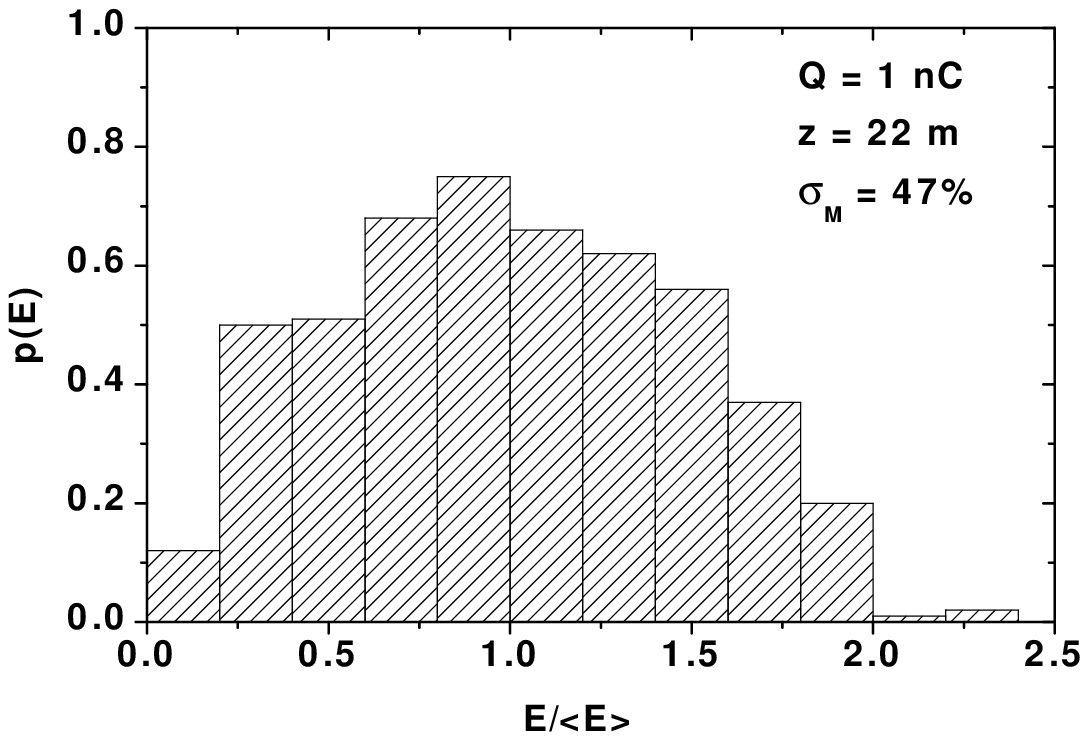}

\caption{
Probability distribution of the energy in the radiation pulse after
narrow band monochromator.
Top and bottom plots
correspond to linear regime and saturation, respectively.
Left column: bunch charge is 0.5~nC.
Right column: bunch charge is 1~nC.
Solid line shows negative exponential distribution
}
\label{fig:mon-prob}
\end{figure}

Figures~\ref{fig:spectr-05} and \ref{fig:spectr-1} show evolution of
the radiation spectra along the undulator. Single-shot spectra are
derived from the data shown in Figs.~\ref{fig:ptemp-05} and
\ref{fig:ptemp-1}. The reference energy for the resonance frequency
corresponds to the maximum of the peak current (see
Fig.~\ref{fig:bunch-ue}). Calculation for a slice corresponding to
maximum of current spike gives the value for the FEL parameter $\rho =
2.5 - 3 \times 10^{-3}$ for maximum of peak current (see
Fig.~\ref{fig:bunch-ue}). Thus, spectrum width generated by a slice of
the electron beam is expected to be about $\Delta \omega /\omega \simeq
2 \rho \simeq 0.5 - 0.6$\%. Calculated spectrum width is visibly wider.
The origin of this phenomenon is in strong energy chirp along the
lasing fraction of the bunch. Using data from Fig.~\ref{fig:bunch-ue}
we can easily estimate extra widening of the radiation spectrum by
$0.5-1$\% according to simple relation $\Delta \omega /\omega = 2
\Delta E/E$. In particular, more wide spectrum for the case of 1~nC is
due to larger energy chirp with respect to 0.5~nC case.

Another subject is statistics of SASE FEL radiation filtered through
narrow-band monochromator. In the linear stage of SASE FEL operation
the value of normalized energy deviation is equal to unity, and energy
fluctuates in accordance with negative exponential distribution:

\begin{equation}
p(E) = \exp \left( -\frac{E}{\langle E\rangle } \right) \ .
\label{neg-exp}
\end{equation}

\noindent This is consequence of the fact that the radiation from SASE
FEL operating in the linear regime is gaussian random process. This
property remains valid for any pulse length. When amplification process
enters nonlinear stage the radiation is not gaussian random process
anymore due to the process of sideband growth in the nonlinear media.
In particular, the probability distribution of the total radiation
energy does not follow gamma distribution anymore as it was shown in
the previous section. Situation with probability distribution of the
radiation energy after narrow band monochromator is more complicated.
Earlier studies have shown that in the case of long radiation pulse
(much longer than coherence length) property of the negative
exponential distribution remains to be valid in the nonlinear regime as
well \cite{book,stat-oc}. Situation changes dramatically when pulse
duration becomes to be short, of about coherence length. Recent studies
have shown that in this case an effect of strong suppression of the
fluctuations of the radiation energy after narrow band monochromator is
expected \cite{short-bunch}. Later on this effect has been measured
experimentally at the VUV FEL, phase I \cite{ttf1-stat}. Our
simulations show that suppression of the fluctuations is expected for
the VUV FEL as well (see Fig.~\ref{fig:mon-prob}). This effect has
simple physical explanation \cite{short-bunch}. Let us consider an
extreme case of SASE FEL driven by an electron bunch of about coherence
length. For this extreme case each radiation pulse consists of a
single spike only. For different shots the radiation pulses have similar
shape, but amplitude fluctuates nearly by negative exponential
distribution. When amplification process enters nonlinear stage,
amplitudes of different pulses are equalized due saturation effects,
while keeping close shape. Spectrum of the radiation pulse is given by
Fourier transform of the radiation field, and at saturation we obtain
nearly similar spectrum envelope for different pulses. As a result, we
can expect that fluctuations of the radiation energy after narrow-band
monochromator should follow fluctuations of the total energy in the
radiation pulse which drop down drastically. For longer radiation
pulses suppression is not so strong and nearly vanishes for the bunches
longer than four coherence length. In the case under study the lasing
part of the bunch is about 2-3 times longer than coherence length.

\section{An overview of the first experimental results}

This work (statistical calculations of the radiation properties) has
been performed in parallel with the first run of the VUV FEL. The range
of optimum tuning of the VUV FEL proposed in \cite{vuvfel-th} and
described here has been chosen for the commissioning of the VUV FEL. A
consistent set of experimental data has been recorded for the case when
VUV FEL has been driven by electron bunches with 1~nC charge.
Average energy in the radiation pulse was equal to 1~$\mu $J
which corresponds to the high gain linear regime \cite{vuvfel-exp}.

\begin{figure}[b]

\includegraphics[width=0.5\textwidth]{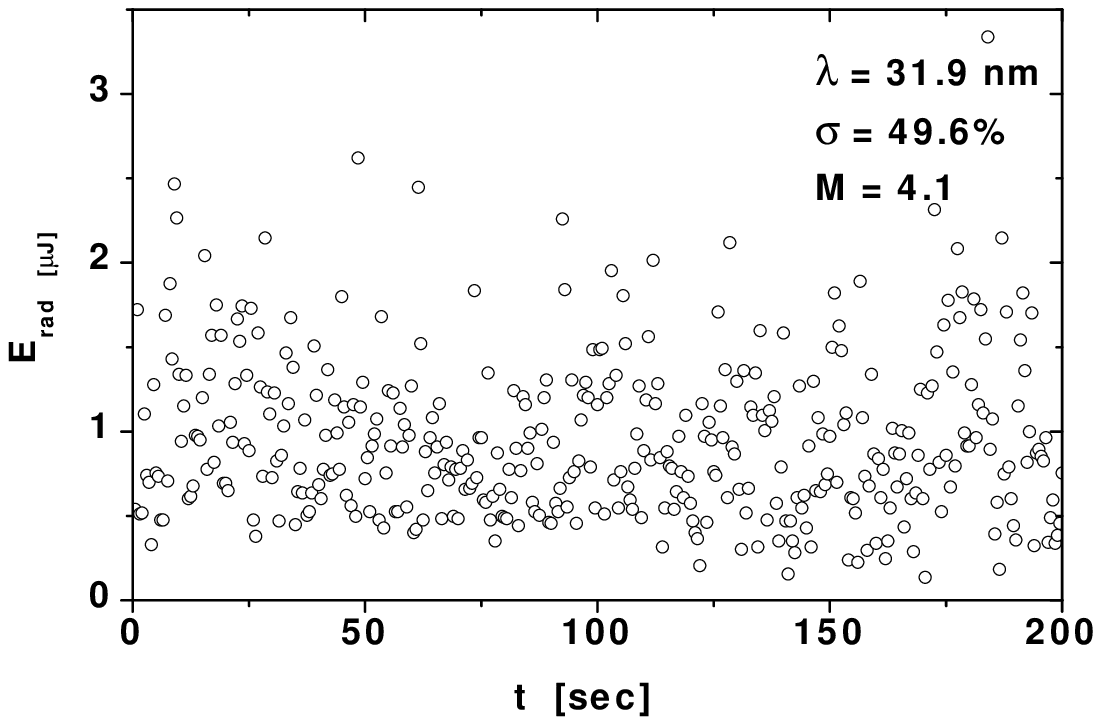}
\includegraphics[width=0.5\textwidth]{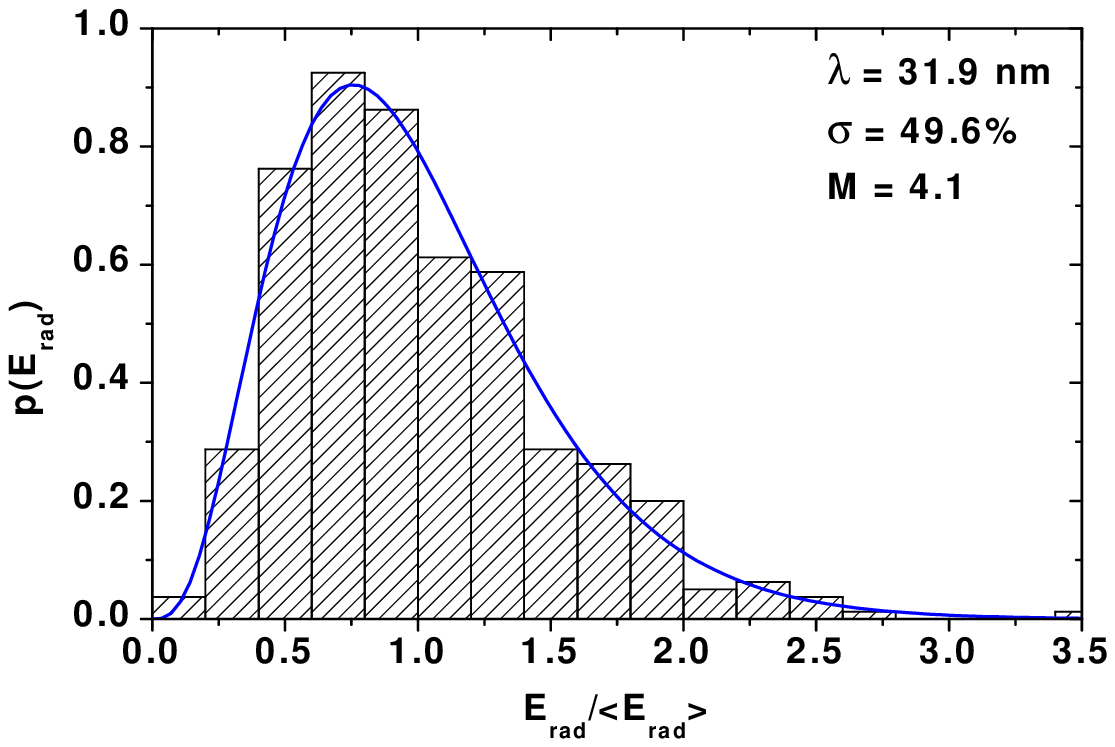}

\caption{
Experimental results from the VUV FEL.
Measurements of the
energy in the radiation pulse.
VUV FEL operates in the linear regime. Left plot -- measured
energy versus time.  Right plot -- probability distribution of the
radiation energy. Solid curve represents gamma distribution
}
\label{fig:hist-lin}
\end{figure}

Plot on the left side of Fig.~\ref{fig:hist-lin} shows measured values
of the radiation energy versus time. Corresponding probability
distribution derived from experimental data is shown on the right plot.
We see that experimentally measured distribution follows well gamma
distribution with parameter $M = 4.1$ derived from experimental data as
$M = 1/ \sigma_E^2$. Note that measured fluctuations are mainly
fundamental fluctuations due to start-up of the FEL process from the
shot noise. Contribution of machine fluctuations is estimated to be
less than 10\%. Measured value for the number of modes in the radiation
pulse is consistent with theoretical prediction (see
Fig.~\ref{fig:pz}).

\begin{figure}[tb]

\includegraphics[width=0.5\textwidth]{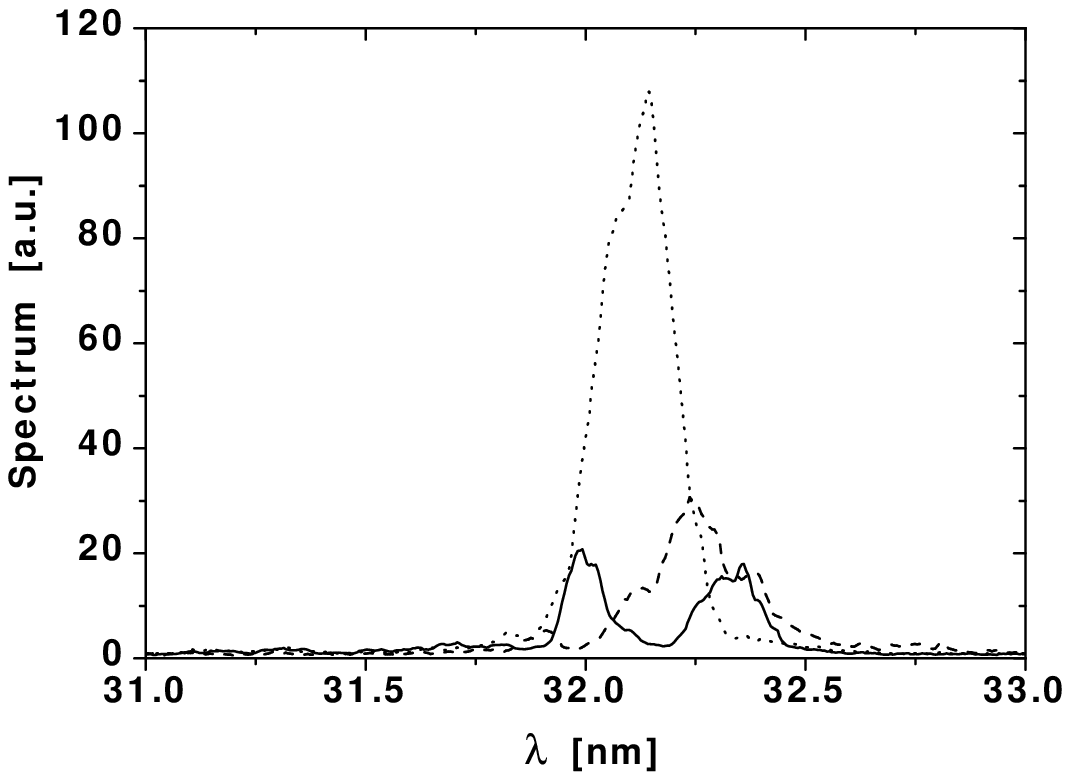}
\includegraphics[width=0.5\textwidth]{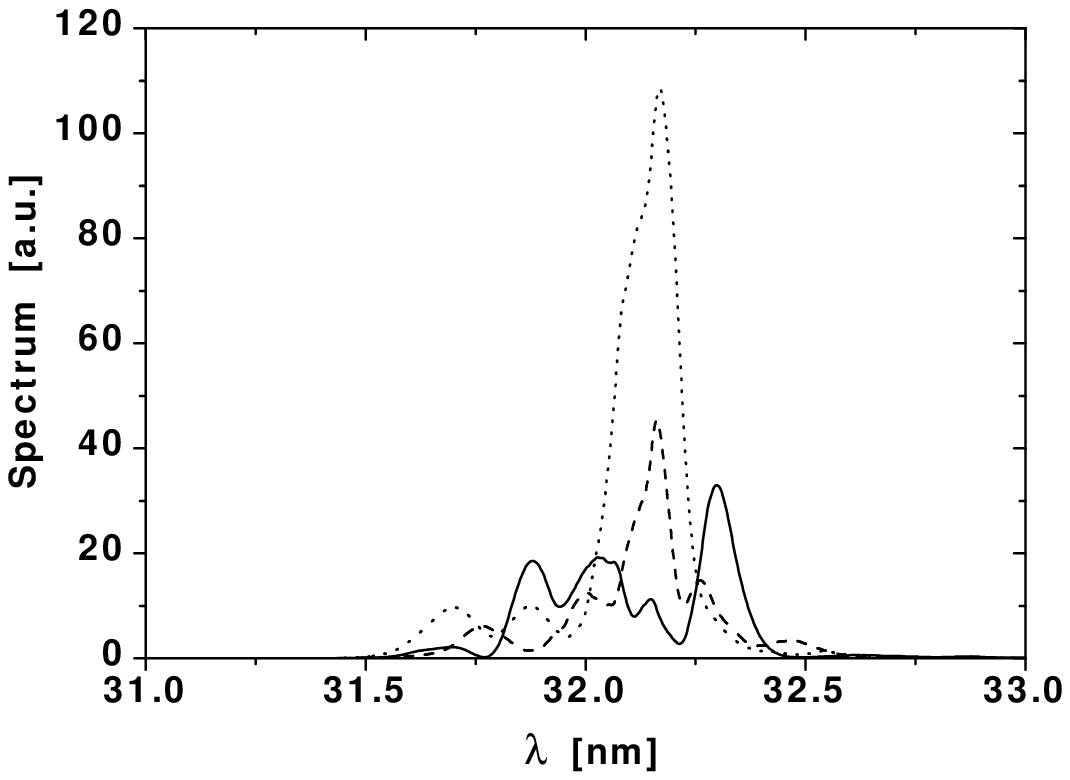}

\caption{
Experimental results from the VUV FEL.
Left: Single-shot spectra of radiation  pulses (experimental results).
Right: simulation of spectral structure of
the radiation pulse
from VUV FEL operating
in the linear regime at the undulator length of 13~m \cite{vuvfel-th}
}

\label{fig:spec-exp}

\end{figure}

Spectral measurements are presented in Fig.~\ref{fig:spec-exp}
\cite{vuvfel-exp}. Single-shot spectra were taken with a monochromator
(0.04~nm resolution) equipped with an intensified CCD camera
\cite{monochr}. Analysis of measured spectra shows that there is good
agreement with early theoretical predictions \cite{vuvfel-th} (see
right plot in Fig.~\ref{fig:spec-exp}). Ultra-short pulse duration does
not allow direct measurements of temporal structure of single pulses,
but measurements of single-pulse spectra are possible. Monochromator
performs Fourier transform of the radiation pulse. Despite the phase is
missed in such measurements, they still contain an essential
information about primary object. In fact, average number of spikes in
the single-shot spectra should correspond to average number of spikes
in the temporal structure. Thus, visual analysis of single shot spectra
allows us to derive a zero-order estimate for the number of
longitudinal modes. Since monochromator acts as a Fourier
transformation, we can state that typical width of the spike in the
spectrum $\Delta \omega$ should be inversely proportional to the
radiation pulse duration. Using the relation $\tau _{\mathrm{rad}} \sim
2\pi /(\Delta \omega )$ we can give an estimation for the pulse
duration. Measured spike widths presented correspond to the radiation
pulse length $20-40$~fs which is in a good agreement with theoretical
results.

\begin{figure}[tb]

\hspace*{3mm}
\includegraphics[width=0.3\textwidth]{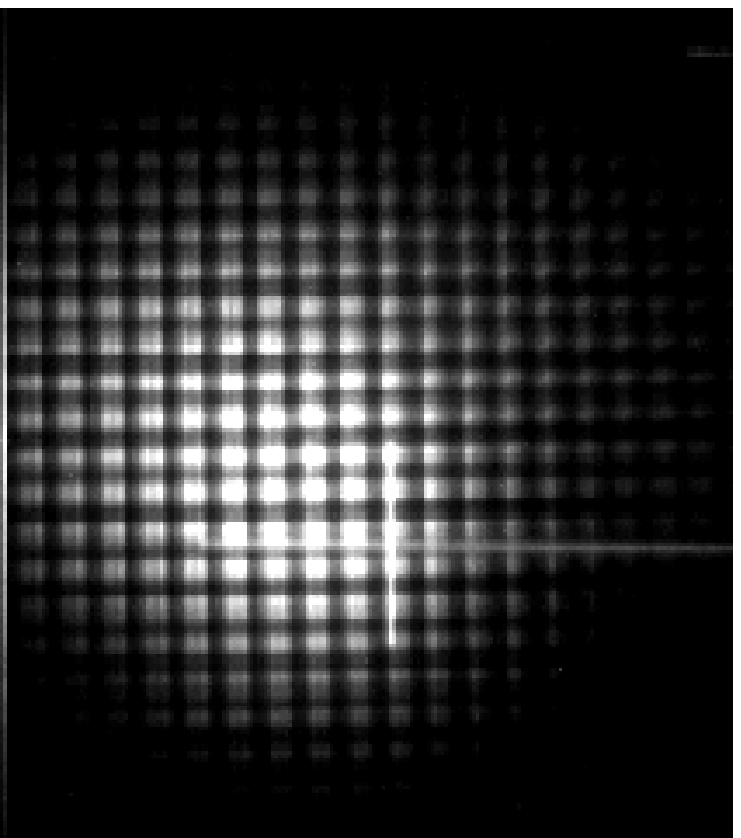}
\hspace*{5mm}
\includegraphics[width=0.115\textwidth]{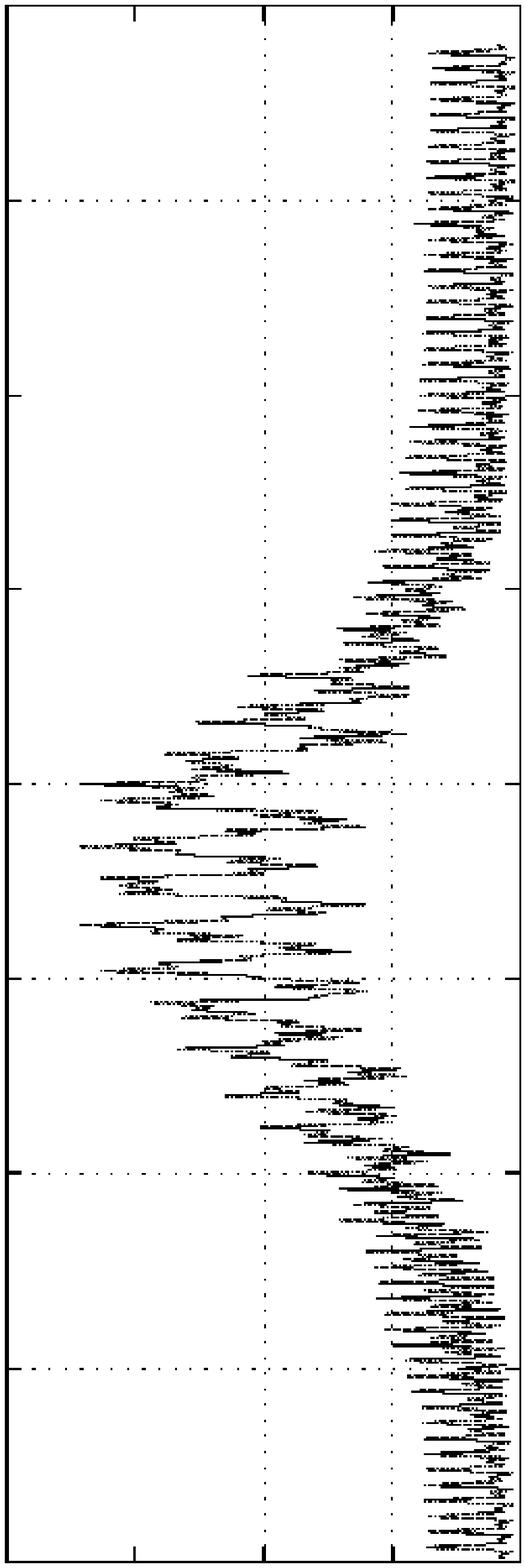}
\hspace*{5mm}
\includegraphics[width=0.5\textwidth]{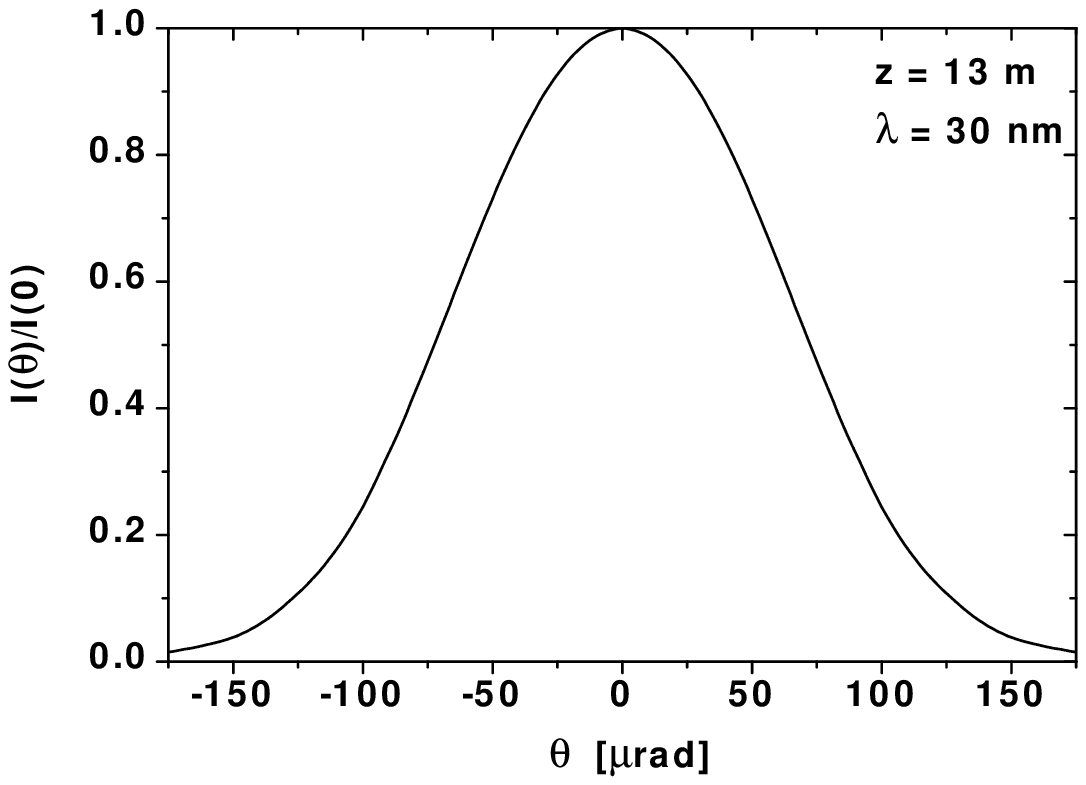}

\caption{
Experimental results from the VUV FEL.
Left plot: image on CeYaG crystal of a single-shot VUV FEL pulse
passed through gold mesh of the radiation detector
placed 18.5~m downstream undulator exit.
Distance between wires in the mesh is 0.25~mm, wire diameter is
60~$\mu $m.
Image analysis (middle
plot) gives FWHM spot size of 3~mm which corresponds to FWHM angular
divergence of 160~$\mu $rad. Right plot: simulated angular distribution
of the radiation intensity in the far zone \cite{vuvfel-th}
}
\label{fig:dir}
\end{figure}

Figure~\ref{fig:dir} shows an image on CeYaG crystal of a single-shot
VUV FEL pulse passed through a gold mesh of the radiation detector
placed 18.5~m downstream undulator exit \cite{mcp-detector}. Distance
between the centers of wires in the mesh is 0.31 mm. Image analysis
gives FWHM spot size of 3 mm which corresponds to FWHM angular
divergence of 160 $\mu $rad. Right plot shows simulated angular
distribution of the radiation intensity in the far zone
\cite{vuvfel-th}. We find good agreement between simulated and
experimental data.

Measurements of angular divergence provide valuable information about
the degree of transverse coherence. FWHM spot size of the photon beam
at the undulator exit is about 250 $\mu $m \cite{vuvfel-th}. One can
easily calculate that product of the spot size of the radiation at the
undulator exit by measured angular divergence results in the value of
about radiation wavelength. This means that the phase volume of the
radiation is close to diffraction limit, and the radiation from the VUV
FEL has high degree of transverse coherence.

\section{Summary}

An important lesson from our study and the first operation of the VUV
FEL is that GW-level, laser-like VUV radiation pulses on a sub-50 fs
scale are produced with a simple and reliable single-pass SASE FEL
scheme. The generation of ultra-short radiation pulses became possible
due to specific tailoring of the bunch charge distribution. Such a
technique is more reliable and effective with respect to sophisticated
HGHG schemes widely discussed as a future alternative to single-pass
SASE FEL in terms of shorter pulse duration and better quality of
output radiation (see, e.g. \cite{hghg-france-japan} and references
therein). Key element of the VUV FEL operating in the femtosecond
regime is nonlinear bunch compression scheme producing tailored
electron bunches with a short high-current leading peak in the density
distribution that produces FEL radiation. An important feature of the
beam formation system is strong influence of collective effects. One of
them, longitudinal space charge, plays an extremely important role:
induced energy chirp along the lasing part of the bunch allows further
shortening of the lasing fraction of the electron bunch. As a result,
it becomes possible to generate ultra-short, down to 20~fs radiation
pulses with GW-level peak power and contrast of 80\%.

\section*{Acknowledgments}

We are grateful to J.R.~Schneider for interest in this work and
stimulating discussions.

\end{document}